\documentclass[amsmath,amssymb]{revtex4}

\usepackage{amsmath}
\usepackage{graphicx}

\usepackage[all]{xy}
\usepackage[section]{placeins}
 \,
 \,
 \,
 \,

\renewcommand{\vec}[1]{\mbox{\boldmath $#1$}}

\newcommand{\g}{\vec{\Gamma}}

\begin{document}

\title{Covariant Lyapunov Vectors for Rigid Disk Systems}

\author{Hadrien Bosetti}
\email{Hadrien.Bosetti@univie.ac.at}
\affiliation{Computational Physics Group, Faculty of Physics, University of Vienna,
 Boltzmanngasse 5, A-1090 Wien, Austria}
\author{Harald A. Posch}
\email{Harald.Posch@univie.ac.at}
\affiliation{Computational Physics Group, Faculty of Physics, University of Vienna,
 Boltzmanngasse 5, A-1090 Wien, Austria}

\date{\today}

\begin{abstract}
      We carry out extensive computer simulations to study the Lyapunov instability of a 
two-dimensional hard disk system in a rectangular box with periodic boundary conditions. 
The system is large enough to allow the formation of Lyapunov modes parallel to the 
$x$ axis of the box. The Oseledec splitting into covariant subspaces of the tangent space is   
considered by computing the full set of covariant perturbation vectors co-moving with 
the flow in tangent-space. These vectors are shown to be transversal, but generally not
orthogonal to each other. Only the angle between covariant vectors associated with
immediate adjacent Lyapunov exponents in the Lyapunov spectrum may become small,
but the probability of this angle to vanish approaches zero. The stable and unstable manifolds
are transverse to each other and the system is hyperbolic.
\end{abstract}
\maketitle
\section{Introduction}

Lyapunov exponents measure the exponential growth, or decay, of infinitesimal phase
space perturbations of a chaotic dynamical system.  For a $D$-dimensional phase space, there are
$D$ exponents, which, if ordered according to size, $\lambda_i \ge \lambda_{i+1}$, are
referred to as the Lyapunov spectrum. The classical algorithm for the computation is based on the fact that almost all volume elements of dimension $d \le D$ in tangent space (with the exception of elements
of measure zero)  asymptotically evolve  with an exponential rate which is equal to the sum of the 
first $d$ Lyapunov exponents.  Such a  $d$-dimensional subspace may be spanned by 
$d$ orthonormal vectors, which may be constructed by the Gram-Schmidt procedure and,
therefore,  are referred to as Gram-Schmidt (GS) vectors. 
The GS-vectors are not covariant, which means that at any point in phase space they are not mapped 
by  the linearized dynamics into the GS vectors at the forward images of that point \cite{Ginelli}.
As a consequence, they are not invariant with respect to the time-reversed dynamics. Due
to the periodic re-orthonormalization of the GS vectors only the radial dynamics is exploited 
for the computation of the exponents, whereas the angular information is discarded. 

Although the angular dynamics is not a universal property and may depend, for example, on the choice of the coordinate system \cite{HHP:1990}, it would be advantageous for many applications, to span the 
subspaces mentioned above by covariant vectors and to study also the angular dynamics of and 
between these vectors. It has the additional advantage to preserve 
the time-reversal symmetry for these tangent vectors, a property not displayed by the GS vectors. 
Recently, an efficient  numerical procedure was developed by Ginelli {\em et al.} \cite{Ginelli}
for the computation of covariant Lyapunov vectors.
Here we apply their algorithm to a two-dimensional system of rigid disks.

The choice of hard elastic particles is motivated by the fact that their dynamics is
comparatively simple, and their ergodic, structural and dynamical properties are well
known and are thought to be typical of more realistic physical systems \cite{Szasz}.  Secondly, 
hard-particle systems in two and three dimensions serve as reference systems for the most 
successful perturbation theories of dense gases and liquids \cite{Barker,Hansen}. Finally, the combination of
a Lyapunov analysis with novel statistical methods for rare events \cite{Dellago} seems
particularly promising for the study of such rare transformations in systems, for which hard core 
interactions are at the root.

The paper is organized as follows. After an introduction of the basic concepts for the
dynamics of phase space perturbations in Section \ref{section_phase},
we summarize  in Section \ref{numerics} the features and our numerical  implementation of the 
algorithm of Ginelli {\em et al.} \cite{Ginelli} for the computation of covariant
vectors and covariant subspaces. In Section  \ref{Henon_map}, the H\'enon map serves as a simple two-dimensional illustration.  The hard-disk model is introduced in Section \ref{Smooth-hard-disks}.
In this work we restrict ourselves to 198 disks, a number which is dictated  by  
computational economy, but still large enough to allow the study of Lyapunov modes.
In Section \ref{versus} we study the relative orientations of Gram-Schmidt and covariant vectors,
which  give rise to the same Lyapunov exponents. Next, in Section \ref{localization},  we
compare the localization properties in physical space for these two sets of perturbation vectors. 
The  configuration and momentum space projections of the perturbation vectors -- Gram-Schmidt 
or covariant -- are the topic of Section \ref{ts_projections}.
The central manifold (or null subspace) and its dependence on the intrinsic continuous symmetries 
-- translation invariance with respect to time and space -- is 
discussed in Section \ref{vanishing_exponents}.
Although the null subspace is completely orthogonal to the unstable and stable subspaces, it
is essential for a proper understanding of the Lyapunov modes \cite{Eckmann:2005,Henk}.
Section \ref{modes} is devoted to a discussion of these modes and how they are represented by the
covariant vectors. In Subsection \ref{hyperbolicity} we compute the angles between 
the covariant modes and test for tangency between covariant Oseledec subspaces.
In Section \ref{summary} we conclude with a summary.


\section{Phase space and tangent space dynamics}
\label{section_phase}

The dynamics of a system of hard disks is that of free flight, interrupted by elastic binary 
collisions. If  ${\bf \Gamma}_0$ denotes the state of the system at time $0$,
the state at time $t$ is given by  ${\bf \Gamma}_t  =  \phi^t ({\bf \Gamma}_0)$, 
where   $\phi^t:  \mathbf{X} \to \mathbf{X}$ defines the flow in the phase space $\mathbf{X}$.
Similarly, if  $\delta{\bf \Gamma}_0$ is a vector  in tangent space  ${\bf TX}$ at ${\bf \Gamma}_0$,
at time $t$ it becomes $\delta{\bf \Gamma}_t =  D\phi^t \vert_{{\bf \Gamma}_0}\cdot 
\delta{\bf \Gamma}_0$, where $D\phi^t$ defines the tangent flow.  It is represented by a
$D \times D$ matrix, where $D$ is the dimension of phase space.
A subspace $\vec E^{(i)}$ of the phase space is said to be covariant if
\begin{equation}
 D \phi^t \vert_{{\bf \Gamma}_0} \vec E^{(i)}({\bf \Gamma}_0) = \vec E^{(i)} (\phi^t ({\g_0})) . 
\label{defcov}
\end{equation}
This definition also applies to covariant vectors, if $\vec E^{(i)}$ is one-dimensional.
Loosely speaking, covariant subspaces (vectors) are co-moving (co-rotating in particular)
with the tangent flow.  An analogous relation holds for the time-reversed flow.

Next we consider the decomposition of the tangent space into subspaces according to the
{\em multiplicative ergodic theorem} of Oseledec \cite{Oseledec:1968,Ruelle:1979,Eckmann:1985}.
Here, we closely follow Ref. \cite{Eckmann:2005}.

        The first part of the multiplicative ergodic theorem asserts that the real and symmetric matrices
\begin{equation}
\Lambda_{\pm}=\lim_{t\rightarrow \pm \infty} \left( \left[ D\phi^t\vert_{\textrm{\smallskip{\g}}} \right]^{T} D\phi^t\vert_{\textrm{\smallskip{\g}}} \right)^{1/2|t|}
\enspace 
\label{Oseledec_matrix}
\end{equation}
exist for (almost all) phase points $\g$. Here, $T$ denotes transposition.  
The eigenvalues of $\Lambda_{+}$ are ordered according to 
$\exp (\lambda^{(1)}) > \cdots > \exp(\lambda^{(\ell)})  $,  where the $\lambda^{(j)}$ are 
the Lyapunov exponents, which appear with multiplicity $m^{(j)}$.
For symplectic systems as in our case,  $\lambda^{(j)}= - \lambda^{(\ell+1-j)}$, which is 
referred to as conjugate pairing. Similarly,  the eigenvalues of $\Lambda_{-}$ are   
$\exp (-\lambda^{(\ell)}) > \cdots > \exp(-\lambda^{(1)})  $.
The  eigenspaces of $\Lambda_{\pm}$ associated with $\exp(\pm \lambda^{(j)})$   are 
denoted by  $\vec U_{\pm}^{(j)}$. They are pairwise orthogonal  but {\em not} covariant. 
If the $\lambda^{(j)}$ are degenerate with  multiplicity $m^{(j)}=\dim \vec U^{(j)}_{\pm}$,
all multiplicities sum to $D$, the dimension of the phase space. Since the matrices
$\Lambda_{\pm}$ are symmetrical, each of the two sets of eigenspaces, $\lbrace \vec U^{(j)}_{\pm} \rbrace$, completely span the tangent space,
\begin{equation}
\mathbf{TX} (\g) = 
\vec U^{(1)}_{\pm} (\g) \oplus \cdots \oplus \vec U^{(\ell)}_{\pm} (\g) .
\end{equation}
The eigenspaces  $ {\vec U}^{(j)}_{\pm} $ are not covariant, but the subspaces
\begin{equation}
\vec U^{(j)}_{+} \oplus \cdots \oplus \vec U^{(\ell)}_{+} \quad \textrm{and} \quad \vec U^{(1)}_{-} \oplus \cdots \oplus \vec U^{(j)}_{-}, \quad j \in \left\lbrace 1,\ldots ,\ell \right\rbrace ,
\label{subspace_CV}
\end{equation}
are. They are, respectively, the most stable subspace of dimension 
\begin{small}$\sum_{i=j}^{\ell} m^{(i)}$\end{small} of $\Lambda_{+}$,
and the most unstable subspace of dimension \begin{small}$ \sum_{i=1}^{j} m^{(i)}$\end{small}
of $\Lambda_{-}$ (corresponding to the most stable subspace of that dimension 
in the past).

   The second part of Oseledec' theorem asserts that for (almost) 
every phase-space point ${\bf \Gamma}$ there exists another decomposition of the 
tangent space into {\em covariant subspaces} $\vec E^{(j)} ({\bf \Gamma})$
referred to as Oseledec splitting,
 \begin{equation}
\mathbf{TX} (\g) = 
\vec E^{(1)} (\g) \oplus \cdots \oplus \vec E^{(\ell)} (\g)  .
\end{equation}
For $\delta {\bf \Gamma} \in \vec E^{(j)} ({\bf \Gamma})$ 
the respective Lyapunov exponent follows from
\begin{equation}
\lim_{t \rightarrow \pm \infty} \dfrac{1}{\vert t \vert} \, \log \, \Vert \, D\phi^t\vert_{\textrm{\smallskip{\g}}} \, \cdot \, \delta \vec \Gamma \, \Vert =
\pm \lambda^{(j)} \qquad \forall \, j \in \left\lbrace 1,\ldots, \ell \right\rbrace .
\label{ccclya}
\end{equation}
The subspaces  $\vec  E^{(j)}$ are covariant (see Eq. ({\ref{defcov}))  but, in general, 
not orthogonal. According to Ruelle
\cite{Ruelle:1979}, they are related to the eigenspaces  $\vec U^{(j)}_{\pm} $ of $\Lambda_{\pm}$:
\begin{equation}
\vec E^{(j)} = \left( \vec U^{(1)}_{-} \oplus \cdots \oplus \vec U^{(j)}_{-} \right)
\cap
\left( \vec U^{(j)}_{+} \oplus \cdots \oplus \vec U^{(\ell)}_{+}\right) .
\label{subspaces}
\end{equation}
This equation is at the heart of the construction of covariant vectors according to
Ginelli {\em et al.} as described in the next section.
Furthermore, one can show that
\begin{equation}
{\vec F}^{(j)} \equiv  
\vec E^{(1)} \oplus \cdots \oplus \vec E^{(j)}   = \vec U^{(1)}_{-} \oplus \cdots \oplus \vec U^{(j)}_{-}
\label{Subspace_backward}
\end{equation}
are covariant subspaces.


\section{Numerical considerations}
\label{numerics}

 Numerical methods probe the tangent space by a set of $D$ tangent vectors,  such that the
 Lyapunov exponents are repeated with multiplicities,
$\lambda_1 \geq \cdots \geq \lambda_D$. Here, the lower index is referred to as the 
Lyapunov index. The relation between the $\lambda^{(j)}$ and $\lambda_i$ is given by
$$
    \lambda^{(j)} = \lambda_{f^{(j-1)} + 1} = \cdots  =  \lambda_{f^{(j)}},
$$
where $f^{(j)} = m^{(1)} + \cdots + m^{(j)} $ is the sum of all subspace dimensions up to $j$.     

For notational convenience in the following, the vectors  
$ {\vec g}_n^{j} , \, j = 1,\dots,D$ 
spanning the tangent space at time $t_n$,  are arranged as  
column vectors of a $D \times D$ matrix ${\vec G}_n \equiv ( {\vec g}_n^1 | \dots | {\vec g}_n^D ) $.
The same convention is used below for other spanning vector sets such as
$\overline{\vec G}_n \equiv (\overline{\vec g}_n^1 | \dots | \overline{\vec g}_n^D ) $  and
 ${\vec V}_n \equiv ({\vec v}_n^1 | \dots | {\vec v}_n^D ) $. 

In the classical algorithm of Benettin {\em et al.}
\cite{Benettin} and Shimada {\em et al.} \cite{Shimada} for the computation of Lyapunov exponents,
an orthonormal set of tangent vectors ${\vec G}_{n-1}$ at time $t_{n-1}$ is evolved to a 
time $t_n \equiv t_{n-1} + \tau$, ($\tau > 0$),
$$  \overline{\vec G}_n  =  {\bf J}_{n-1}^{\tau} {\vec G}_{n-1},$$ where ${\bf J}_{n-1}^{\tau}$
is the Jacobian of  the evolution map taking the phase space point ${\bf \Gamma}_{n-1}$
at time $t_{n-1}$ to ${\bf \Gamma}_n$ at time $t_n$.
The column vectors of $ \overline{\vec G}_n$ at time $t_n$ generally are not orthonormal 
any more and need to be re-orthonormalized with a Gram-Schmidt procedure.  This gives
the matrix ${\vec G}_n$ with column vectors $\{{\vec g}^j\}_n$, which form  the next orthonormal 
Gram-Schmidt (GS) basis at time $t_n$. These vectors are pairwise orthogonal but not covariant.
Each GS renormalization step is equivalent to a so-called QR decomposition of the matrix
$ \overline{\vec G}_n$,  $ \overline{\vec G}_n  = {\vec G}_n {\vec R}_n$, where the 
matrix ${\vec R}_n$ is upper triangular \cite{recipes}.
The diagonal elements of ${\vec R}_n$ are required for the accumulative 
computation of the Lyapunov exponents.
This procedure is iterated until convergence for the Lyapunov exponents is obtained.  

For the computation of a covariant set of vectors $\{{\vec v}^j \}_0$ spanning the tangent space
for the phase point ${\bf  \Gamma}_0 \equiv {\bf \Gamma}(0)$ at, say, time $t_0$,
Ginelli {\it et al.} \cite{Ginelli} start with a well-relaxed set of GS vectors 
at $t_0$ and follow the dynamics forward for a sufficiently long time up to  
$t_{\omega} = t_0 + \omega \tau$,
storing ${\vec G}_n$ and  $\overline{\vec G}_n$ (or, equivalently, ${\vec R}_n$) for 
$t_n = t_0 + n\tau, \;n = 0, \cdots, \omega$ along the way. At $t_{\omega}$ a set of unit tangent vectors
$\left\lbrace {\vec v}^j \right\rbrace_{\omega}$ is  constructed according to
\begin{equation}
\vec v^{j}_{\omega} \in \vec S^j_{\omega} \equiv \textrm{span}\left\lbrace \vec g^{1}_{\omega} , \ldots , 
\vec g^{j}_{\omega} \right\rbrace 
\qquad \forall \, j \in \left\lbrace 1,\ldots,D\right\rbrace 
\enspace ,
\label{initial_condition}
\end{equation}
which serve as starting vectors for a backward  iteration from $t_{\omega}$  to time $t_0$. 
The vector ${\vec v}^j_{n}$ will stay in $\vec S^j_{n}$ at any intermediate time $t_n$,
because  ${\vec S}_{n}^j$ is the most stable subspace of dimension $j$ for the time-reversed
iteration. Arranging these vectors again as column vectors of a matrix  ${\vec V}_n$ and
expressing them in the GS basis at time $t_n$, one has ${\vec V}_n =  {\vec G}_n {\vec C}_n$,
where the matrix ${\vec C}_n$ is again upper triangular with elements  
$\left[ \mathbf{C}_{n} \right]_{i,j} = {\vec g}^i_n \cdot {\vec v}^j_n$. If, at any step $n$, ${\vec C}_{n-1}$
is constructed from ${\vec C}_n$ according to
${\vec C}_{n-1} = \left[{\vec R}_n\right]^{-1} {\vec C}_n$ , 
Ginelli {\it et al.} have shown that 
${\vec V}_n = J_{n-1} {\vec V}_{n-1}$ and, hence, the respective  column vectors of this matrix 
follow the natural tangent space dynamics without re-orthogonalization.  They are covariant
but not orthogonal in general. At this stage of the algorithm, renormalization 
of ${\vec v}_{n-1}^j$ is still required to escape the exponential divergence of the vector norms
without affecting their orientation. After reaching $t_0$ at the end of the iteration, 
the vectors  ${\vec v}_0^j$ point into their proper 
orientations in tangent space such that, according to Eq. (\ref{subspaces}),
$\textrm{span}\left({\vec v}_0^1, \cdots , {\vec v}_0^ {f^{(j)}}\right) = {\vec E}^{(1)}({\bf \Gamma}_0)
 \oplus \cdots \oplus {\vec E}^{(j)}({\bf \Gamma}_0)$ is the most-unstable 
 subspace of dimension $f^{(j)} \equiv m^{(1)} + \cdots + m^{(j)}$
of the tangent space at the space point ${\bf \Gamma}_0$, going forward in time. 
If there are degeneracies (as in the presence of Lyapunov modes to be discussed
below), the Oseledec subspace ${\vec E}^{(j)}$ is spanned according to
\begin{equation}
          {\vec E}^{(j)} = {\vec v}^{f^{(j-1)}+1} \oplus \cdots \oplus {\vec v}^{f^{(j)}},
\label{spanning}          
\end{equation}           
where, as in the following, we omit the arguments for the phase-space point. 
 If there are no degeneracies, ${\vec v}^{f^{(j)}} = {\bf E}^{(j)}$. 
Similarly, the Gram-Schmidt vectors may be expressed in terms of the eigenspaces of $\Lambda_{-}$,
\begin{equation}
        {\vec U}_{-}^{(j)} = {\vec g}^{f^{(j-1)}+1} \oplus \cdots \oplus {\vec g}^{f^{(j)}}.
\nonumber 
\end{equation}
For nondegenerate subspaces one finds ${\vec U}_{-}^{(j)}  = {\vec g}^{f^{(j)}} 
$ \cite{Ershov,Legras,Eckmann:2005}.        
 
  The drawback of this algorithm for many-particle systems is the large storage requirement 
 for the matrices  ${\vec G}_n$ and  $\overline{\vec G}_n$ (or, equivalently, ${\vec R}_n$) for 
the intermediate times $t_n = t_0 + n\tau, \;n = 0, \cdots, \omega$, because $\tau$ must not be 
chosen too large (containing not more than, say, 20 particle collisions). At the expense of computer time, 
this can be bypassed by storing the matrices only for times separated by, say, $ 100 \tau$ 
intervals and recomputing the forward dynamics in between when required during the time-reversed iteration. In this case, also the phase-space trajectory needs to be stored.


\section{A simple example: The H\'enon map}
\label{Henon_map}

\begin{figure}[htb]
\centering
\includegraphics[angle=-90,width=.65\textwidth]{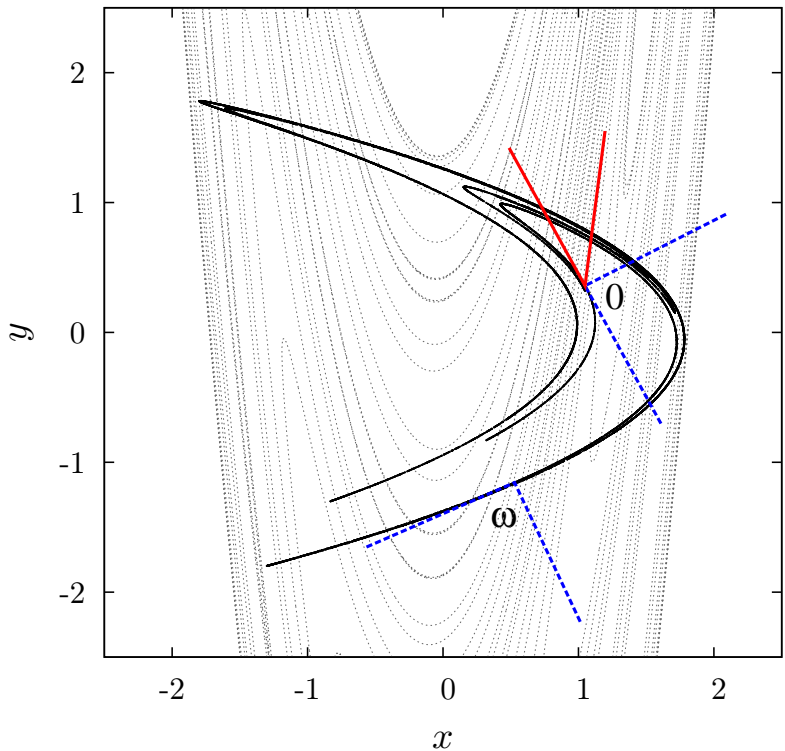}
\caption{The H\'enon attractor (black line) and a finite-length approximation of its stable manifold (dotted line)
are shown. The red vectors are the covariant vectors at the phase point $0$ as 
explained in the main text. The blue vectors are Gram-Schmidt vectors.}
\label{Figure_1}
\end{figure}
To illustrate the foregoing algorithm, we apply it to a simple two-dimensional example, 
the H\'enon map \cite{Henon}, 
\begin{eqnarray}
x_{n+1} &=& a - x_n^2 + b \, y_n
\nonumber
\enspace , \\
y_{n+1} &=& x_n \enspace ,
\nonumber
\end{eqnarray}
with $a=1.4$ and $b=0.3$. In Fig. \ref{Figure_1}  
the H\'enon attractor is shown (black line),
which is known to coincide with its unstable manifold. An approximation of the 
stable manifold is shown by the dotted lines.  At the point $0$ 
the initial GS basis is indicated by the two orthogonal vectors in blue, where one, as required,  
points into the  direction of the unstable manifold. If these vectors are evolved forward 
in time with the GS method for a  few hundred steps, the two orthogonal GS vectors at 
the point $\omega$ are obtained. Taking these vectors as 
the initial vectors ${\vec v}_{\omega}^1$ and   ${\vec v}_{\omega}^2$, the consecutive backward iteration yields
the covariant vectors at point $0$ indicated in red. As expected, one is parallel to the
unstable manifold, the other parallel to the stable manifold at that point.


\section{Systems of hard disks }
\label{Smooth-hard-disks}


Now we turn to the study of a two-dimensional system of hard disks in a box with periodic 
boundaries, where the particles suffer elastic hard collisions (without roughness), and
move along straight lines in between collisions. The case of rough hard disks is the topic of 
a forthcoming publication \cite{BP2010}.

The Lyapunov instability of hard disk systems has been studied in detail in the past
\cite{DPH1996,PH2000,FHPH2004,TM2003a,TM2003b}.  Here we are  mainly concerned with the
differences encountered with the GS and covariant vectors, which, as we have seen, 
give rise to identical
Lyapunov spectra. To facilitate comparison with our previous work, we consider 
reduced units for which the particle diameter $\sigma$, the particle mass $m$ and the
kinetic energy per particle, $K/N$, are unity. Here, $K$ is the total energy, which is purely kinetic, 
and $N$ denotes the number of particles.
Lyapunov exponents are given in units of $\sqrt{K/N m \sigma^2}$. If not otherwise stated,
our standard system consists of  $N = 198$ particles at a density $\rho \equiv N/(L_x L_y) = 0.7$ and a
simulation box with an aspect ratio  $ L_y/L_x = 2/11$, which is periodic  in $x$ and $y$.
The choice of such a small aspect ratio facilitates the observation of 
the Lyapunov modes to be discussed later. As usual, the total momentum is set to zero. 

The state of the system is given by the coordinates and momenta of all the particles,
$$
    {\vec \Gamma} = \{ {\vec q}_n, {\vec p}_n; \; n = 1,\cdots, N\}.
$$    
Similarly,  an arbitrary tangent vector $\delta {\bf \Gamma}$ - either a Gram-Schmidt vector ${\vec g}$ or 
a covariant vector ${\vec v}$ - consists of the respective coordinate and momentum perturbations,
\begin{equation}
         \delta {\bf \Gamma} = \{ \delta{\vec q}_n, \delta{\vec p}_n; \;  n = 1,\cdots, N \}.
         \label{state}
\end{equation}
The time evolution of these vectors and the construction of the map  from one Gram-Schmidt step 
to the next has been discussed before \cite{DPH1996,DP1997}.      
 
\begin{figure}[ht]
\centering
\includegraphics[angle=-90,width=0.74\textwidth]{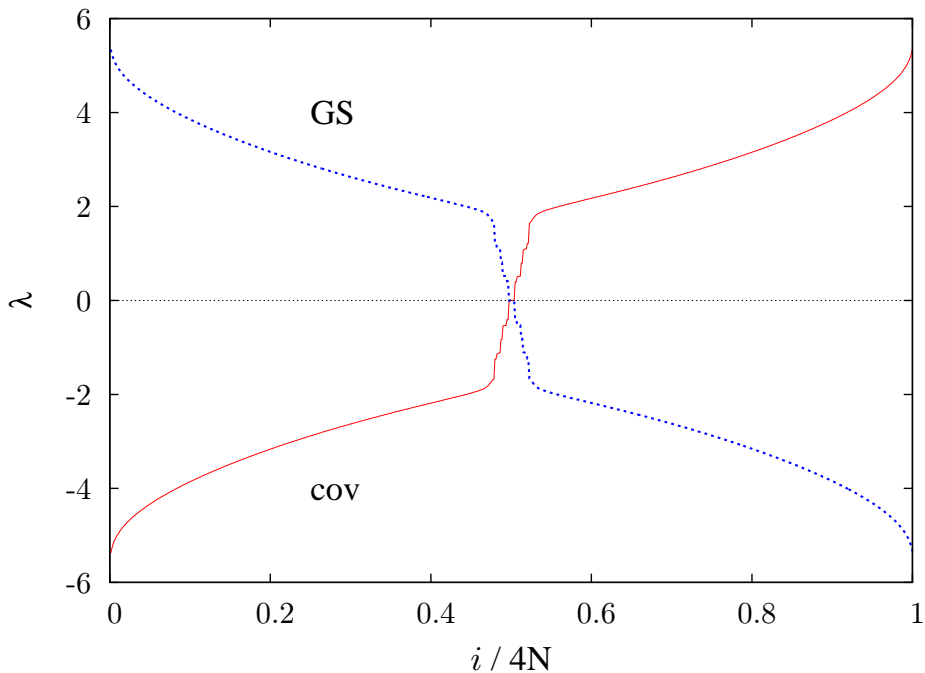}
\caption{Lyapunov spectrum for the 198 disk system described in the main text.
The spectrum calculated in the forward direction  with the GS method is shown by the blue 
 line, the one calculated in the backward direction with the covariant vectors by the red line. 
 Reduced indices $i/4N$ are used on the abscissa. Although the spectrum is defined only for
integer $i$, solid lines are drawn for clarity.}
\label{spectrum}
\end{figure}
Fig. \ref{spectrum} shows the Lyapunov spectrum for this system computed both in forward 
direction with the GS vectors  (blue line) and in backward direction with the covariant vectors (red line). The time of the simulation in the forward direction is  for 
$t_0 + t_{\omega} = 2.5 \times  10^5 \tau $, where $\tau = 0.6$ is the largest interval between two 
successive Gram-Schmidt  re-orthonormalizations, which does not affect the spectrum.
The backward simulation is for a time $t_{\omega} - t_0= 2.5  \times 10^4 \tau \,$. The time $t_0$ 
(usually of the order of $1 \times 10^4 \tau$) is required
for the preparation of the relaxed initial state at $t_0$.
It can be observed in the figure that the unstable directions in the future correspond well to the stable directions in the past and vice versa. Of course, if the sequence of covariant vectors is 
followed in the forward direction of time, the spectrum is identical to the classical GS results
(blue line in Fig. \ref{spectrum}).

\subsection{Covariant versus Gram-Schmidt vectors}
\label{versus}
Whereas the time evolution of the GS vectors is determined by  the exponential growth
of  infinitesimal volume elements  belonging to subspaces
${\vec g}^1 \oplus \cdots \oplus {\vec g}^{i}$ for $i \in \left\lbrace 1,\ldots, D\right\rbrace$ 
according to $\exp\left( t\,\sum_{j=1}^{i} \lambda_j \right) $, the growth of an infinitesimal
perturbation representing a covariant vector  ${\vec v}^i $  is directly proportional to
$\exp\left( t \, \lambda_i \right) $, for all $i$. Thus, it is interesting to compare the
relative orientations of respective vectors giving rise to the same exponent.
\begin{figure}[t]
\begin{tabular}{cc}
\centering
\begin{minipage}[c]{.45\linewidth}
\includegraphics[angle=-90,width=1\textwidth]{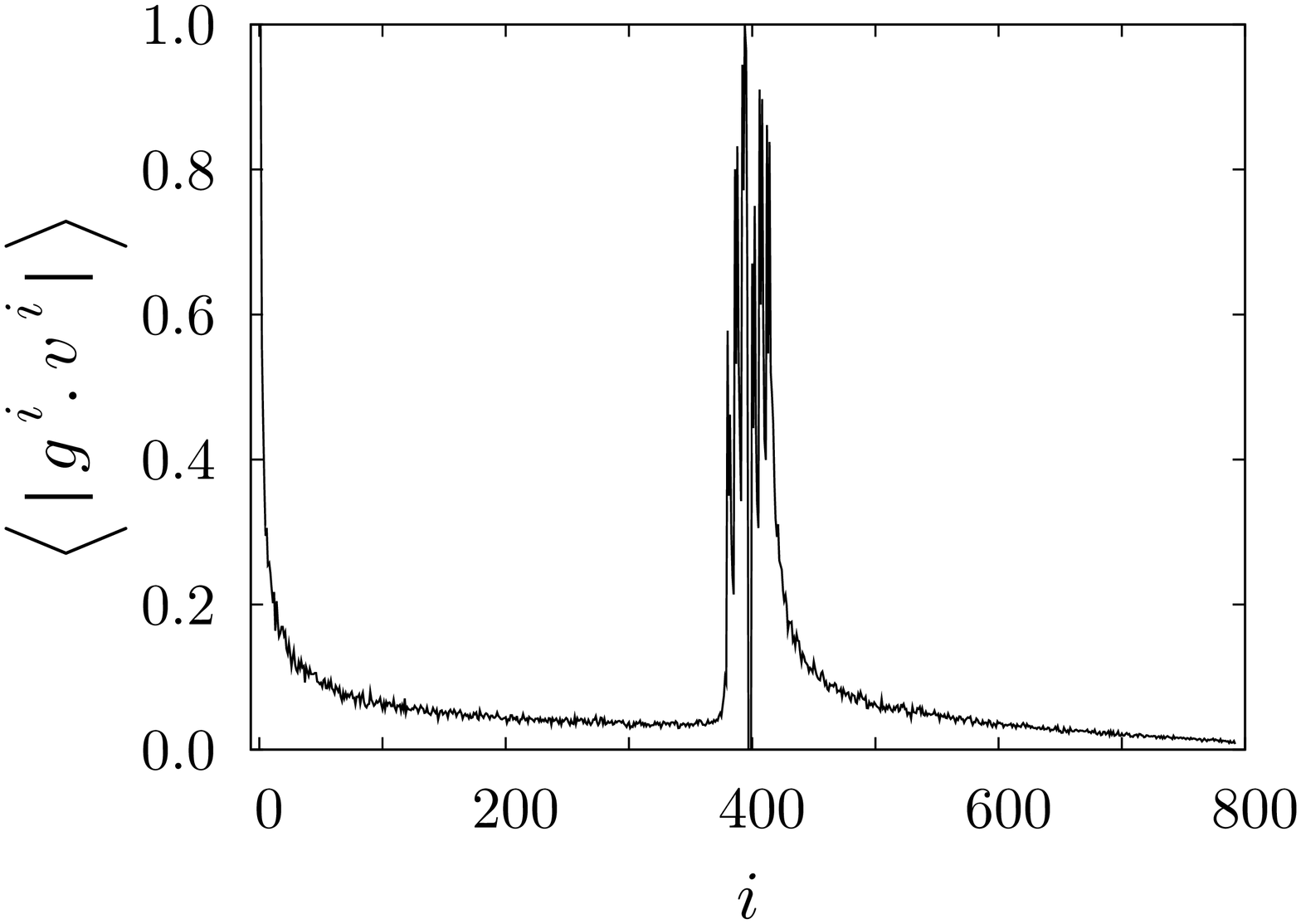}
\end{minipage}  &
\begin{minipage}[c]{.45\linewidth}
\includegraphics[angle=-90,width=1\textwidth]{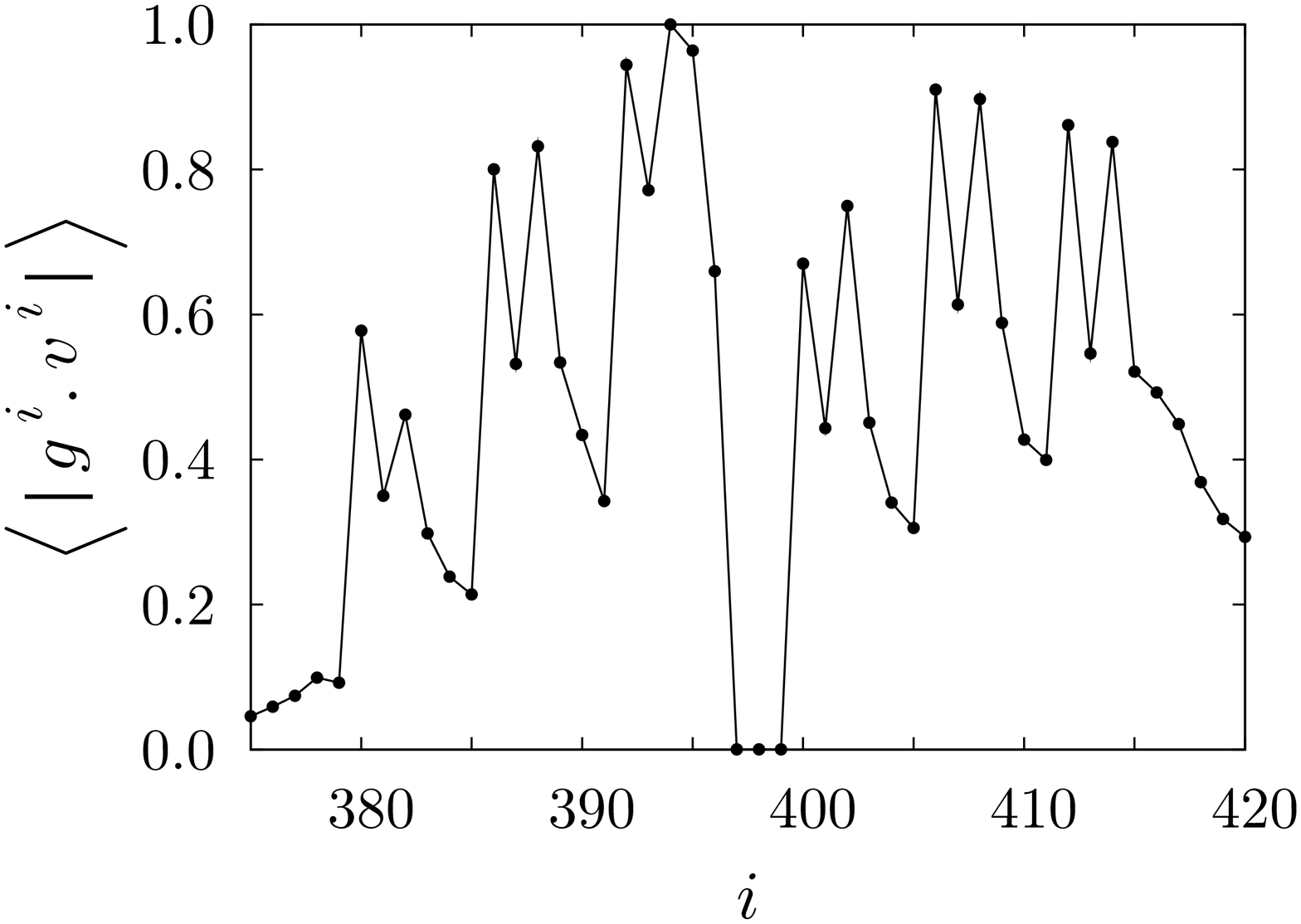}
\end{minipage}
\end{tabular}
\caption{ Plot of the scalar product norm, $\langle | {\vec g}^i \cdot {\vec v}^i | \rangle$,
for GS and covariant vectors giving rise to the same Lyapunov exponent
$\lambda_i$, as a function of $i$. The line is a time average as discribed in the main text. 
Left panel: full range of Lyapunov exponents;  Right panel: enlargement of the central part.}
\label{Fig3}
\end{figure}
In the left panel of  Fig. \ref{Fig3} the difference in orientation 
of the two types of vectors is demonstrated by  a plot of $|{\vec g}^i \cdot {\vec v}^i|$  as a function of $i$. 
The black line is an average over 100 frames separated by time intervals of $250 \tau$.  
Since for tangent vectors only their direction and not the sense of direction is important,
an absolute value is taken (here and for analogous  cases below), otherwise the
scalar product might average to zero over long times, with equal numbers of vectors pointing
into opposite directions. 
For the unstable directions in the left half of the left panel, one observes a rapid decrease 
of the scalar product with $i$ and, hence a rapid increase of the angle between respective 
covariant and GS vectors. This decrease is repeated for the stable directions in the right half of the figure. 
These two parts are separated by the mode region, an enlargement of which is shown in
the right panel of Fig.  \ref{Fig3} and which will be dealt with in more detail below.

\begin{figure}[h]
\begin{tabular}{c c}
\\
\begin{minipage}[c]{.45\linewidth}
\includegraphics[angle=-90,width=1\textwidth]{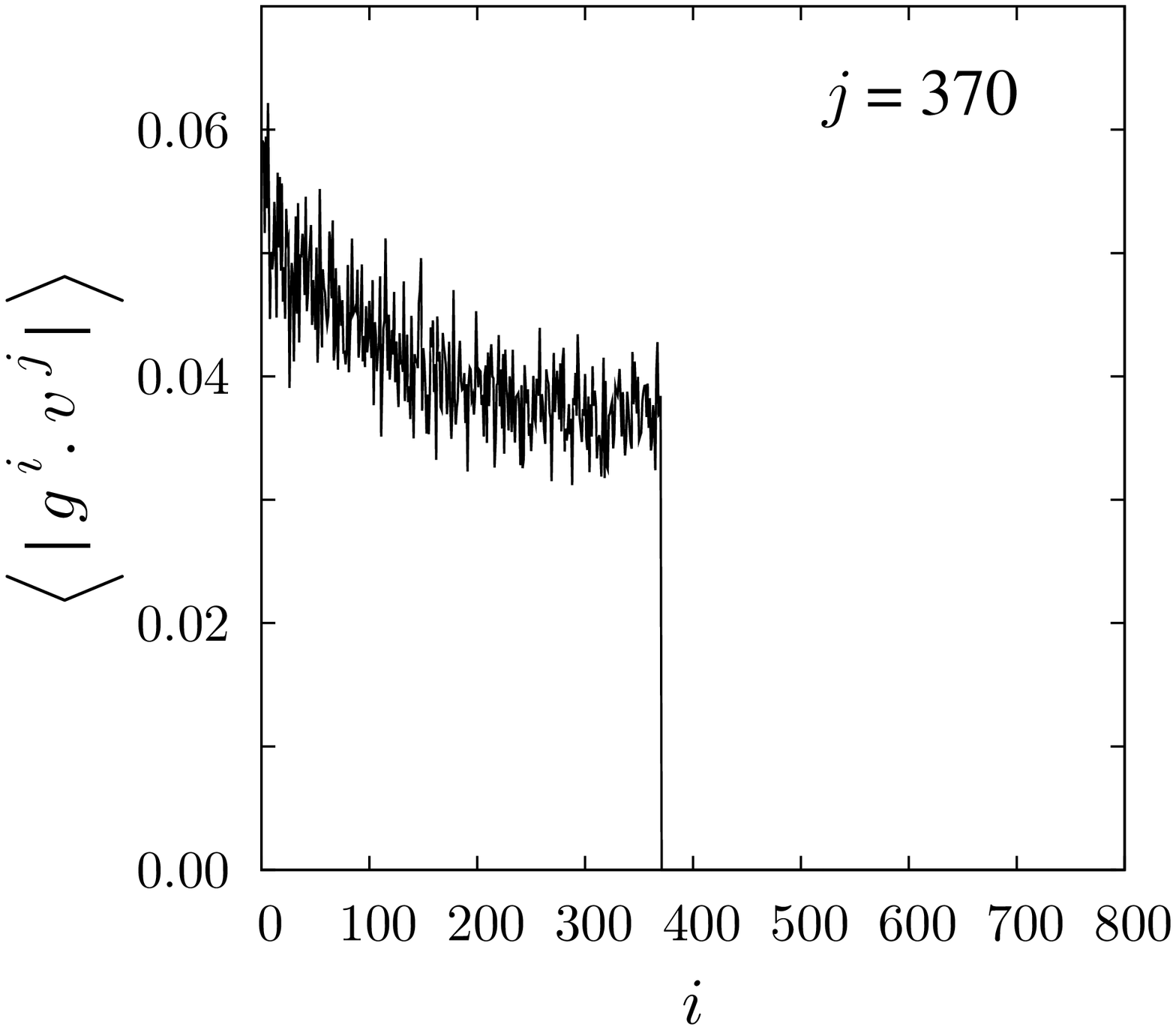}
\end{minipage} &
\begin{minipage}[c]{.45\linewidth}
\includegraphics[angle=-90,width=1\textwidth]{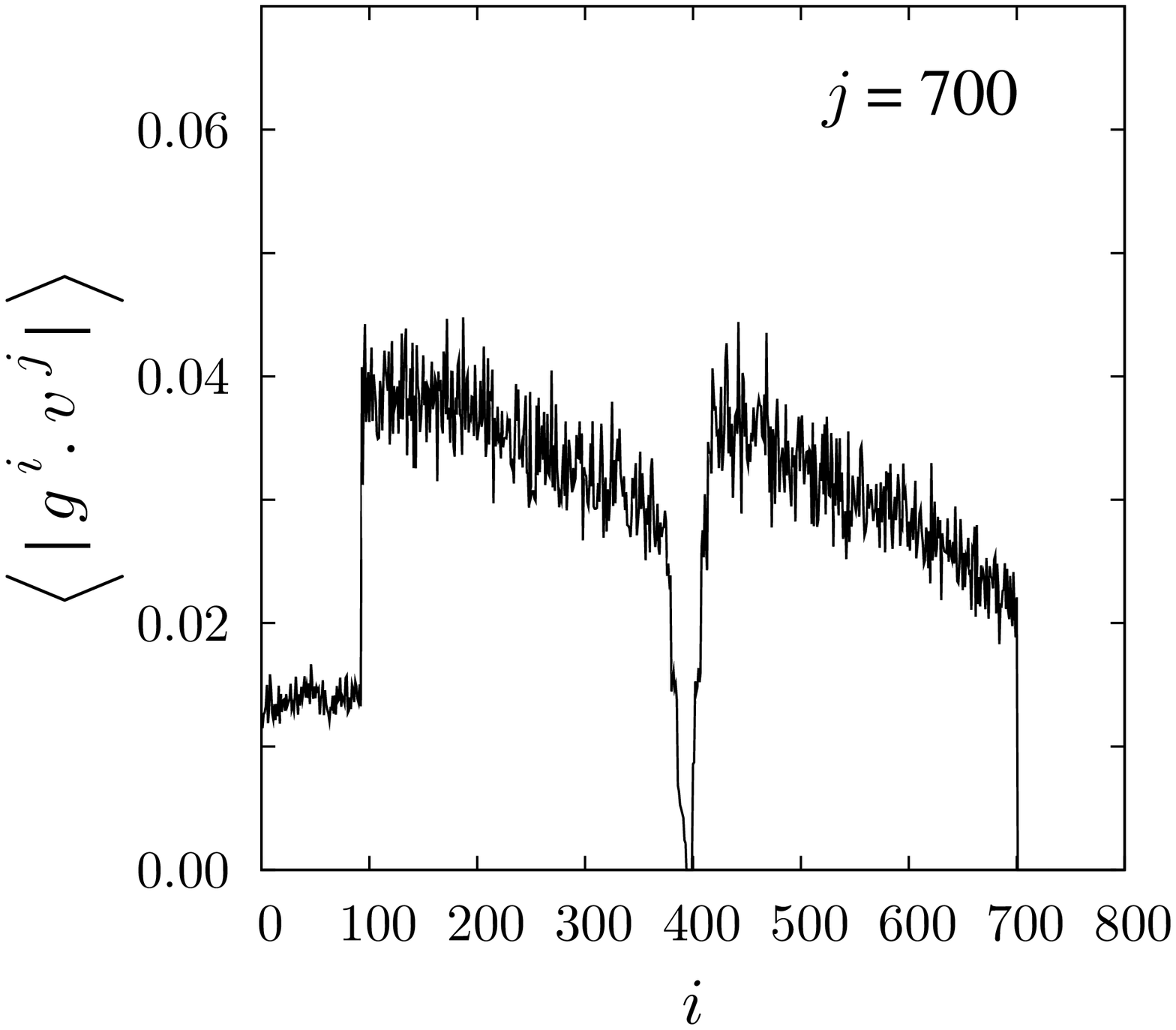}
\end{minipage} \\
\begin{minipage}[c]{.45\linewidth}
\includegraphics[angle=-90,width=1\textwidth]{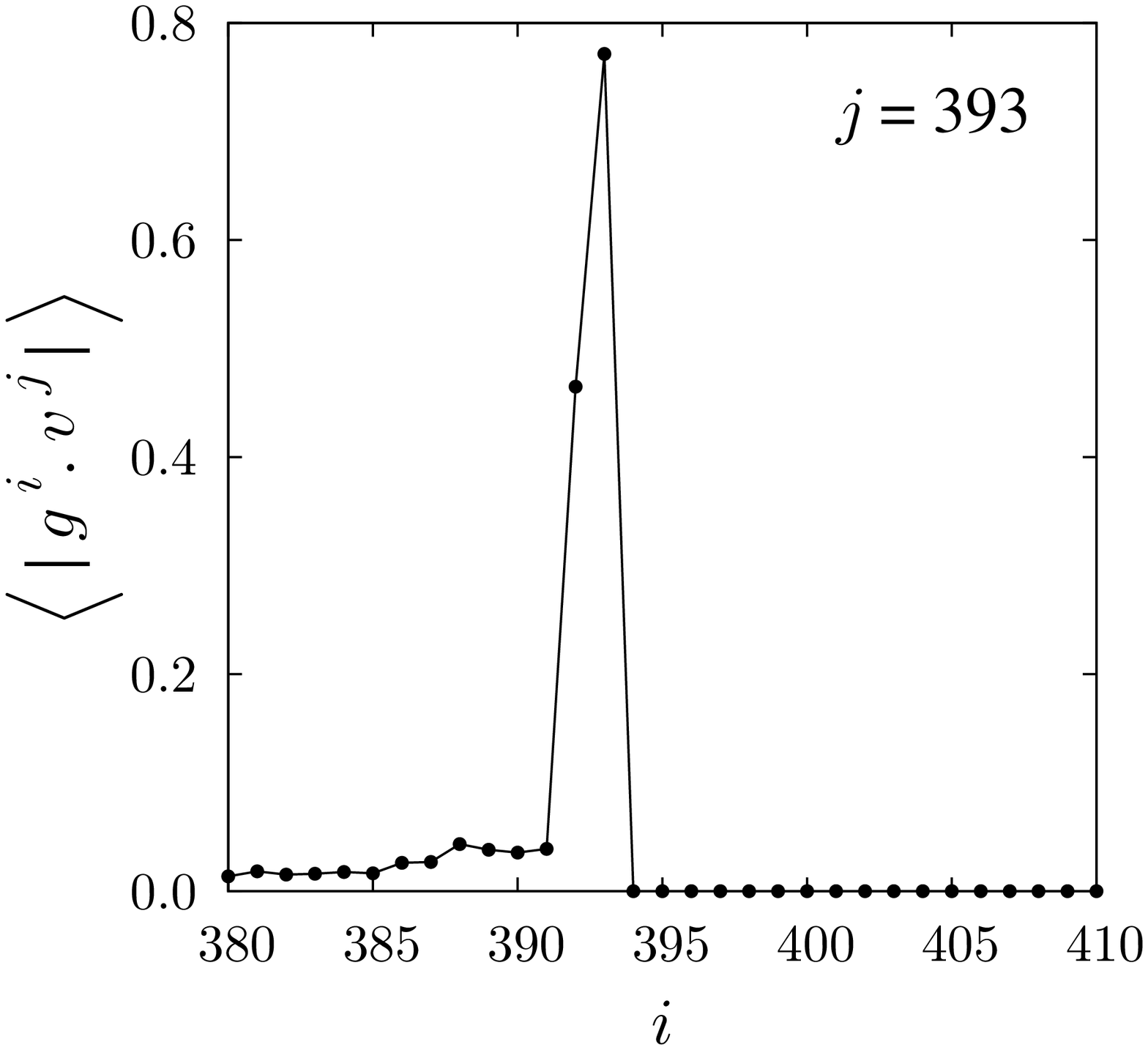}
\end{minipage}&
\begin{minipage}[c]{.45\linewidth}
\includegraphics[angle=-90,width=1\textwidth]{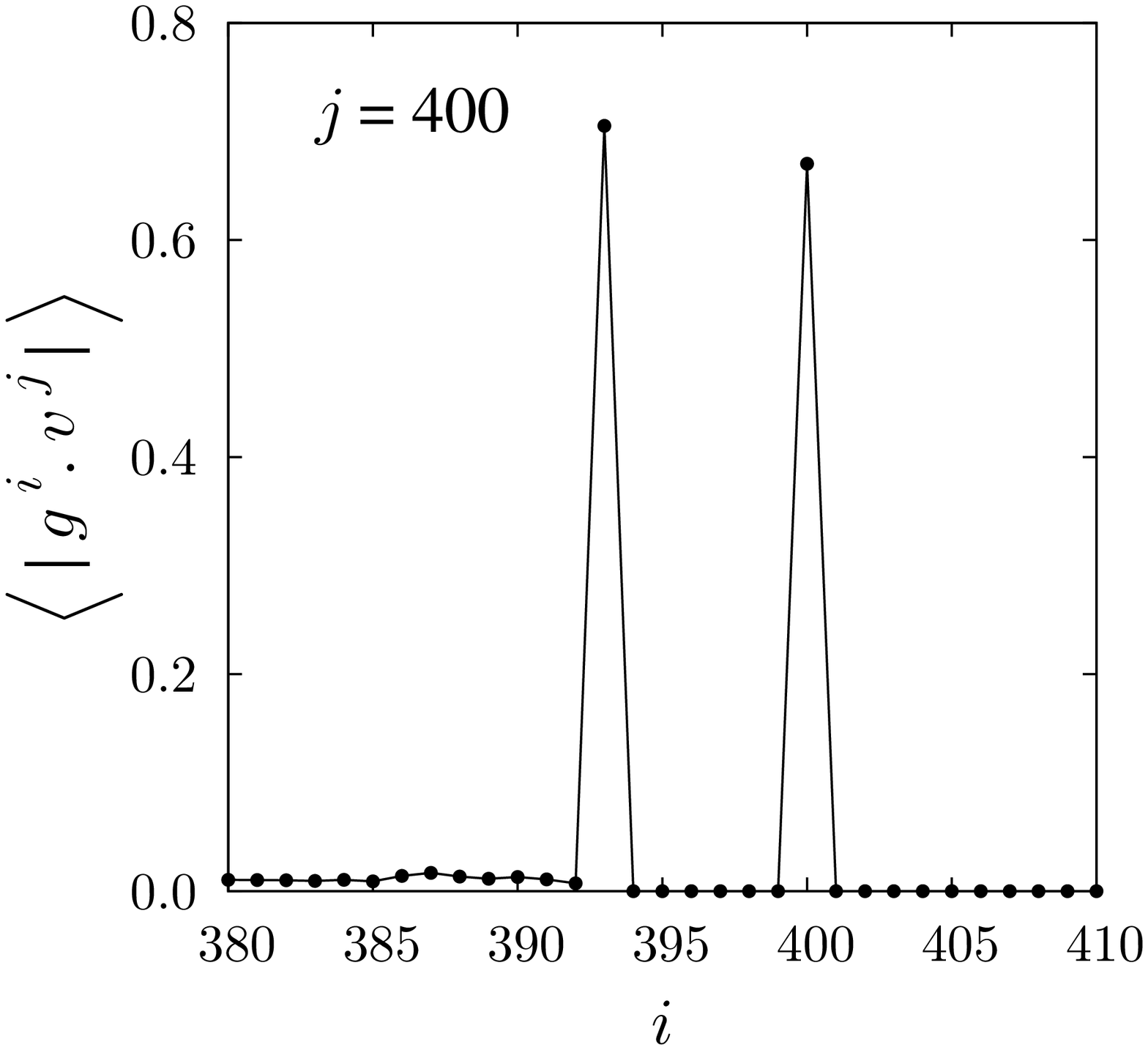}
\end{minipage}
\end{tabular}
\caption{Time averaged  absolute value of the scalar product of selected covariant vectors
${\vec v}^j$ (as indicated by the labels) with the whole set of Gram-Schmidt vectors ${\vec g}^i$
as a function of $i$.}
\label{Figure_4}
\end{figure}

In Fig. \ref{Figure_4} we show similar projections (time averages of absolute values 
of scalar products as before) for selected covariant vectors with the whole Gram-Schmidt vector set.
One observes that the covariant vectors ${\vec v}^{j}$ belong to the GS subspace 
${\vec g}^1 \oplus\cdots\oplus {\vec g}^j $, for all 
$j \in \left\lbrace 1,\ldots,4N\right\rbrace $ and, thus, give rise to the upper-triangular property of
the matrix $\mathbf{R}$ in the QR-decomposition mentioned above.
The curves in the figure strongly depend on the choice of $j$: \\
$\bullet$ If it belongs to the unstable subspace ${\vec E}^{u}$ but does not represent a 
Lyapunov mode
(top-left  panel for $j =370$),  there is no obvious orientational correlation with any of the GS vectors with index 
$i < j$.  For $i=j=1$ corresponding to the maximum exponent, 
the covariant and GS vectors are identical. If, however, the covariant vector represents
a Lyapunov mode as in the bottom-left panel for $j=393$, then its angle with the respective GS vector 
may become smaller, giving rise to a scalar product closer to unity.  \\
$\bullet$ If the covariant vector belongs to the stable subspace ${\vec E}^s$ but does 
not represent a mode as for $j = 700$ in the top-right panel of Fig. \ref{Figure_4}, 
it has non-vanishing components in the GS basis
for all $i \le j$ with the exception of the zero subspace $2N-2 \le i  \le 2N+3$, which is strictly orthogonal.
With the exception of the step at  the conjugate index $ i = 4N+1-j = 93$, the origin of which 
is not fully understood, there is no indication of orientational correlations
between the covariant vector with any of the GS vectors for $i \le j$.
If, however, the covariant vector represents a mode as for $j = 400$ in the lower-right
panel of the figure,  there is strong orientational correlation not only with the
respective GS vector with $i=400$, but also with its conjugate pair at $4N+1 - i$  ($= 393$ in our example).

It is interesting to note that the leading GS and covariant vectors in the null subspace are always identical
(up to an irrelevant sign):  ${\vec v}^{2N-2} = {\vec g}^{2N-2}$. 

\subsection{Localization}
\label{localization}

The maximum (minimum) Lyapunov exponent is the rate constant for the
fastest growth (decay) of a phase-space perturbation and is dominated by the
fastest dynamical events, a locally-enhanced collision frequency. It is not too surprising that the associated
tangent vector components are significantly different from zero for only a few
strongly-interacting particles at any instant of time. Thus, the respective perturbations 
are strongly localized in physical space. This property persists in the thermodynamic 
limit such that the fraction of tangent-vector components contributing to the generation 
of $\lambda_1$ follows a power law $\propto N^{-\eta}, \eta > 0$, and
converges to zero for $N \to \infty$ \cite{MP2002,PF2002,FHPH2004,FP2005}.   
The localization becomes gradually worse for larger indices $i > 1$, until it ceases to
exist and (almost) all particles collectively contribute to the coherent Lyapunov modes
to be discussed below. Similar observations for spatially extended systems 
have been made by various authors \cite{Manneville,LR1989,FMV1991,TM2003a,TM2003b},
which were consequently explained in terms of simple models \cite{Astrid,TMXXX}. 
We also mention Ref. \cite{Pikovsky}, where the tangent-space dynamics of the
first Lyapunov vector ${\vec g}^1$ for various one-dimensional Hamiltonian lattices is 
compared to that for the Kardar-Parisi-Zhang model of spatio-temporal chaos.
The unexpected differences found for  the scaling properties  are traced back
to  the existence of long-range correlations, both in space and time,  in the 
Hamiltonian chains, the origin of which, however, could not be fully disclosed.
The same correlations are conjectured to be responsible for a slow
$1/\sqrt{N}$ convergence of $\lambda_1$ towards its thermodynamic limit
\cite{Pikovsky}, which is also observed for hard-disk systems \cite{DPH1996}.

\begin{figure}[t]
\centering
\includegraphics[angle=-90,width=0.7\textwidth]{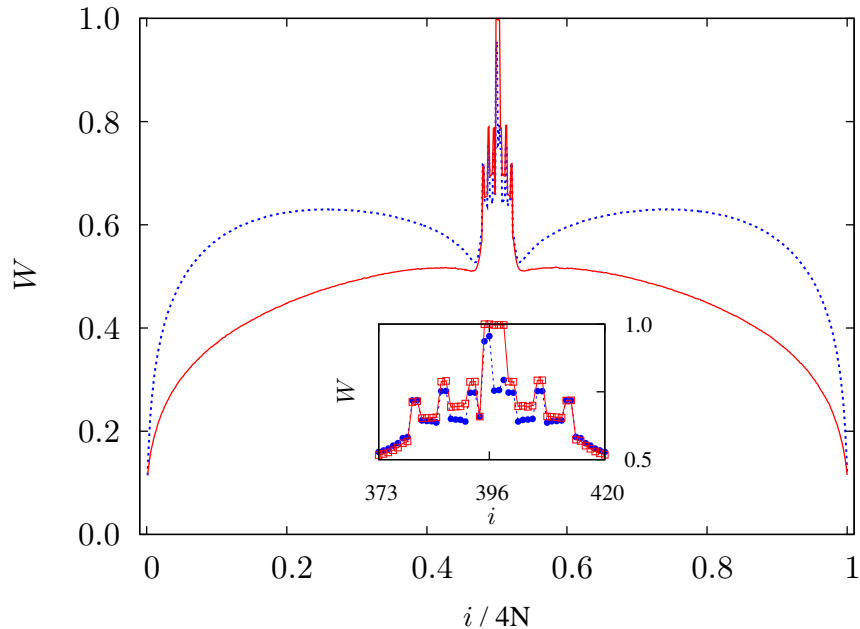}
\caption{Localization spectra $W$ for the complete set of Gram-Schmidt vectors (blue)  
and covariant vectors (red). The details of the hard-disk system are given in Section
\ref{Smooth-hard-disks}.
Reduced indices $i/4N$ are used on the abscissa. In the inset a magnification of the
central mode region is shown.    }
\label{Figure_5}
\end{figure}
     Up to now, all considerations concerning localization were based on the Gram-Schmidt
vectors. Here, we demonstrate the same property for the covariant vectors.
According to Eq. (\ref{state})   
we define the contribution of an individual  disk $n$ to a particular perturbation vector 
as the square of the projection of   $\delta {\vec \Gamma}$
onto the subspace pertaining to this disk,
 $$\mu_n = (\delta {\vec q}_n)^2 + (\delta {\vec p}_n)^2. $$
 Since $ \delta {\vec \Gamma}$ is either a GS vector or a covariant vector both of which are
 normalized,   one has $\sum_{n=1}^N \mu_n = 1$, and 
$\mu_n$ may be interpreted as a kind of action probability of particle $n$ 
contributing to the perturbation in question.  It should be noted that
for the definition of $\mu_n$ the Euclidean norm is used and that all localization measures
depend on this choice. Qualitatively, this is still sufficient to demonstrate localization.
From all the localization measures introduced \cite{FMV1991,MP2002},
 the most common is due to Taniguchi and Morriss \cite{TM2003a,TM2003b},  
 \begin{equation}
 W  =\frac{1}{N} \exp[S] , \;\;
S  = \left\langle - \sum_{n=1}^N \mu_n \ln \mu_n\right\rangle.
\nonumber
 \end{equation} 
Here, $S$  is the Shannon entropy for the ''probability''  distribution  $\mu_n$, and
   $\langle \cdots \rangle$ denotes a time average. $W$
 is bounded according to $1/N \le W \le 1$,  where the lower and upper bounds apply
 to complete localization and delocalization, respectively. 
 In Fig. \ref{Figure_5}, we compare $W$ obtained 
 for the full set of Gram-Schmidt vectors (blue curve) to  that of all the covariant} vectors (red curve). The spectra
 are obtained by identifying $\delta {\vec \Gamma}$ with all vectors of the respective sets,
 $i = 1,\cdots,4N$. 
 Not too surprisingly, the localization is stronger for the covariant vectors, whose
 direction in tangent space is solely determined by the tangent flow and is not
 affected by renormalization constraints. Another interesting feature is the symmetry
 $W_i = W_{4N+1 -i}$, which is a direct consequence of the symplectic nature of the flow
 \cite{Hadrien}.

\subsection{Tangent space projections}
\label{ts_projections}
\begin{figure}[tbp]
\centering
\includegraphics[angle=-90,width=0.8\textwidth]{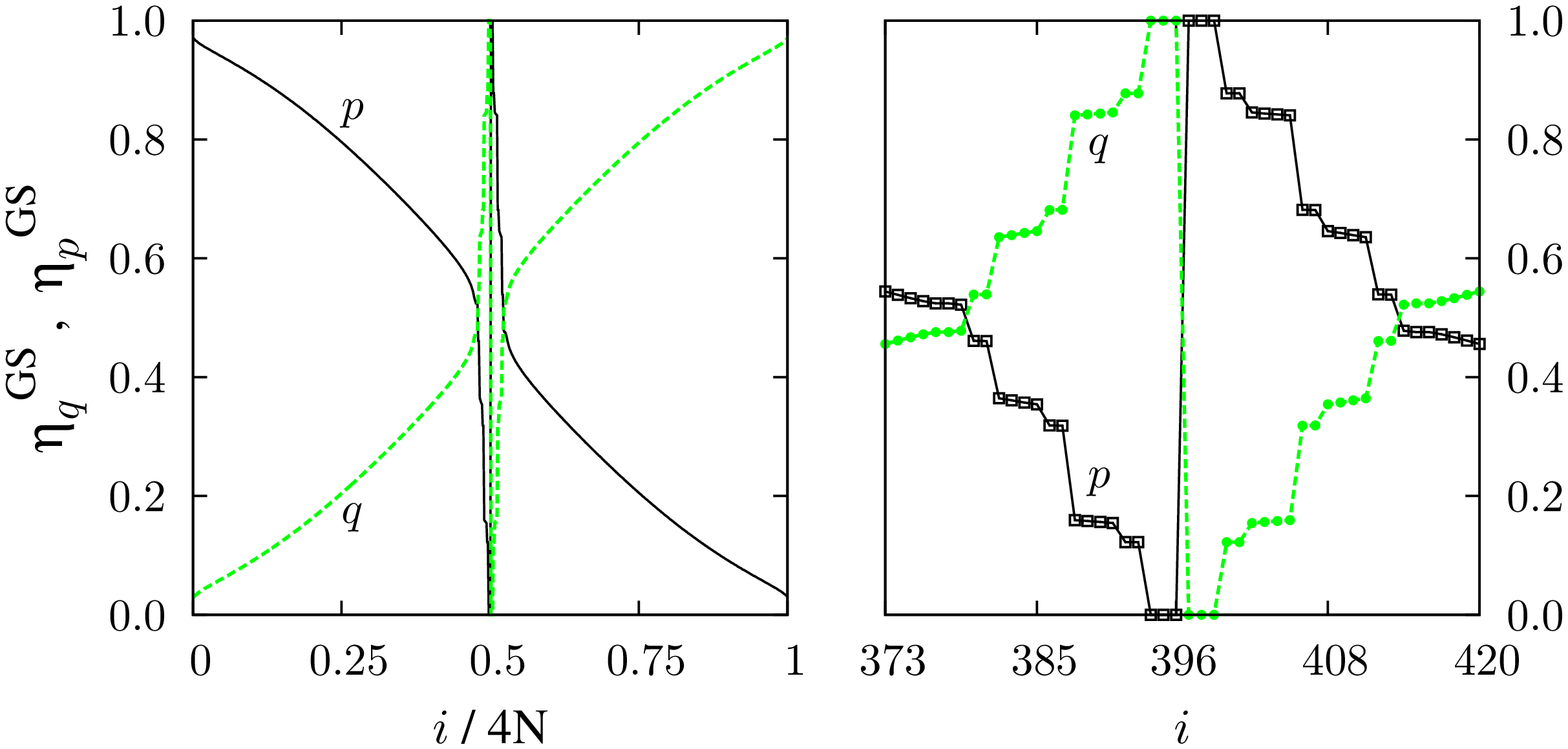}
\caption{Mean squared projections  for the full  Gram-Schmidt vector set, $i=1,\cdots, 4N$, 
onto the coordinate subspace ${\vec Q}$, $\eta_q^{GS}$ (green line), and 
the momentum subspace ${\vec P}$, $\eta_p^{GS}$ (black line), for the 198-particle system
defined above. Left panel: full spectrum;  Right panel: magnification of the central 
mode-carrying region.}
\label{Figure_6}
\includegraphics[angle=-90,width=0.8\textwidth]{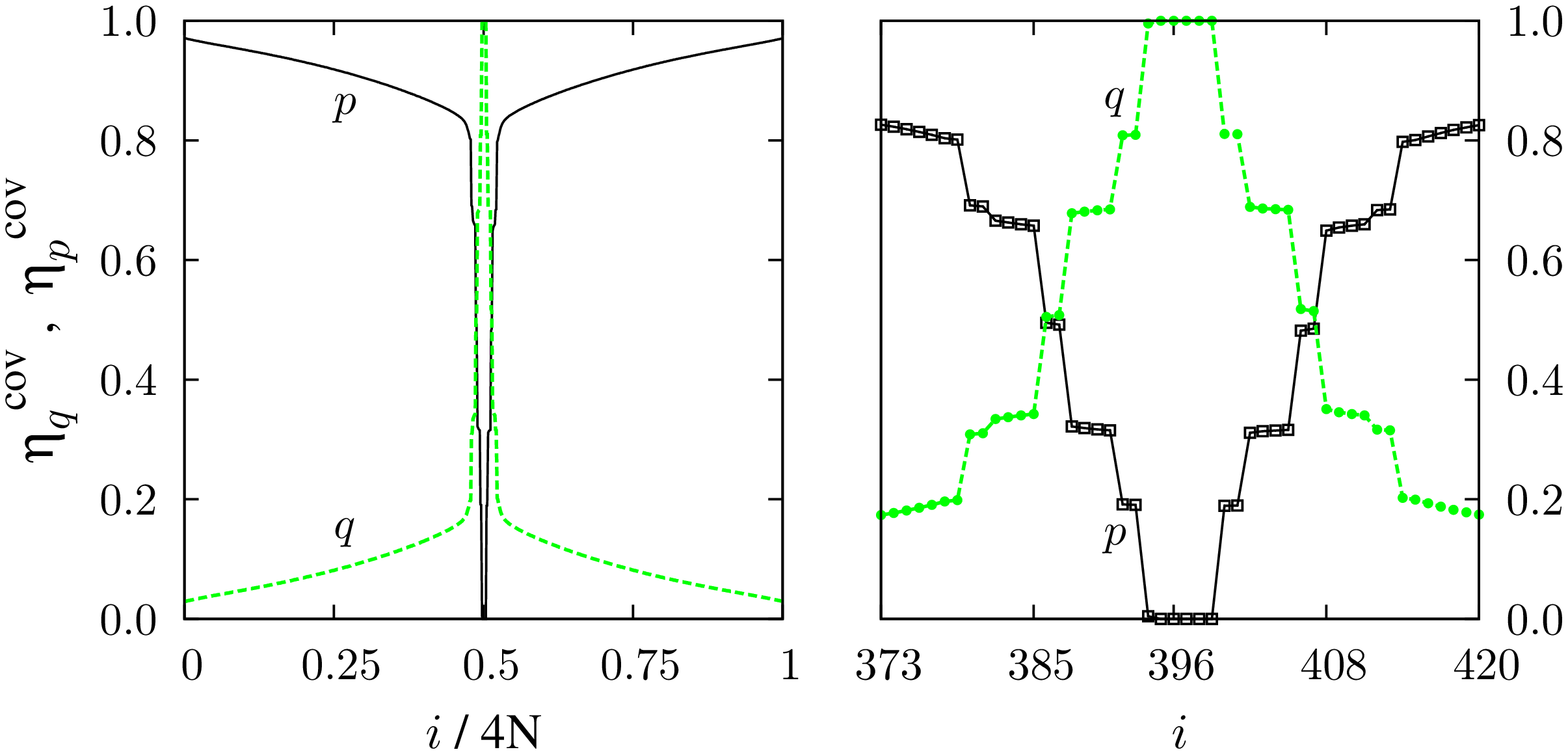}
\caption{Mean squared projections  for the full covariant vector set, $i=1,\cdots, 4N$, 
onto the coordinate subspace ${\vec Q}$, $\eta_q^{cov}$ (green line), and 
the momentum subspace ${\vec P}$, $\eta_p^{cov}$ (black line), for the 198-disk system
defined above.
Left panel: full spectrum;  Right panel: magnification of the central mode-carrying region.}
\label{Figure_7}
\end{figure}
It is interesting to see how much the coordinate and momentum subspaces contribute 
to a particular tangent vector $\delta {\vec \Gamma}$ (see Eq. \ref{state}), which may be a
Gram-Schmidt vector ${\vec g}^i$
or a  covariant vector ${\vec v}^i$, both associated with the same Lyapunov exponent $\lambda_i$.
The time-averaged squared projections of $\delta {\vec \Gamma}$ 
onto the coordinate  and momentum subspaces ${\vec Q}$ and ${\vec P}$, respectively, are given by 
\begin{equation}
\eta_q = \left\langle \sum_{n=1}^N \delta {\vec q}_n^2 \right\rangle
\qquad , \qquad
\eta_p = \left\langle \sum_{n=1}^N \delta {\vec p}_n^2 \right\rangle
\end{equation}
and are plotted in Fig. \ref{Figure_6} for the whole set of Gram-Schmidt vectors,
and in Fig.  \ref{Figure_7} for the whole set of covariant vectors,
$i=1,\cdots,4N$. One notes that for the Gram-Schmidt case the contributions of
${\eta_q}$ and ${\eta_p}$ to a vector ${\vec g}^i$ and its conjugate ${\vec g}^{4N+1-i}$ are interchanged, whereas for the covariant vectors ${\vec v}^i$ and ${\vec v}^{4N+1-i}$
they are the same. This is particularly noticeable for the expanded central regions
in the respective right panels of Figs. \ref{Figure_6} and \ref{Figure_7}.


\subsection{Central manifold and vanishing exponents}
\label{vanishing_exponents}
The dynamics of a closed particle system such as ours is strongly affected by the
inherent continuous symmetries, which  leave the Lagrangian and, hence, the equations of motion
invariant. The symmetries relevant for our two-dimensional system with periodic
boundaries are the homogeneity of time
(or invariance with respect to time translation), and the homogeneity of space (or
invariance with respect to space translations in two independent directions). Each
of these symmetries is associated with two vector fields with sub-exponential growth (or decay) and, therefore, gives rise to two vanishing Lyapunov exponents \cite{Gaspard}.  At any phase-space point
${\vec \Gamma}$, the six vectors span a six-dimensional subspace ${\cal N}({\vec \Gamma})$
of the tangent space  ${\bf TX}({\vec \Gamma})$, which is referred to as null space
or central manifold. This subspace is covariant. 
If the $4N$ components of the state vector are arranged as
\begin{equation}
\g\,=\,\left( q_x^1,q_y^1,\ldots,q_x^N,q_y^N \,;\, p_x^1,p_y^1,\ldots,p_x^N,p_y^N \right),
\end{equation}
the six orthogonal spanning vectors, which are the
generators of the elementary symmetry transformations, are given by \cite{FHPH2004,Eckmann:2005}
\begin{eqnarray}
\vec e_1 &=& \dfrac{1}{\sqrt{2 K}\,} \, ({p_x^1},{p_y^1},\ldots,{p_x^N},{p_y^N}\,;\,0,0,\ldots,0,0)\enskip ,\\
\vec e_2 &=& \dfrac{1}{\sqrt{N}\,} \, (1,0,\ldots,1,0\,;\,0,0,\ldots,0,0)\enskip ,\\
\vec e_3 &=& \dfrac{1}{\sqrt{N}\,} \, (0,1,\ldots,0,1\,;\,0,0,\ldots,0,0)\enskip, \\ 
\vec e_4 &=& \dfrac{1}{\sqrt{2 K}\,} \, (0,0,\ldots,0,0\,;\,{p_x^1},{p_y^1},\ldots,{p_x^N},{p_y^N})\enskip ,\\
\vec e_5 &=& \dfrac{1}{\sqrt{N}\,} \, (0,0,\ldots,0,0\,;\,1,0,\ldots,1,0)\enskip ,\\
\vec e_6 &=& \dfrac{1}{\sqrt{N}\,} \, (0,0,\ldots,0,0\,;\,0,1,\ldots,0,1) \enskip .\\ \nonumber
  \end{eqnarray}
$\vec  e_1$ corresponds to a change of the  time origin, 
$\vec e_4$ to a change of energy, $\vec e_2$ and $\vec e_3$  to an (infinitesimal) uniform translation 
of the origin in the $x$ and $y$ directions, respectively, and $\vec e_5$ and $\vec e_6$ to a
perturbation of the total momentum in the $x$ and $y$ directions, respectively.   
The six vanishing Lyapunov exponents are located in the center of the Lyapunov spectrum
with indices $2N-2 \le i \le 2N+3$.
The first three of these vectors have non-vanishing components only for the position perturbations in
the $2N$-dimensional configuration subspace ${\vec Q}$,
the remaining only for the momentum perturbations in the $2N$-dimensional momentum subspace ${\vec P}$.  
They are related by  $\vec e_{k} = \mathbf{J} \, \vec e_{k+3}$ for $ k \in \left\{ 1,2,3 \right\}$, 
where ${\bf J}$ is the symplectic (skew-symmetric) matrix.

Let us  consider the projection matrices  $\alpha$ and $\beta$ of the GS and
covariant vectors, respectively,  onto the natural basis,
\begin{equation}
\alpha_{i,k} =  {\vec g}^i \cdot {\vec e}_k;  \quad  
\beta_{i,k} =  {\vec v}^i \cdot {\vec e}_k,  \quad
 k \in \left\lbrace1,\ldots,6 \right\rbrace \quad i \in \{2N-2, \cdots, 2N+3\}.
 \nonumber
\end{equation}
For $ i \notin \left\{ 2N-2 ,\ldots, 2N+3 \right\}$ these components vanish.
Without loss of generality, we consider in the following example a system with only $N=4$ particles 
in a periodic box, which is relaxed for $t_r = 10^6$ time units, followed by a forward and backward
iteration lasting for $t_{\omega} - t_0 = 10^5$ time units. Very special initial conditions for the
backward iteration ${\vec v}_{\omega}^i = {\vec g}_{\omega}^i$ for $ i = 1,\cdots , 4N (= 16)$ are used.  
The projections at the time $t_0$ are given in Table \ref{table_alpha} for the 
GS vectors, in Table \ref{table_beta} for the covariant vectors.
\begin{center}
\begin{small}
\begin{table}[htbp]
\caption{Instantaneous projection matrix $\alpha$ of  Gram-Schmidt vectors 
(for $i \in \{2N-2,\cdots,2N+3\}$) onto the 
natural basis $\left\{ {\vec e}_k , 1\leq k \leq 6 \right\} $ of the central manifold.
The system contains $N = 4$ disks. The powers of 10 are given in square brackets.}
\label{table_alpha}
\begin{tabular*}{1\textwidth}{@{\extracolsep{\fill}}p{1.4cm} |r r r r r r}
$i$ & $\alpha_{i,1}$ & $\alpha_{i,2}$ & $\alpha_{i,3}$ & $\alpha_{i,4}$ & $\alpha_{i,5}$ & $\alpha_{i,6}$\\
\hline
$2N-2$ &     -0.766                           &    0.582 &      0.273                               &  $-0.766[-6] $& $0.582 [-6]$&  $0.273[-6]$\\
$2N-1$ &       0.256                          &   -0.114 &      0.960                               &   $0.256 [-6]$& $-0.114[-6]$ &  $0.960[-6]$\\
$2N$    &       0.590                           &    0.805 &     -0.062                               &  $0.590 [-6]$& $0.805  [-6]$ & $-0.062[-6]$\\
$2N+1$ & $-0.611 [-6]$& $ 0.782 [-6]$ &  $-0.121 [-6]$&    0.611 &     -0.782 &     0.121\\
$2N+2$ & $ 0.575 [-6]$& $ 0.544 [-6]$   &  $0.611 [-6]$ &    -0.575 &     -0.544 &     -0.611\\
$2N+3$ & $-0.543[-6]$&  $-0.304 [-6]$ &  $ 0.783 [-6]$&     0.543 &      0.304 &     -0.783\\
\end{tabular*}
\end{table}
\begin{table}[htbp]
\caption{Instantaneous projection matrix matrix $\beta$ for the the six
central covariant vectors onto the 
natural basis $\left\{ {\vec e}_k , 1\leq k \leq 6 \right\} $ of the central manifold.
The system contains of $N=4$ particles. The powers of 10 are given in square brackets.}
\label{table_beta}
\vspace{4mm}
\begin{tabular*}{1\textwidth}{@{\extracolsep{\fill}}p{1.4cm} | r r r r r r}
$i$ & $\beta_{i,1}$ & $\beta_{i,2}$ & $\beta_{i,3}$ & $\beta_{i,4}$ & $\beta_{i,5}$ & $\beta_{i,6}$\\
\hline
$2N-2$ &     -0.766    &    0.582 &      0.273         &  $-0.766[-6] $& $0.582 [-6]$&  $0.273[-6]$\\
$2N-1$ &       0.256   &   -0.114 &      0.960         &   $0.256 [-6]$& $-0.114[-6]$ &  $0.960[-6]$\\
$2N$    &       0.590    &    0.805 &     -0.062         &  $0.590 [-6]$& $0.805  [-6]$ & $-0.062[-6]$\\
$2N+1$ & -0.611 &  0.782  &  -0.121&      0.611 [-5] &   -0.782 [-5]&     0.121 [-5]\\
$2N+2$ &  0.575 &  0.544   &  0.611 &    -0.575  [-5] &   -0.544[-5] &     -0.611[-5]\\
$2N+3$ & -0.543 &  -0.304 &   0.783 &     0.543  [-5]&     0.304 [-5]&     -0.783[-5]\\

\end{tabular*}
\end{table}
\end{small}
\end{center}
A comparison of the two tables reveals the following: \\
$\bullet$ The six orthogonal GS 
vectors ${\vec g}^i; i=2N-2, \cdots, 2N+3$  completely span the null subspace (the squared
elements for each rows add up to unity in Table \ref{table_alpha}). The same is true for the six non-orthogonal
covariant vectors   ${\vec v}^i; i=2N-2, \cdots, 2N+3$ in Table \ref{table_beta}.\\
$\bullet$ The first three covariant and Gram-Schmidt vectors completely agree. This is
a consequence of the special initial conditions for the former at the  time $t_{\omega}$ as mentioned above.
During the backward iteration the three covariant vectors stay in their respective subspaces and
remain parallel to the GS vectors  (which were stored during  the forward phase of the algorithm). 
At $t_0$ they are still identical to their GS counterparts. The first vectors always agree,
${\vec v}_0^{2N-2} = {\vec g}_0^{2N-2}$, if less special  initial conditions conforming 
to Eq. \ref{initial_condition}  are used. \\
$\bullet$  Equivalent components have the same mantissa but may differ by a
factors of $10^5$ or $10^6$, which are related to  the duration of the relaxation phase $t_r$ and
of the forward-backward iteration time $t_{\omega} - t_0$.\\
The explanation for this behavior \cite{Hadrien} is obtained by a repeated explicit application of the
linearized maps for the free streaming and consecutive  collision of particles \cite{DP1997,DPH1996} 
to the six basis vectors ${\vec e}^i$. 
 One finds  that
\begin{eqnarray}
 D \phi^t _{{\bf \Gamma}_0} \cdot {\vec e}^j \left({\bf \Gamma}_0\right) &=& {\vec e}^j \left({\bf \Gamma}_t\right), 
 \label{first} \\
 D \phi^t _{{\bf \Gamma}_0}  \cdot {\vec e}^{j+3} \left({\bf \Gamma}_0\right)  &=& t \; {\vec e}^j \left({\bf \Gamma}_t 
 \right) + {\vec  e}^{j+3}\left({\bf \Gamma}_t\right) ,  \label{second}
 \end{eqnarray}
for $j \in \{1,2,3\}$.
 Eq. (\ref{second}) implies that any perturbation vector with
non-vanishing components parallel to ${\vec e}_4$, ${\vec e}_5 $, or ${\vec e}_6$ will rotate towards
${\vec e}_1$, ${\vec e}_2 $, and ${\vec e}_3$, respectively.  
It follows i) that the null subspace ${\cal N}({\bf \Gamma}) $  is covariant; 
ii) that the subspaces ${\cal N}_1 = \mbox{span}\{{\vec e}_1\}$,  ${\cal N}_2 = \mbox{span}\{{\vec e}_2\}$
and ${\cal N}_3 = \mbox{span}\{{\vec e}_3\}$ are separately covariant (from Eq. (\ref{first}));
that, as was already noted in Ref. \cite{Eckmann:2005},  ${\cal N}$ can be further decomposed into the three two-dimensional covariant subspaces
 ${\cal N}_p = \mbox{span}\{{\vec e}_1, {\vec e}_4\}$, 
${\cal N}_x = \mbox{span}\{{\vec e}_2, {\vec e}_5\}$, and ${\cal N}_y = \mbox{span}\{{\vec e}_3, {\vec e}_6\}$.


\subsection{Lyapunov modes}
\label{modes}
We have seen in Section \ref{localization} that the perturbation vectors are  less
and less localized, the smaller the Lyapunov exponents become, until they are 
coherently spread out over the physical space and form periodic spatial patterns with a
well-defined wave vector ${\vec k}$.  This collective patterns are referred  to as Lyapunov modes. 
The modes were observed for hard  particle systems in 
one, two and three dimensions \cite{PH2000,HPFDZ2002,Eckmann:2005,TM2003a,TM2003b},
for hard planar dumbbells \cite{MPH1998a,MPH1998b,MP2002} and  for one and
two-dimensional soft particles \cite{RY2004,YR2004,FP2005}. A formal classification of the 
modes is given in Ref. \cite{Eckmann:2005}. Physically, they are interpreted as
periodic modulations with wave number $(k \ne 0)$ of the null modes associated with the elementary 
continuous symmetries and conservation laws. Since this modulation involves the
breaking of such symmetries, the modes have been interpreteted as Goldstone modes
\cite{Henk}. Theoretical approaches are  based on random matrix theory
\cite{EG2000,TM2002}, periodic orbit expansion \cite{TDM2002}, and kinetic theory
\cite{NM2001,MN2004,Henk}. 

So far the numerical work on Lyapunov modes has been exclusively concerned with the
orthonormal Gram-Schmidt vectors $\{ {\vec g}^i \}, \; i = 1,\cdots , 4N$.
The purpose of this section is to point out some differences one encounters, if the modes
for the Gram-Schmidt and covariant vectors are compared.
\begin{figure}[htbp]
\centering
\includegraphics[angle=-90,width=0.7\textwidth]{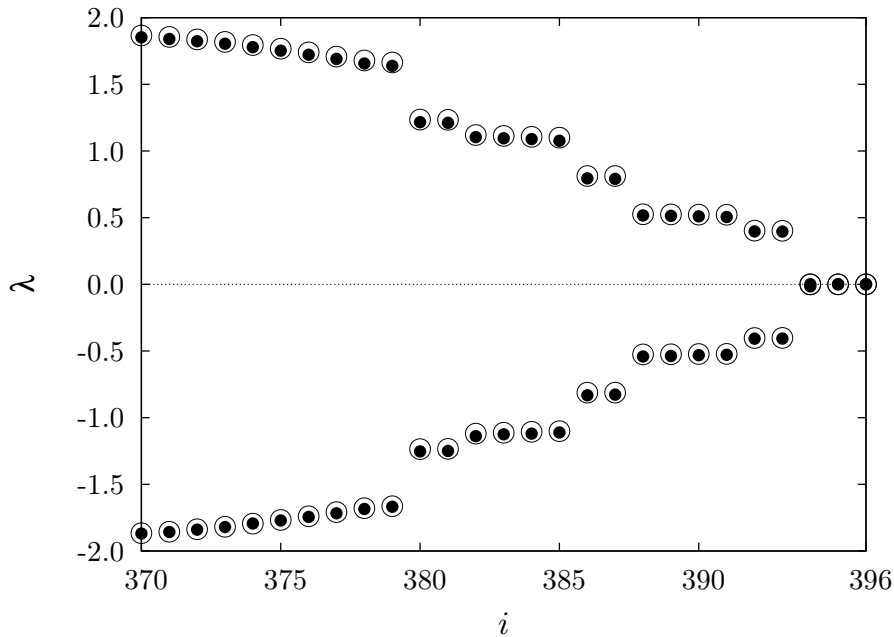}
\caption{Enlargement of the mode regime for the Lyapunov spectrum depicted in Fig. \ref{spectrum}.
The open symbols indicate exponents computed from the Gram-Schmidt vectors, 
the full dots are for exponents obtained from the covariant vectors.}
\label{Figure_8}
\end{figure}

Fig. \ref{Figure_8} shows an enlargement of the mode-carrying region for the 
Lyapunov spectrum of Fig. \ref{spectrum}. In order to emphasize the conjugate
pairing symmetry $\lambda_i = - \lambda_{4N+1 - i}$ for symplectic systems, 
conjugate exponent pairs are plotted with the same index $i$ on the abscissa, 
where now $i \in \{1,\cdots,2N\}.$ The open circles are computed from the Gram-Schmidt
vectors in the forward direction of time, the dots from the covariant vectors 
during the time-reversed iteration. Considering the size of the system $(N =198)$, 
the agreement is excellent. 

The steps in the spectrum due to degenerate exponents
is a clear indication for the presence of Lyapunov modes.  According to the
classification in our previous work \cite{Eckmann:2005}, the steps with a
two-fold degeneracy are transverse (T) modes  -- T(1,0), T(2,0) and T(3,0)  from 
right to left in Fig. \ref{spectrum}.
Similarly, the steps with a four-fold degeneracy of the exponents are 
longitudinal-momentum (LP) modes -- LP(1,0), LP(2,0) and LP(3,0) again from the right.
The arguments $(n_x,n_y)$ account for  the number of periods of the sinusoidal
perturbations in the $x$ and $y$ directions.  Since our simulation cell is rather narrow, only wave 
vectors ${\vec k}$ parallel to the $x$ axis of the (periodic) cell appear, leaving 
$0$ for the second argument
\cite{Eckmann:2005}. As usual, ``transverse'' and ``longitudinal'' 
refer to the spatial polarization with respect to ${\vec k}$ 
of the wave-like pattern.

\begin{figure}[htbp]
\centering
\includegraphics[angle=-90,width=0.7\textwidth]{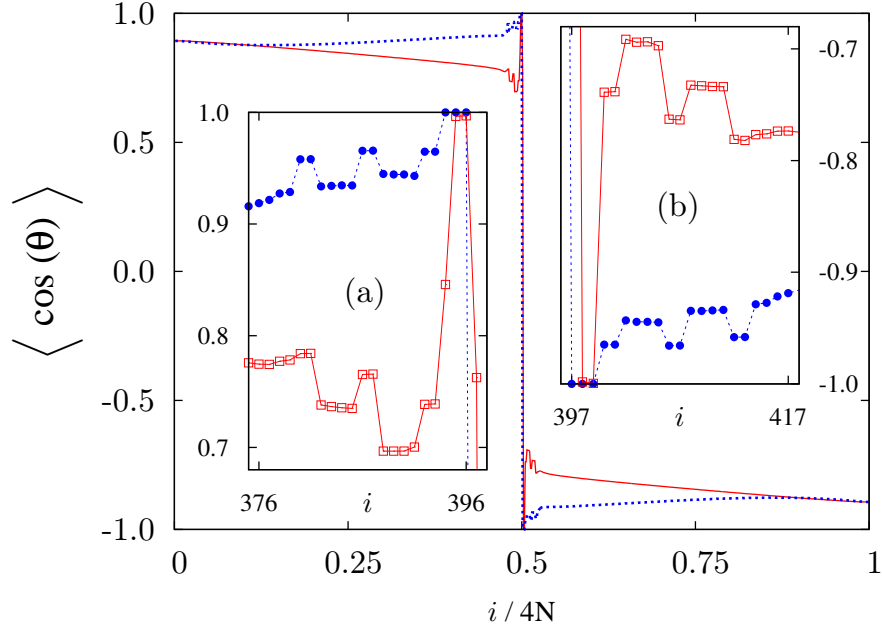}
\caption{Time averaged value of $\cos(\Theta) = (\delta {\vec q} \cdot \delta {\vec p}) /
  ( | \delta {\vec q}|   |\delta {\vec p}|)   $ as a function of the Lyapunov index $i$
for a system with $N = 198$ hard disks. Here, $\delta {\vec q} \in {\vec Q} $ and 
$ \delta {\vec p} \in {\vec P}$ are the $2N$-dimensional vectors of all position perturbations respective 
all momentum perturbation for the Gram-Schmidt vectors ${\vec g}_i$ (blue line) and the covariant 
vectors  ${\vec v}_i$  (red line).  The insets are magnifications of the
mode-carrying region. }
\label{Figure_9}
\end{figure}

One of our early observations, which greatly facilitates the classification of the modes 
for the Gram-Schmid vectors \cite{Eckmann:2005}, is that in the limit $N \to \infty$ 
the cosine of the angle $\Theta$ between the $2N$-dimensional vectors of the position perturbations 
and  momentum perturbations converges to +1 for the smallest positive,
 and to -1 for the smallest negative exponents. 
See the blue line in Fig. \ref{Figure_9}. Furthermore, the relation
\begin{equation}
\delta{\vec p} = C_{\pm} \delta{\vec q}
\label{C}
\end{equation}
holds with known constants $C_{\pm}$. 
This means that these vectors  are nearly parallel or anti-parallel for large $N$ and that 
the mode classification may be based solely on $\delta{\vec q}$. 
Somewhat surprisingly, this property does not strictly
hold anymore for the covariant vectors. This is shown by the red line in Fig. \ref{Figure_9}, where
$\cos(\Theta)$ is seen to differ significantly from $\pm 1$ for all $i$ outside of the null subspace
(for which $394 \le i \le 399$).
Unfortunately, this has dire consequences for the representation of the covariant vector modes,
since they cannot be purely understood as a vector field of the position perturbations only as in the
GS case. For the purpose of this paper, however,  we restrict to the GS-based
classification of Ref. \cite{Eckmann:2005}.

\begin{table}[bt]
\caption{Basis vectors of $(n_x,0)$ modes for a hard disk system in a rectangular box with periodic boundaries. We use the notation  $c_x=\cos (k_x \, x)$, and $s_x=\sin (k_x \, x)$, where the wave vector 
is given by ${\vec k} = (k_x, k_y) = ( 2 \pi n_x/ L_x , 0)$. Here $n_x \in \{ 1,2,3\}.$}
\begin{center}
\begin{tabular}{c|c|c c}
\hline
\hline
$\vec n$ & \text{Basis of } $\textbf{T}(\vec n)$ & \text{Basis of } $\textbf{L}(\vec n)$& \text{Basis of } $\textbf{P}(\vec n)$ \\
\hline
\begin{math} \left( \begin{array}{c}
 n_x \\
 0
\end{array}\right)\end{math}
 &
\begin{math}\quad \left( \begin{array}{c}
 0 \\
 c_x
\end{array}\right)\end{math} ,
\begin{math} \left( \begin{array}{c}
 0 \\
 s_x
\end{array}\right) \quad\end{math}
&
\begin{math}\quad \left( \begin{array}{c}
 c_x \\
 0
\end{array}\right)\end{math} ,
\begin{math} \left( \begin{array}{c}
 s_x \\
 0
\end{array}\right) \quad \end{math}
& 
\begin{math} \quad \left( \begin{array}{c}
 p_x \\
 p_y
\end{array}\right) s_x \end{math},
\begin{math} \left( \begin{array}{c}
 p_x \\
 p_y
\end{array}\right) c_x \quad \end{math}
\\
\hline
\hline
\end{tabular}
\end{center}
\label{mode_table}
\end{table}
\begin{figure}[htbp]
\centering
\begin{minipage}[c]{.45\linewidth}
\includegraphics[angle=-90,width=1\textwidth]{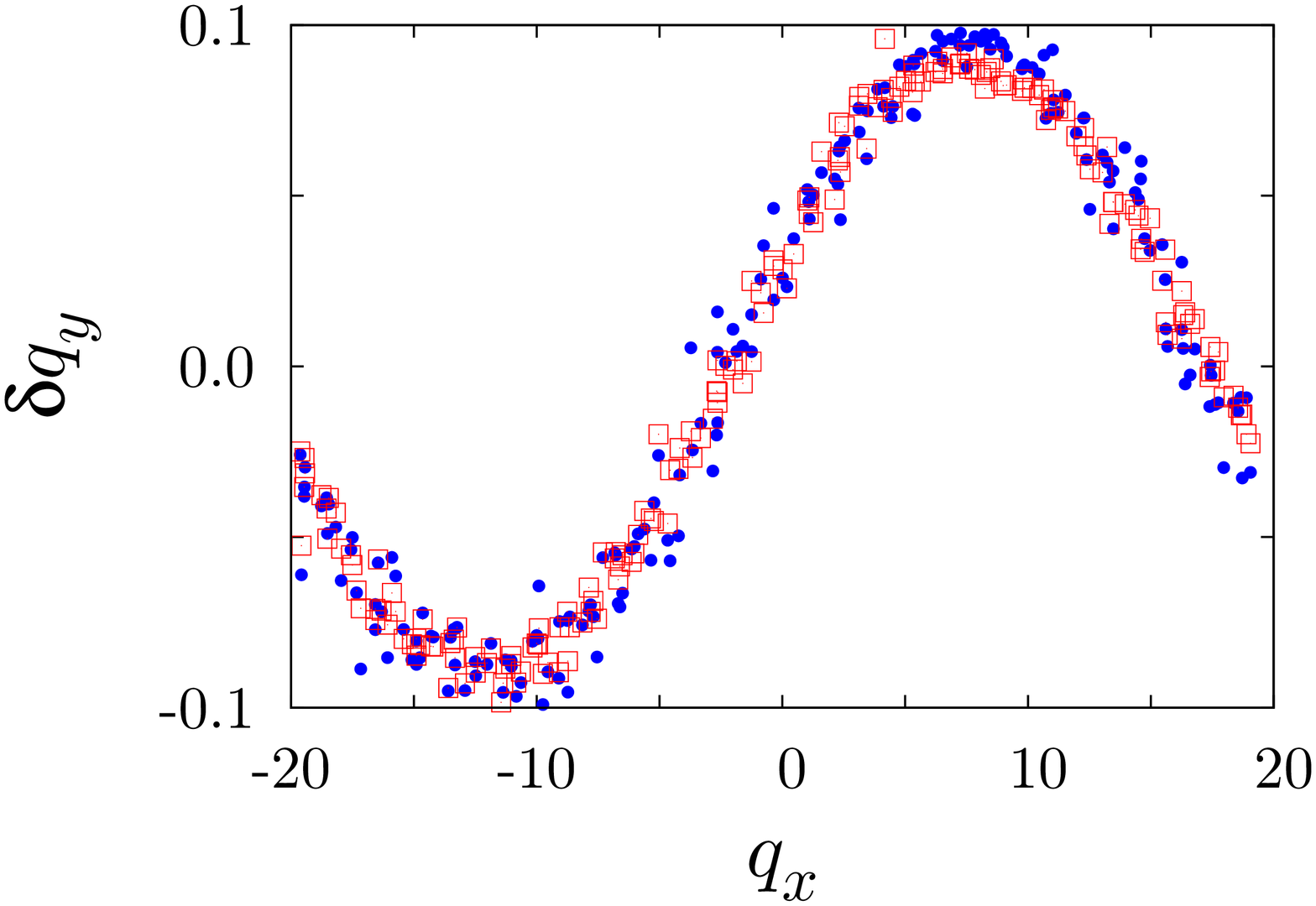}\\
\includegraphics[angle=-90,width=1\textwidth]{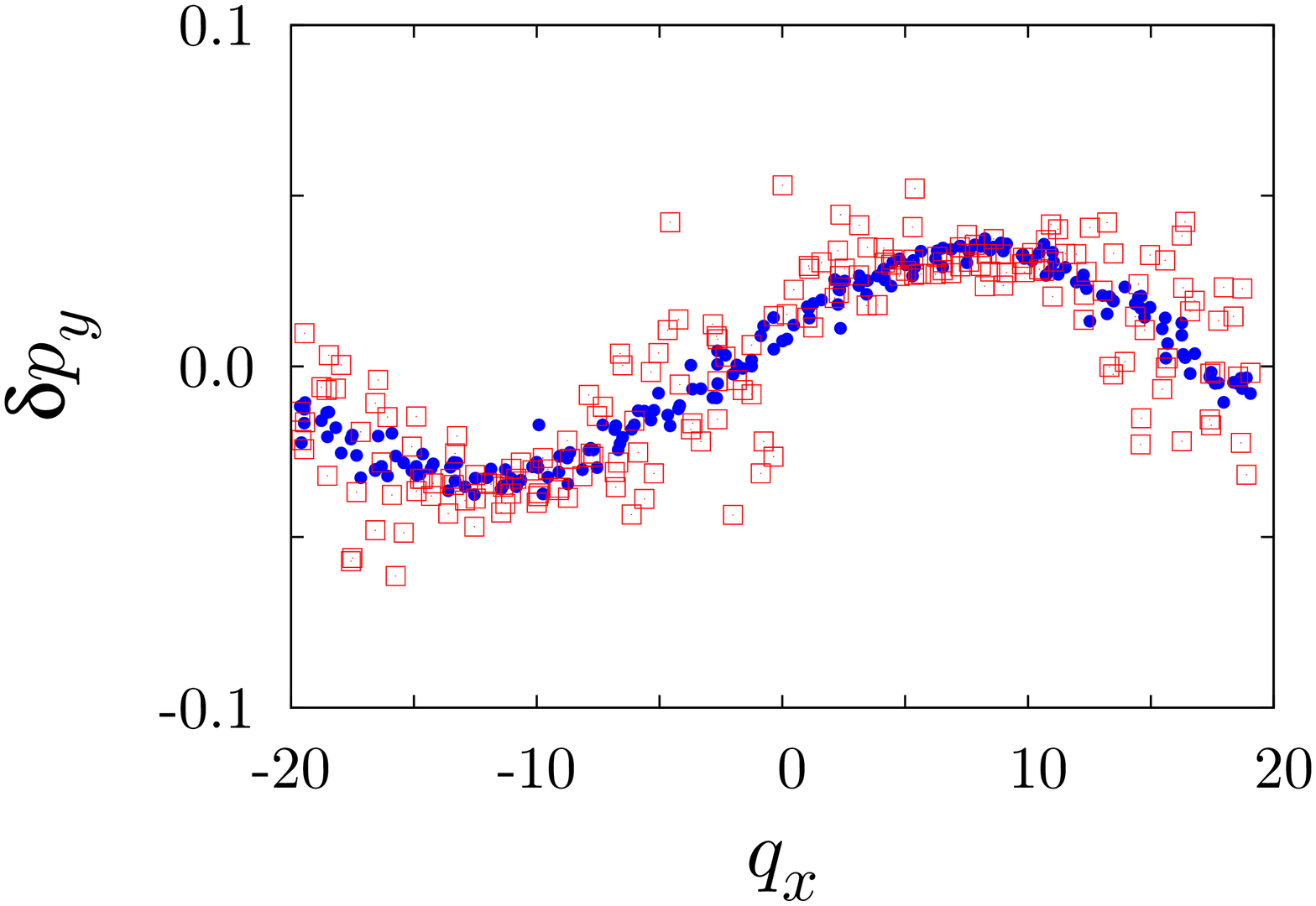}
\end{minipage}\hfil
\begin{minipage}[c]{.45\linewidth}
\includegraphics[angle=-90,width=1\textwidth]{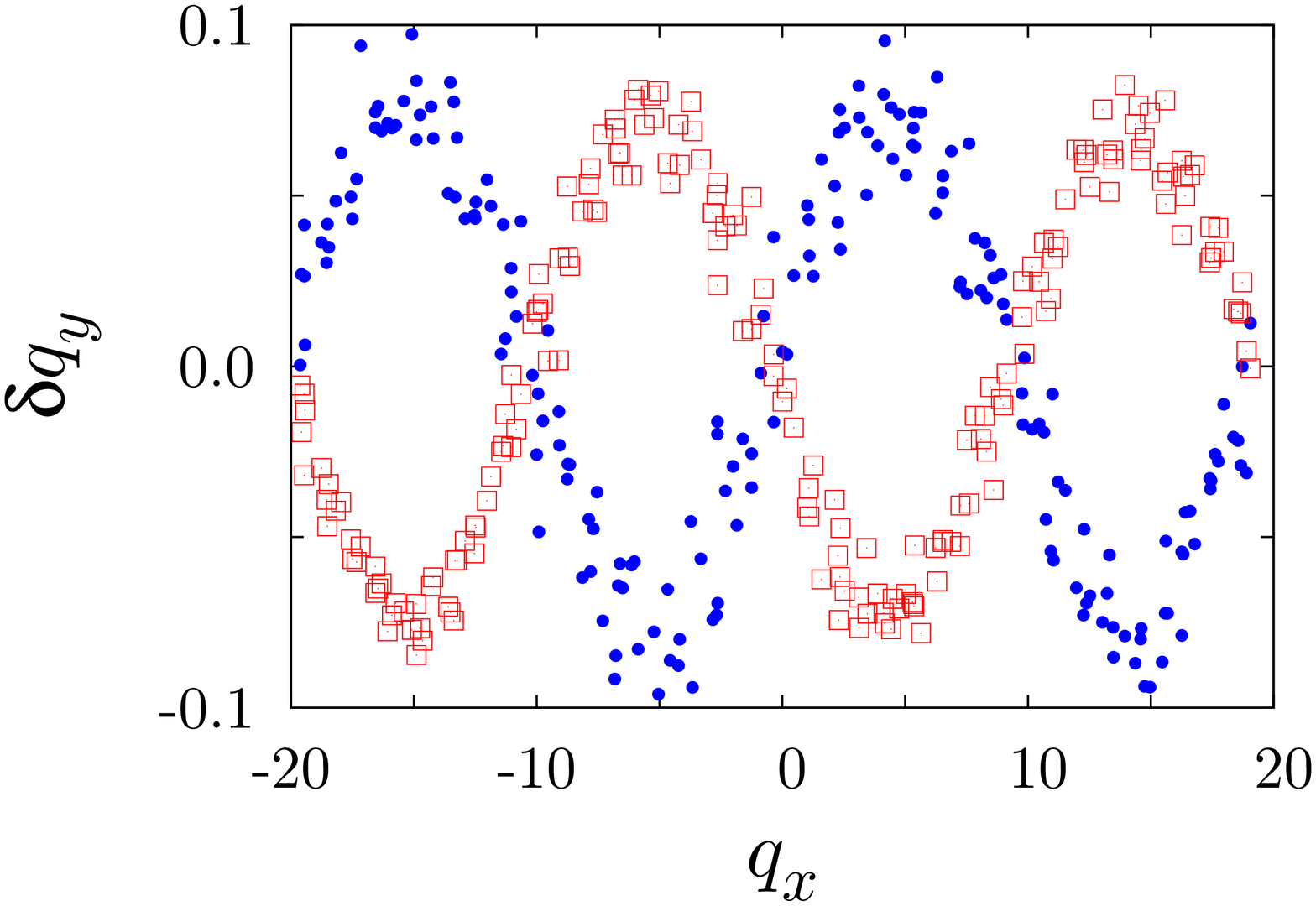}\\
\includegraphics[angle=-90,width=1\textwidth]{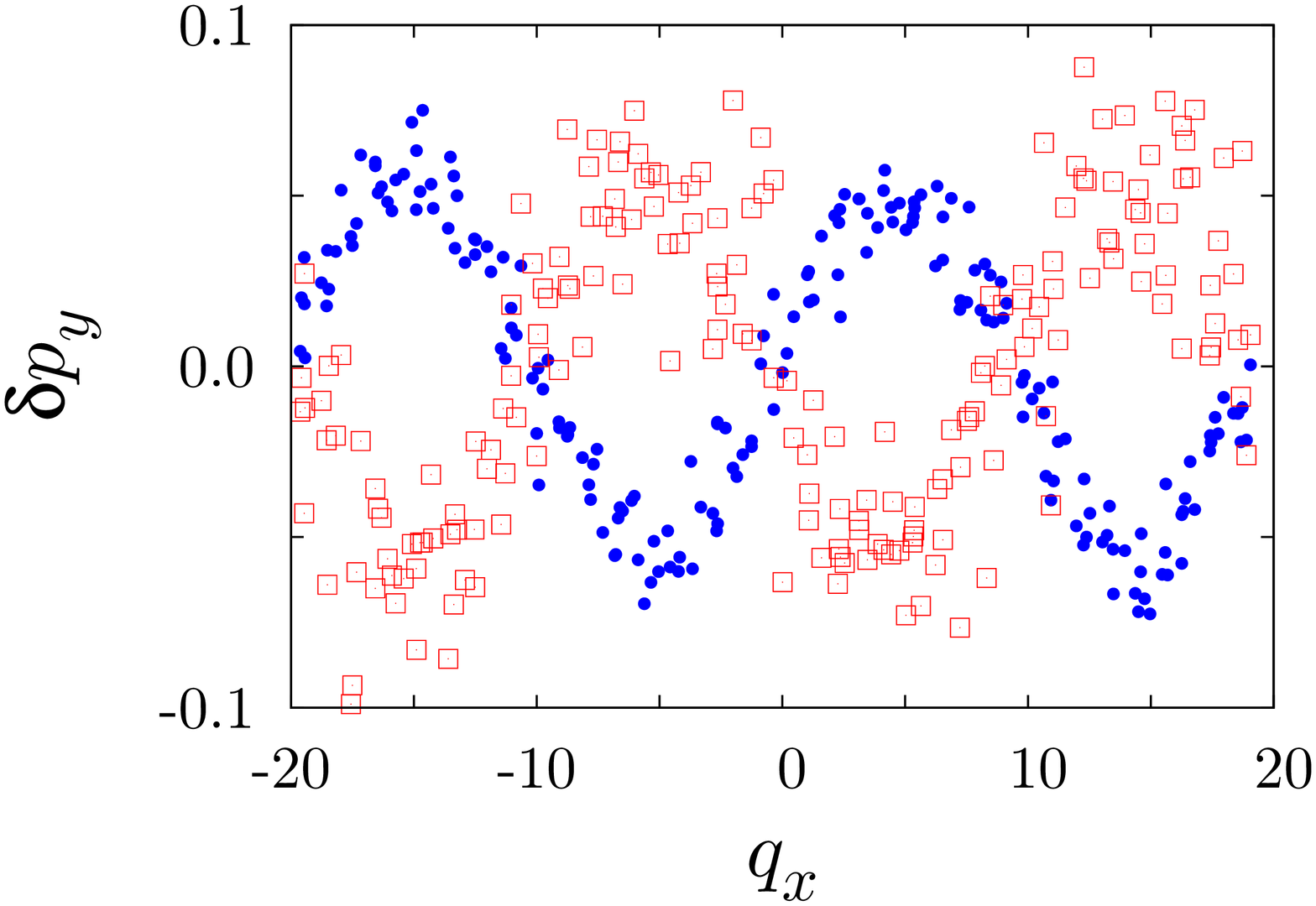}
\end{minipage}
\caption{Instantaneous transverse Lyapunov modes T(1,0) for index $i=393$ (left panels),  and T(2,0) for index
$i=387$ (right panels)  for the 198-disk system. In the panels at the top (bottom) the $y$-coordinate
perturbations $\delta q_y$
($y$-momentum perturbations $\delta p_y$) of all particles 
are plotted as a function of  their $x$ coordinate, $q_x$, in the simulation cell. The wave vector is 
parallel to the $x$ axis.
The blue dots are for Gram-Schmidt vectors, the red squares for covariant vectors. }
\label{Figure_10}
\end{figure}
Transverse modes are two-dimensional subspaces (for the periodic boundary conditions 
and a rectangular box), for which two orthogonal basis vectors are given in 
Table  \ref{mode_table}.
As an example, we show in the panels on the left-hand side of  Fig. \ref{Figure_10} 
snapshots of  the mode T(1,0) for the index $i = 393$, namely plots of $\delta q_y$ as a function of $q_x$ 
(top left), and of $\delta p_y$ as a function of $q_x$ (bottom left).
The respective plots for the $x$ components
fluctuate around  zero, as expected,  and are not shown. Analogous plots for the
mode T(2,0) with  $i=387$ are shown in the panels on the right-hand side.
The blue points are for  GS vectors, the red squares  for the respective covariant vectors.
It is interesting to note that the scatter of the points 
for the position perturbations is smaller  for the covariant modes (red squares) than for 
the GS modes (blue dots).  A fit shows that the residuals for the covariant modes are smaller
by about a factor of two in comparison to Gram-Schmidt. 
Quite the opposite is true for the momentum perturbations in the bottom row of panels.
Although the proportionality of Eq. (\ref{C}) still holds,
the scatter of the red squares for the covariant vectors is larger than that of the  blue dots
for the GS vectors. Such a behavior is always observed and is not simple numerical noise.
The reason for this behavior is related to the previous discussion in connection with Fig. \ref{Figure_9}
and needs further clarification. 
\begin{figure}[htbp]
\centering
\begin{minipage}[c]{.45\linewidth}
\includegraphics[angle=-90,width=1\textwidth]{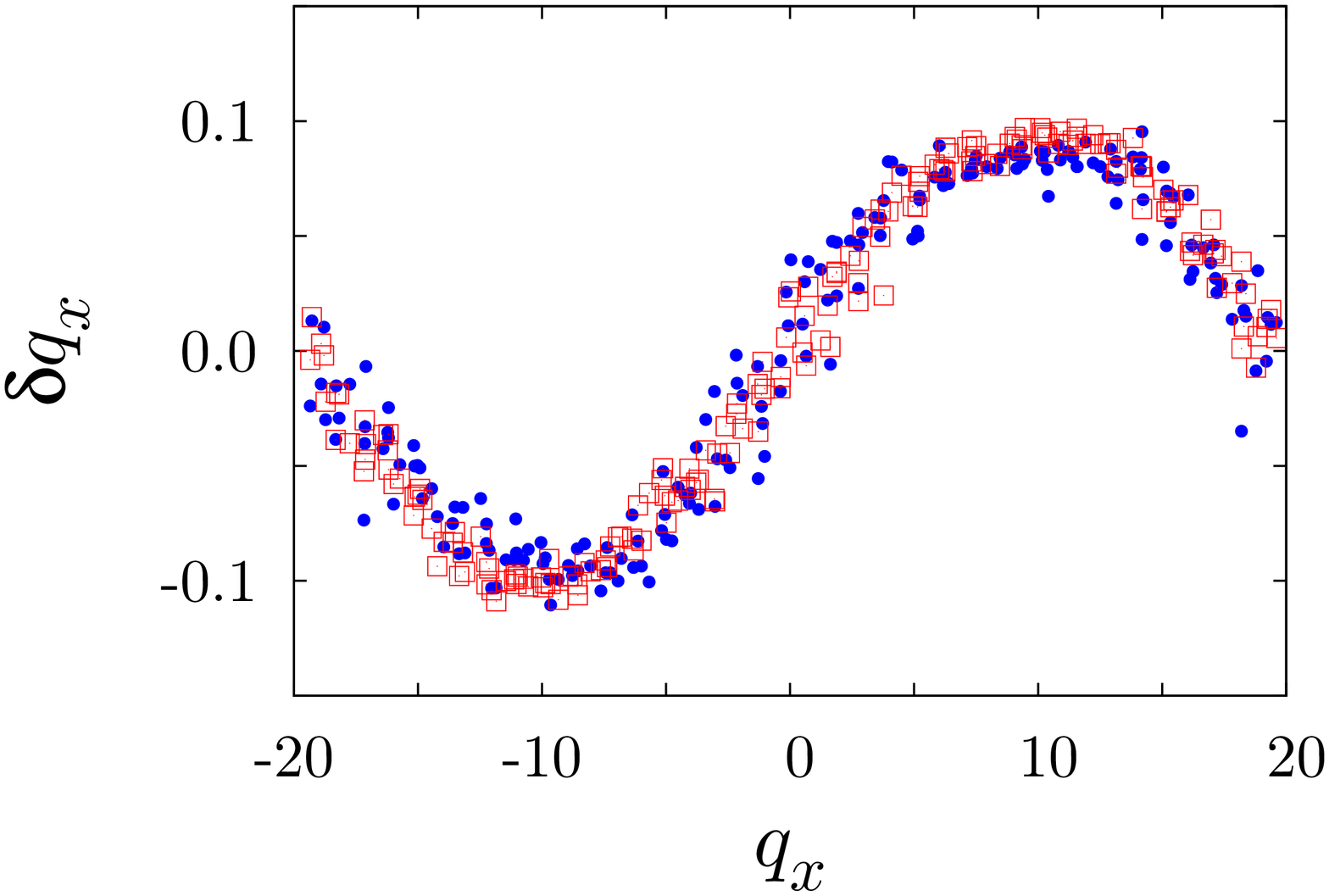}
\end{minipage} \hfil
\begin{minipage}[c]{.45\linewidth}
\includegraphics[angle=-90,width=1\textwidth]{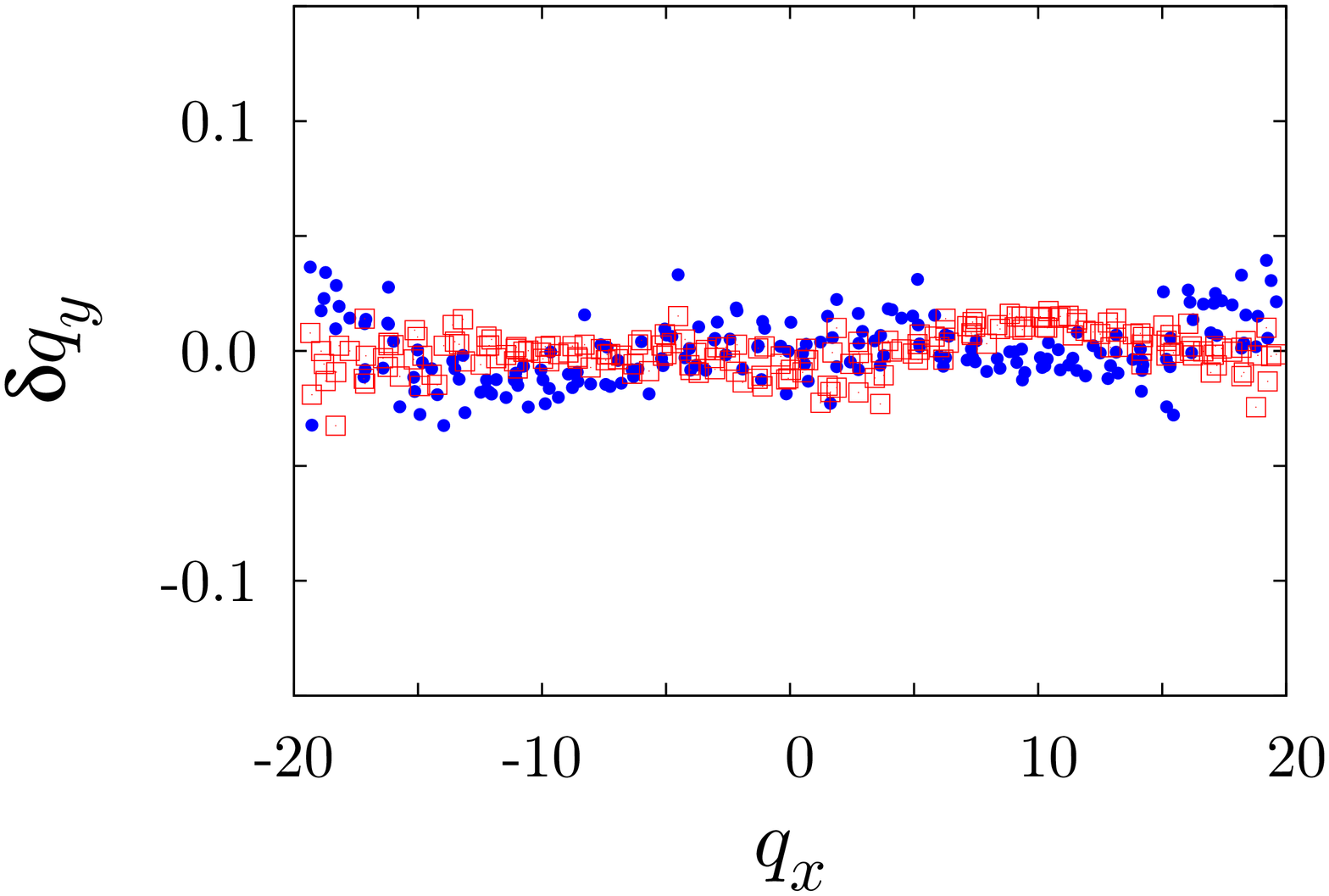}
\end{minipage} \hfil
L(1,0)\\
\begin{minipage}[c]{.45\linewidth}
\includegraphics[angle=-90,width=1\textwidth]{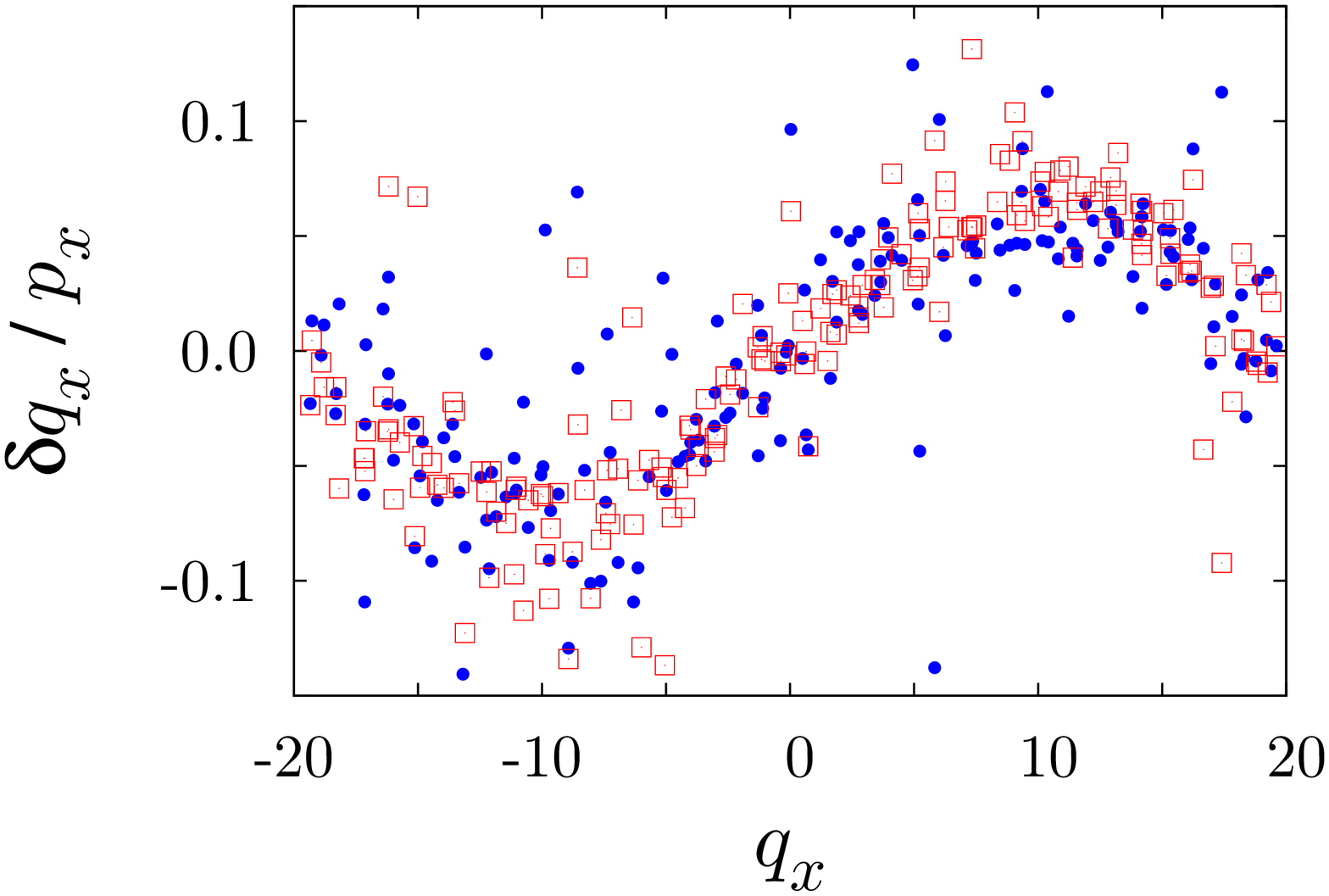}
\end{minipage} \hfil
\begin{minipage}[c]{.45\linewidth}
\includegraphics[angle=-90,width=1\textwidth]{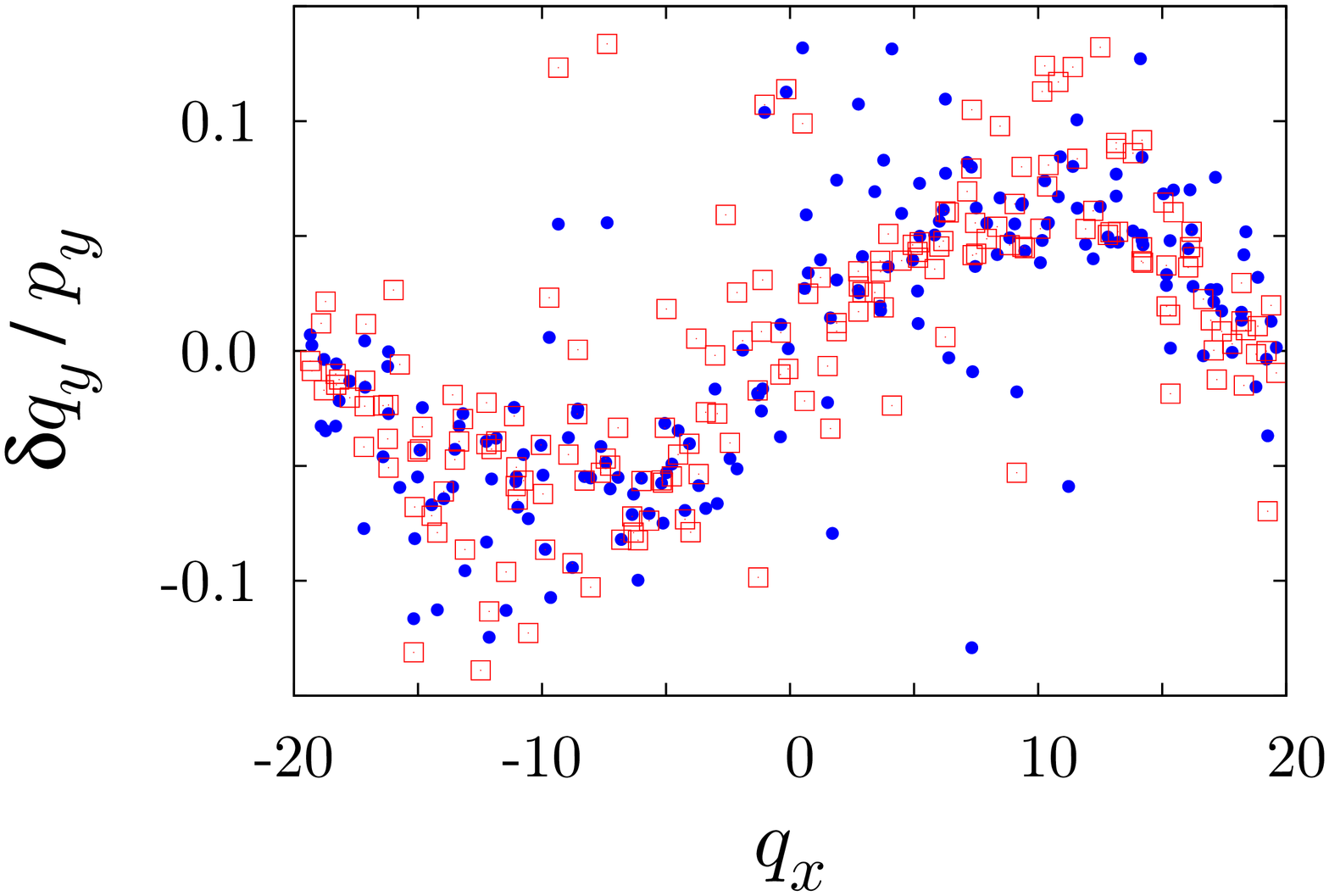}
\end{minipage} \hfil
P(1,0)\\
\caption{Reconstructed position perturbations of the pure L(1,0) mode 
(top panels) and P(1,0) mode (bottom panels). Only the patterns
proportional to $\sin(2 \pi q_x/ L_x)$ are shown.  
Blue dots: Gram-Schmidt vectors; Red squares: covariant vectors.
For details we refer to the main text.}
\label{Figure_11}
\end{figure}

\begin{figure}[htbp]
\centering
\begin{minipage}[c]{.45\linewidth}
\includegraphics[angle=-90,width=1\textwidth]{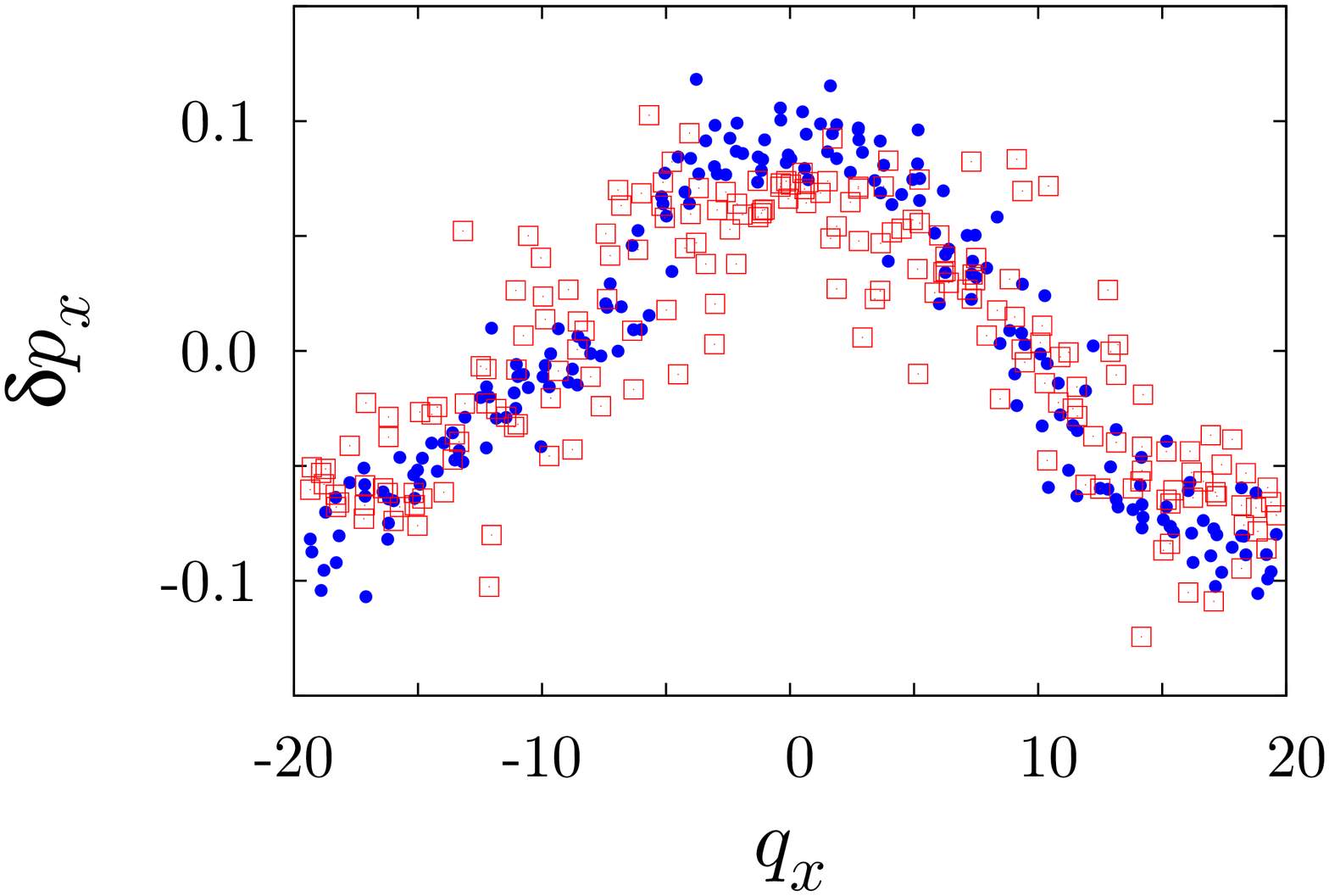}
\end{minipage} \hfil
\begin{minipage}[c]{.45\linewidth}
\includegraphics[angle=-90,width=1\textwidth]{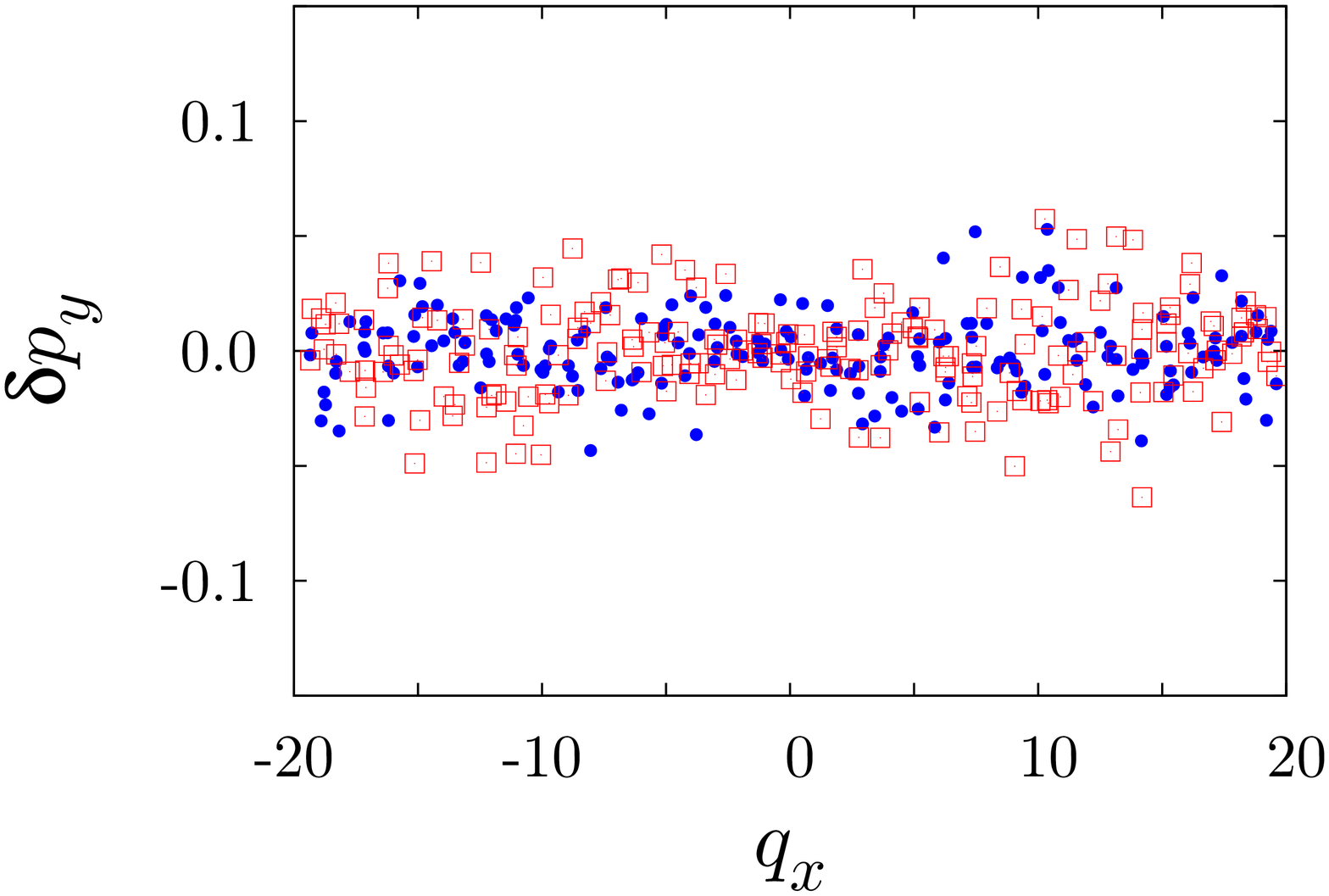}
\end{minipage} \hfil
L(1,0)\\ 
\begin{minipage}[c]{.45\linewidth}
\includegraphics[angle=-90,width=1\textwidth]{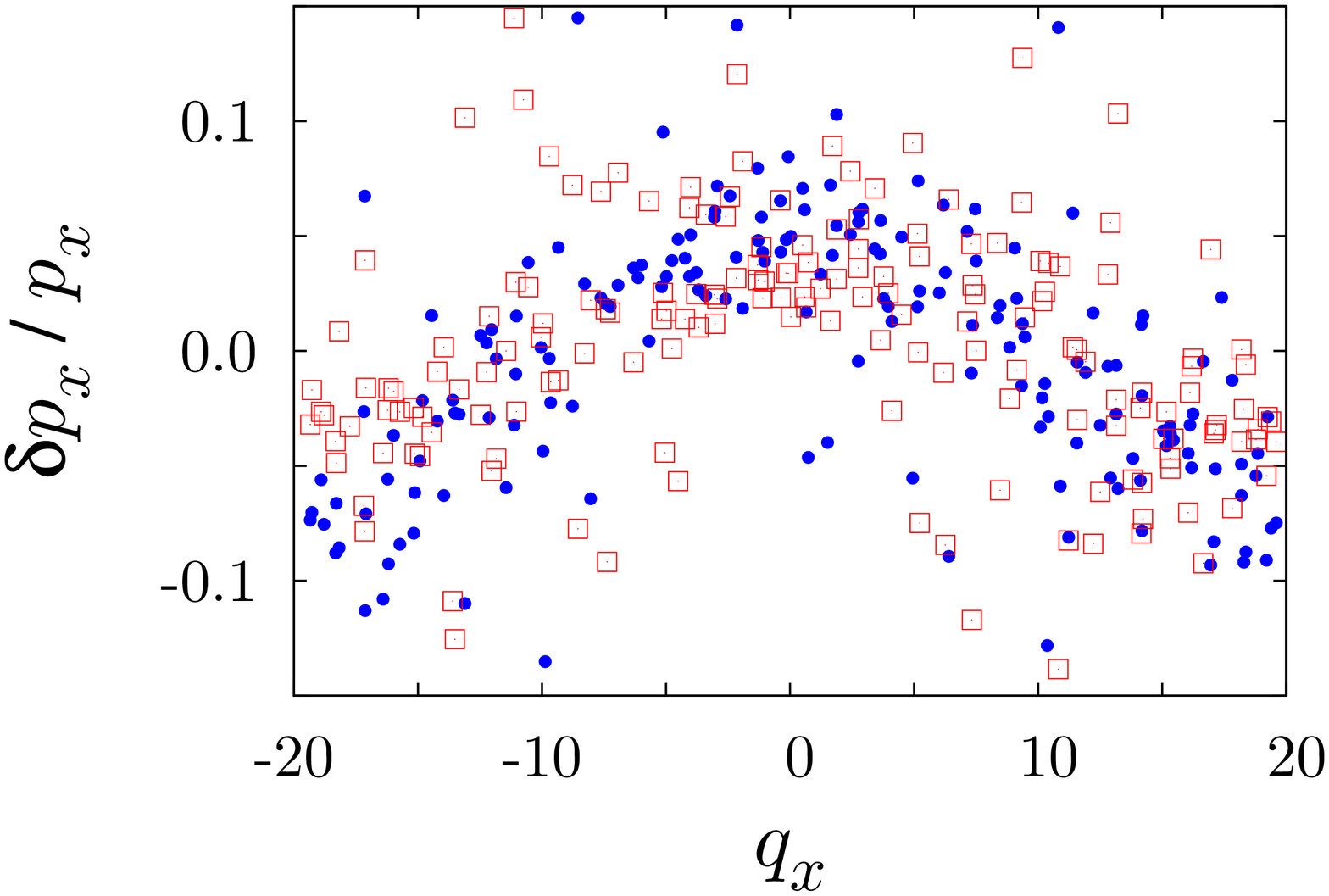}
\end{minipage} \hfil
\begin{minipage}[c]{.45\linewidth}
\includegraphics[angle=-90,width=1\textwidth]{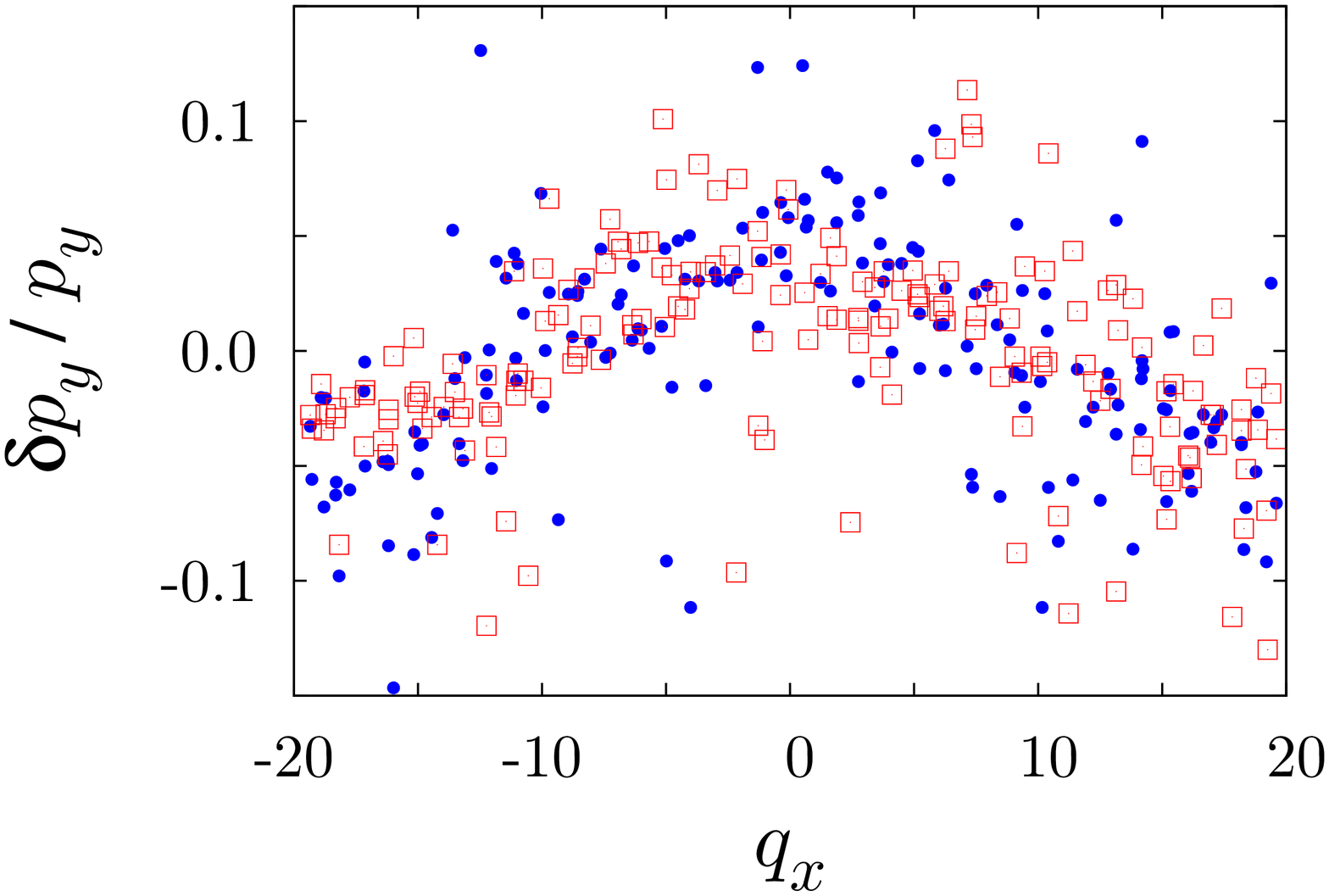}
\end{minipage} \hfil
P(1,0)\\
\caption{Reconstructed momentum perturbations for the pure L(1,0) mode 
(top panels) and  P(1,0) mode (bottom panels) depicted already in Fig. \ref{Figure_11}.
Only the patterns proportional to $\cos( 2 \pi q_x / L_x )$ are shown.
Blue dots: Gram-Schmidt vectors; Red squares: covariant vectors.}
\label{Figure_12}
\end{figure}
Longitudinal (L) and associated momentum (P) modes share the same degenerate Lyapunov exponent
$\lambda^{(i)}$, and generally appear superimposed in experimental vectors. With a rectangular box 
and periodic boundaries, they form four-dimensional LP perturbations. The 
superposition varies periodically with time. This ``dynamics'' has been identified as a rotation of the 
pure L and P vectors in the standard frame. For details we refer to previous work in Ref. \cite{Eckmann:2005}.
The patterns for the pure L mode are easily recognizable as sine and cosine functions, 
but those for the P modes are not.
As is evident from the spanning vectors for L(1,0) and P(1,0) also listed in Table \ref{mode_table}, 
the P modes are proportional
to the  instantaneous velocities of all particles, which does not at all constitute a smooth vector field.
For a pattern to be recognizable, these velocities need to be ``divided out''.  A full mode reconstruction is
required as is described for the case of Gram-Schmidt vectors  in Ref {\cite{Eckmann:2005}. Here
we carry out an analogous reconstruction in terms of the covariant vectors and compare them to the
GS modes. In Fig. \ref{Figure_11} two of the reconstructed patterns for L and P modes belonging to the
four-dimensional LP(1,0) subspace with indices $i \in \{388, 389, 390, 391\}$ are shown.   
The blue dots are for GS modes, the red squares for covariant modes.  To judge the
quality of the reconstruction, we have included in the top-right panel also the 
$\delta q_y$-versus-$q_x$ curve, which vanishes nicely as required. 

For comparison, Fig. \ref{Figure_12} gives results for a completely analogous reconstruction,
where instead of the position perturbations as in Fig. \ref{Figure_11},  the corresponding 
momentum perturbations are used. For this example cosine patterns were selected,
whereas in Fig. \ref{Figure_11} sine patterns were used. As before, blue dots refer to
GS vectors, red squares to covariant vectors.

\subsection{Transversality}
\label{hyperbolicity}

From the Lyapunov spectrum of Fig. \ref{spectrum} and the magnification of its central part
in Fig. \ref{Figure_8}, the following inequalities are read off,
\begin{equation}
\lambda_1>\cdots \ge \lambda_{2N-3} >\left[ \lambda^{(0)} \right] > \lambda_{2N+4}  \ge \cdots>\lambda_{4N}
\enspace ,
\end{equation}
where the equal sign applies for the degenerate exponents belonging to modes.
$\left[\lambda^0\right] = 0$ is sixfold degenerate in our case. Conjugate pairing assures that
 $\lambda_i = - \lambda_{2N+1 - i}$.
The Oseledec splitting provides us with the following structure of the tangent space,
\begin{equation}
\mathbf{TX}=
\vec{E^u} \oplus \; {\cal N} \oplus \vec{E^s},
\end{equation}
where  ${\vec E}^u = \vec v^{1} \oplus \cdots \oplus \vec v^{2N-3}$ and 
${\vec E}^s = \vec v^{2N+4} \oplus \cdots \oplus \vec v^{4N}$ are the covariant stable and unstable subspaces, 
respectively, and ${\cal N}$ is the null subspace or central manifold. The question arises whether the
system is hyperbolic, which implies that the  angles between the
stable manifold ${\vec E}^s$ and the unstable manifold ${\vec E}^u$  are bounded away
from zero for all phase points (Due to the existence of a central manifold this is referred to as 
partial hyperbolicity in the mathematical literature \cite{Bochi:2002}). Even more, we may ask 
whether the angles between all Oseledec subspaces and, hence between all covariant vectors,  
are bounded away  from zero for all phase space points. To find an answer to that question, we
compute in the following the scalar products for all covariant vector pairs and present representative results.
(This procedure reminds us of the so-called coherence angles introduced by 
d 'Alessandro and Tenenbaum \cite{Alessandro:1995,Alessandro:2000},
measuring the angular distance between a physically interesting direction and the direction of 
maximum perturbation expansion).

The lines in Figure \ref{Figure_13} depict the product  norms
$\langle | {\vec v}^j\cdot {\vec v}^i | \rangle $ for selected covariant vectors ${\vec v}^j$ with all other covariant vectors ${\vec v}^i,\; i \ne j$. As before, a time average is performed.
The panels on the left-hand side provide three examples for ${\vec v}^j$ from the
unstable manifold outside of the mode regime ($j = 1, 200$, and $370$ from top-left to 
bottom-left, respectively),  and similarly on the right-hand side from the stable manifold 
outside of the mode regime ($j = 420, 600$, and $792$ from 
bottom-right to top-right, respectively). 
\begin{figure}[htbp]
\centering
\begin{minipage}[c]{.45\linewidth}
\includegraphics[angle=-90,width=1\textwidth]{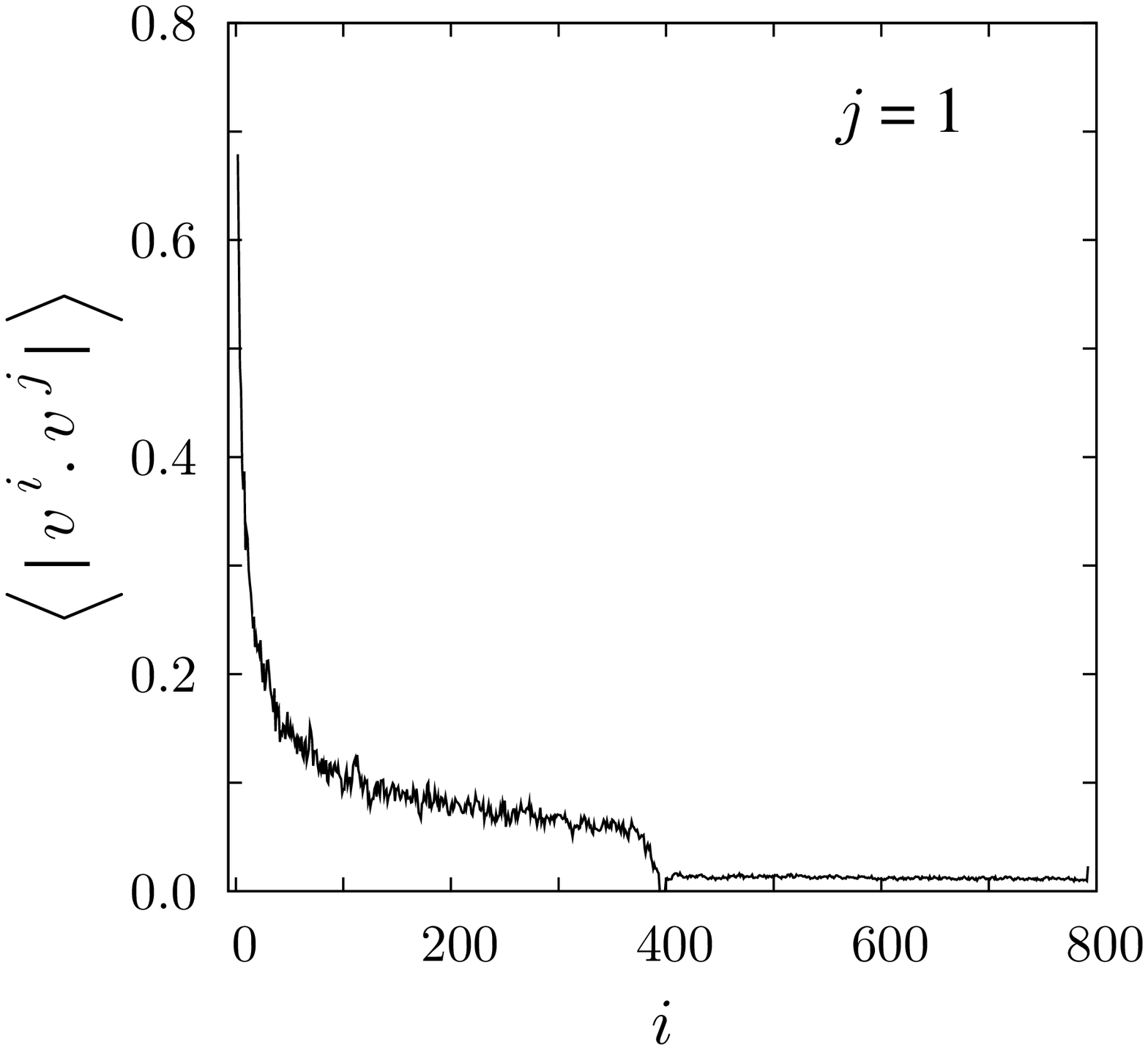}
\end{minipage} \hfil
\begin{minipage}[c]{.45\linewidth}
\includegraphics[angle=-90,width=1\textwidth]{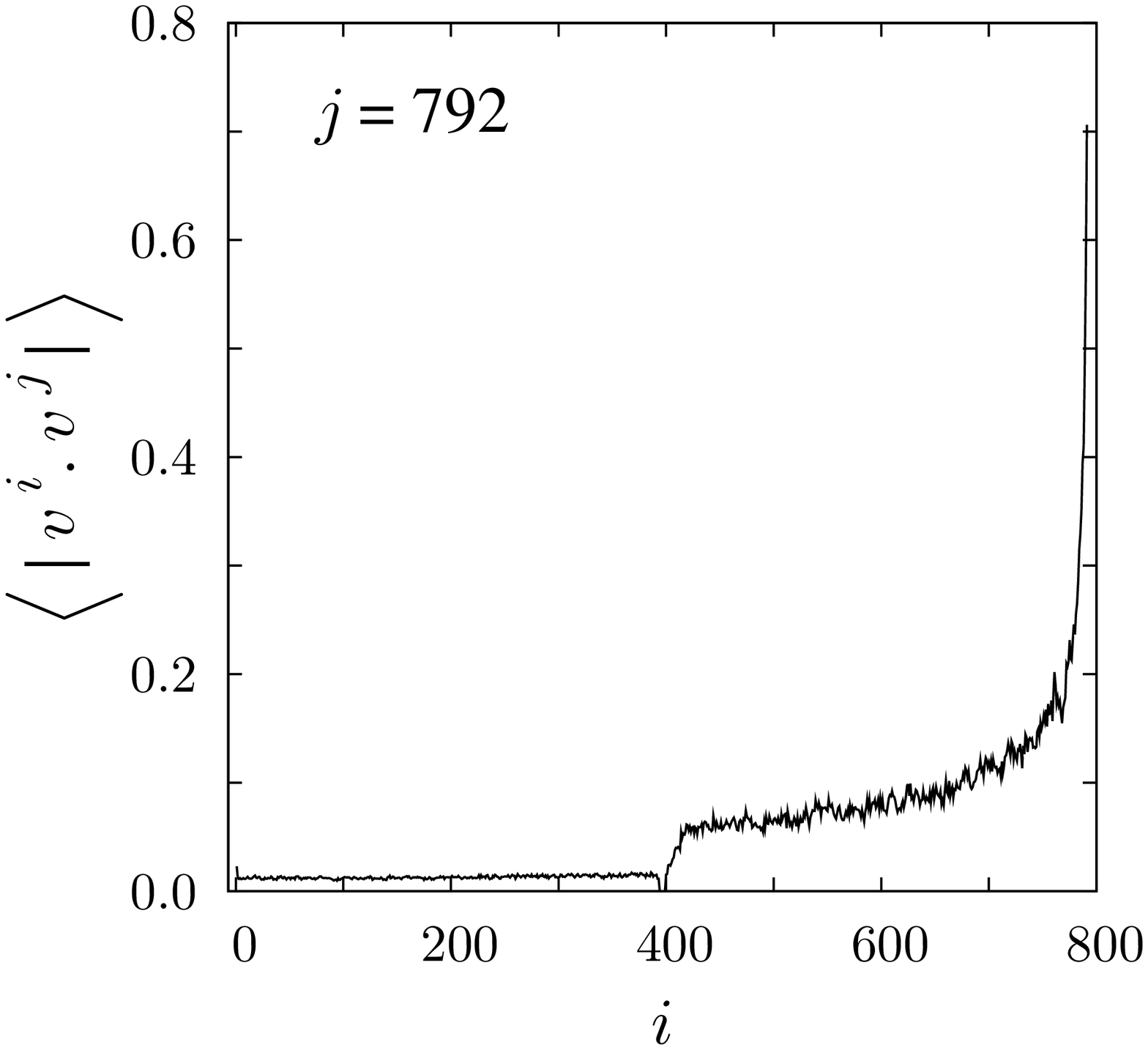}
\end{minipage} \\
\begin{minipage}[c]{.45\linewidth}
\includegraphics[angle=-90,width=1\textwidth]{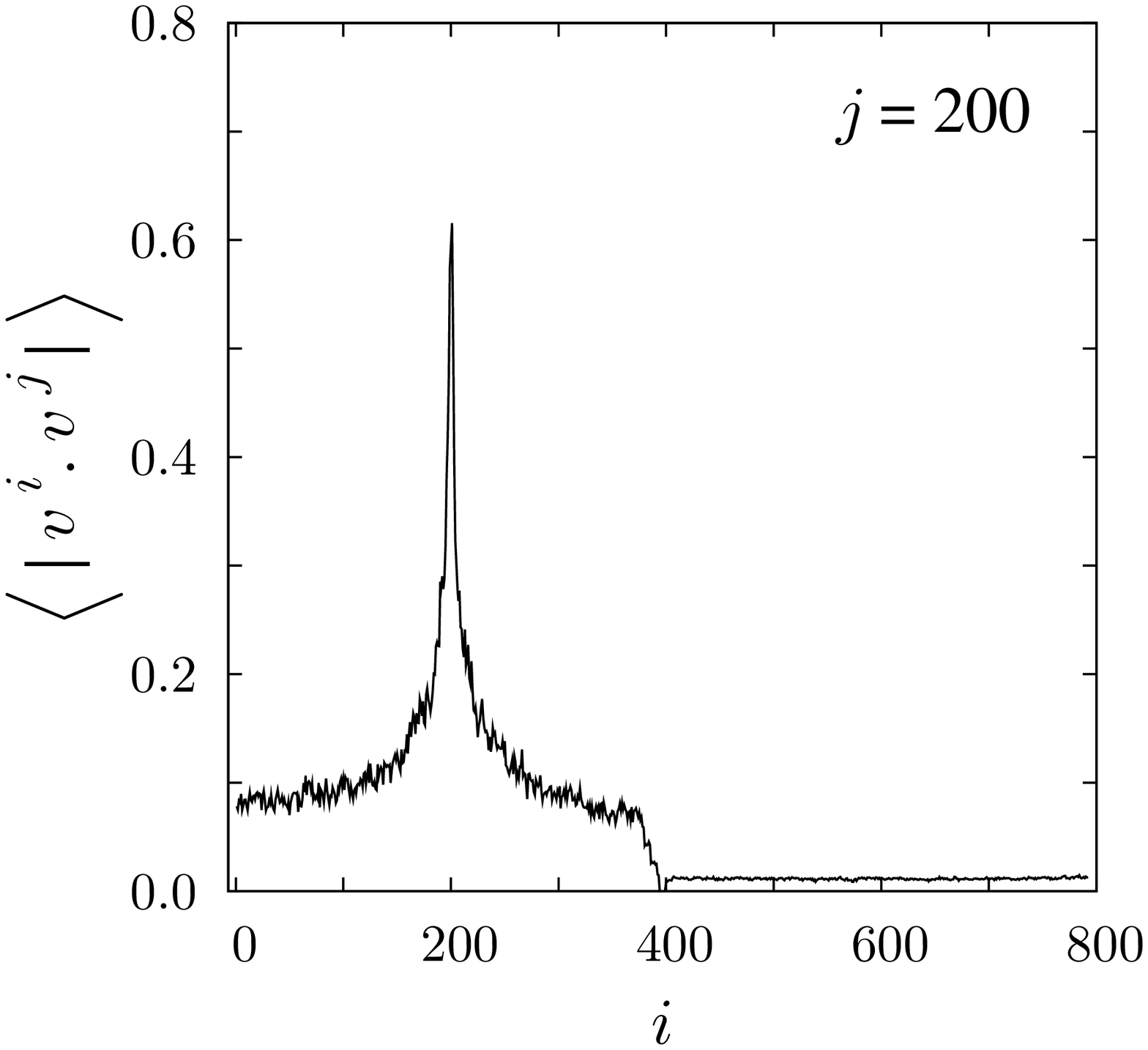}
\end{minipage}\hfil
\begin{minipage}[c]{.45\linewidth}
\includegraphics[angle=-90,width=1\textwidth]{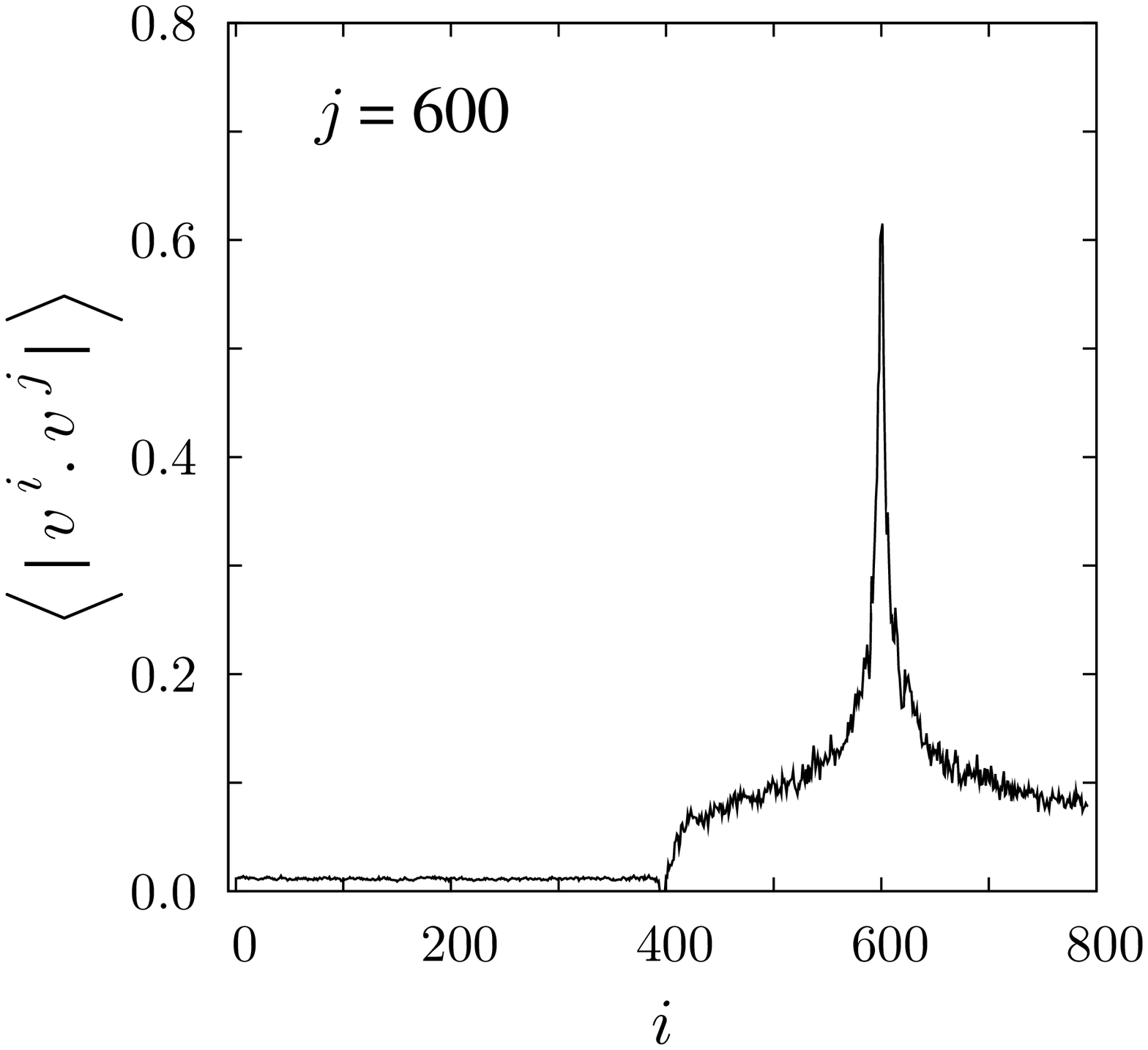}
\end{minipage} \\
\begin{minipage}[c]{.45\linewidth}
\includegraphics[angle=-90,width=1\textwidth]{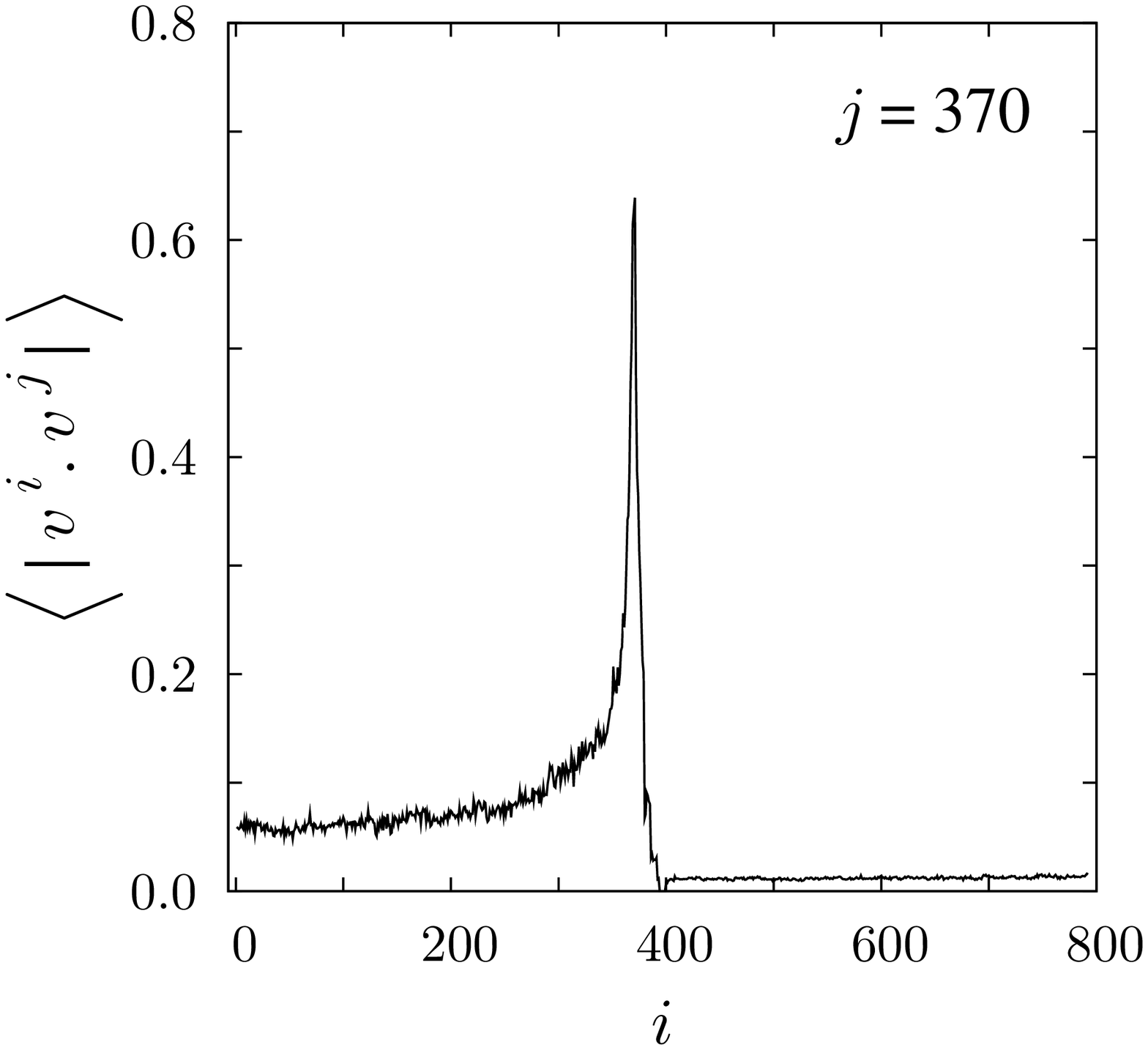}
\end{minipage}\hfil
\begin{minipage}[c]{.45\linewidth}
\includegraphics[angle=-90,width=1\textwidth]{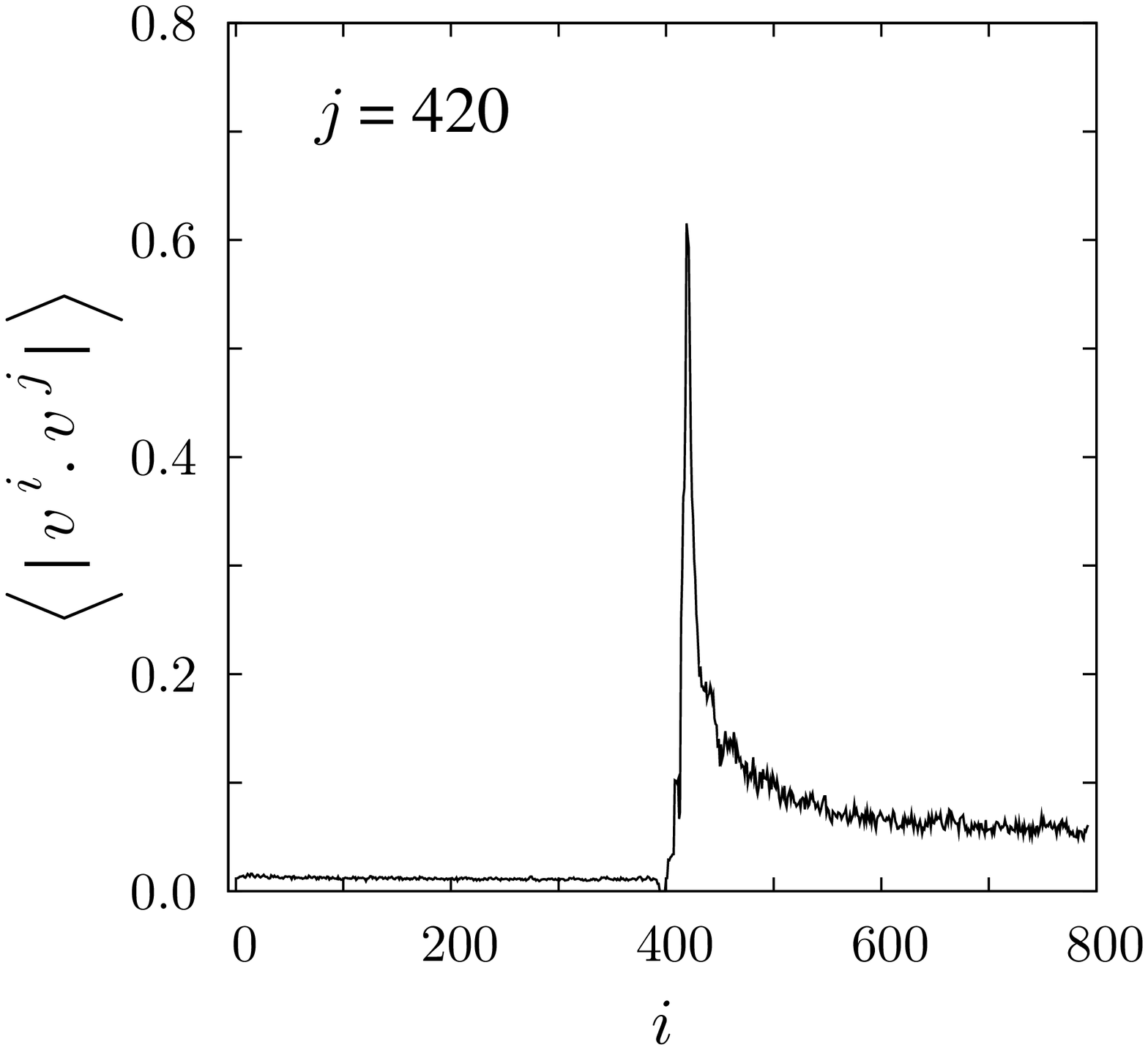}
\end{minipage}
\caption{The  lines are time averaged ($100$ frames separated by 150 time units) 
absolute values of the scalar product of a specified covariant vector ${\vec v}^j$ 
with all the remaining covariant vectors ${\vec v}_{i\ne j}$ as a function of $i$. 
Left panels from top to bottom:
$j = 1, 200$, and $370$ belonging to the unstable manifold; Right panels from bottom to top:
$j = 420, 600$, and $792$ from the  stable manifold.}
\label{Figure_13}
\end{figure}
One immediately observes that the stable and unstable subspaces are not orthogonal.  As has
been mentioned in   Section \ref{vanishing_exponents} and is also convincingly demonstrated in
the following Fig. \ref{Figure_14}, the null subspace ${\cal N}$ is orthogonal 
to both ${\vec E}^u$ and ${\vec E}^s$. For two covariant vectors from the same
subspace, ${\vec E}^u$ or ${\vec E}^s$, however,  the scalar product does not vanish indicating 
considerable nonorthogonality. But at the same time it is also well bounded away from 
unity, which means that the two vectors do not become parallel either. However, one 
possible exception may be the
covariant vector pairs $({\vec v}^j, {\vec v}^{j+1})$ for adjacent Lyapunov exponents
in the spectrum. In these cases, the scalar product reaches a pronounced maximum in all of the 
six panels of Fig. \ref{Figure_13} which may still allow these vectors to become
parallel occasionally. This will be discussed further below.

So far we have considered only vectors ${\vec v}^j$ outside of the mode regime.
The case of covariant vectors representing modes is treated separately 
in Fig. \ref{Figure_14},
\begin{figure}[htbp]
\centering
\begin{minipage}[c]{.45\linewidth}
\includegraphics[angle=-90,width=1\textwidth]{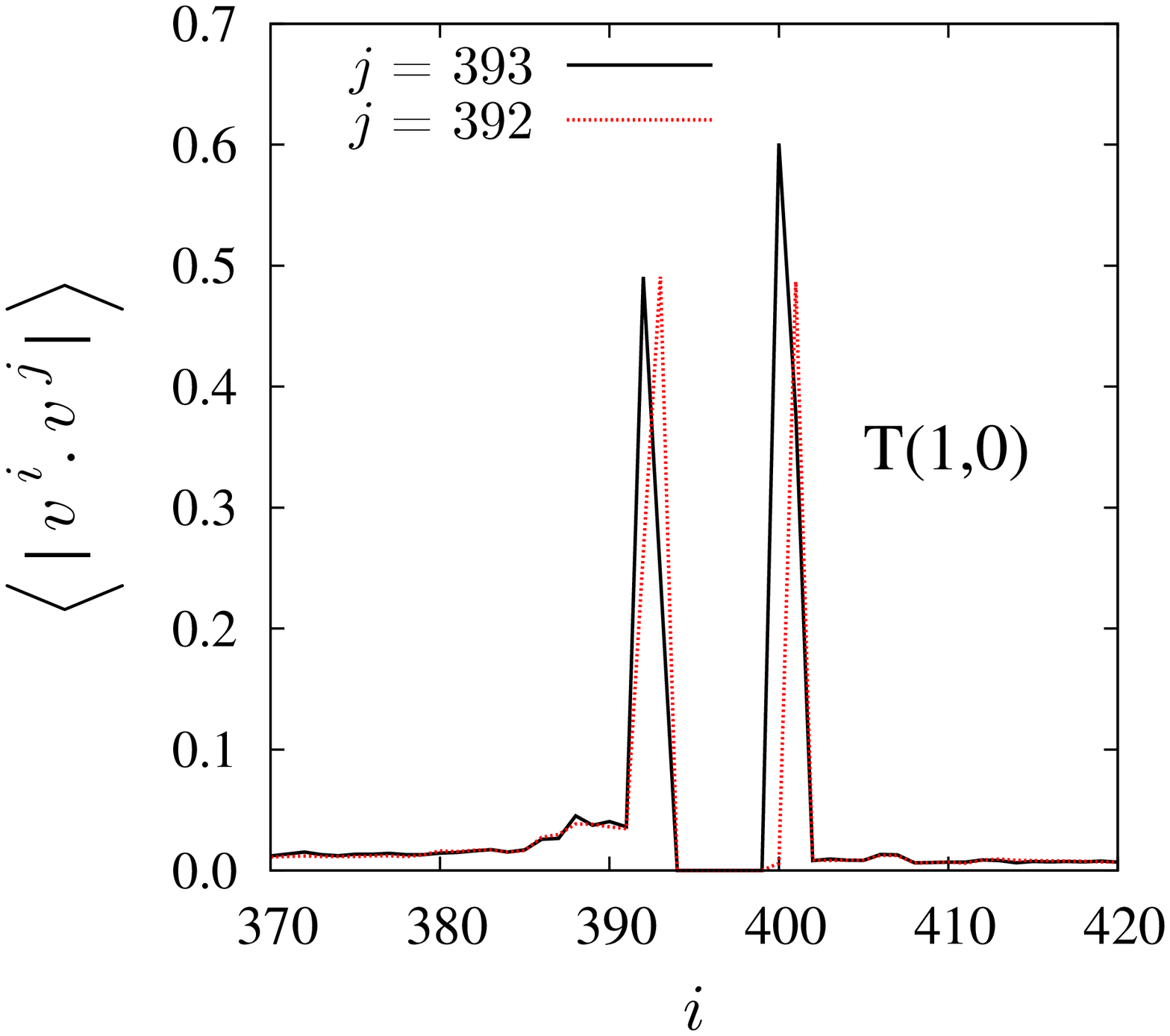}
\end{minipage}\hfill
\begin{minipage}[c]{.45\linewidth}
\includegraphics[angle=-90,width=1\textwidth]{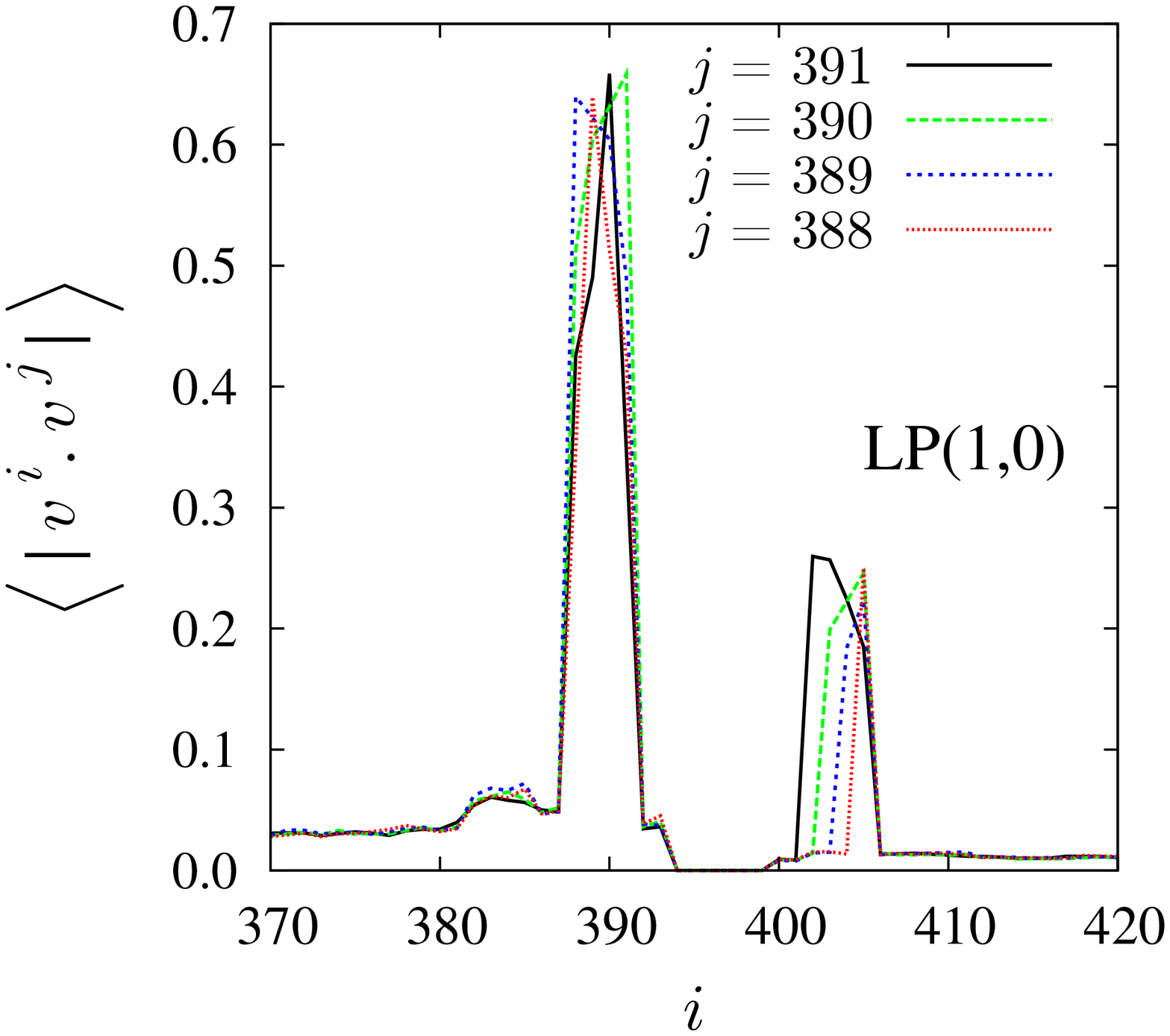}
\end{minipage}\\
\begin{minipage}[c]{.45\linewidth}
\includegraphics[angle=-90,width=1\textwidth]{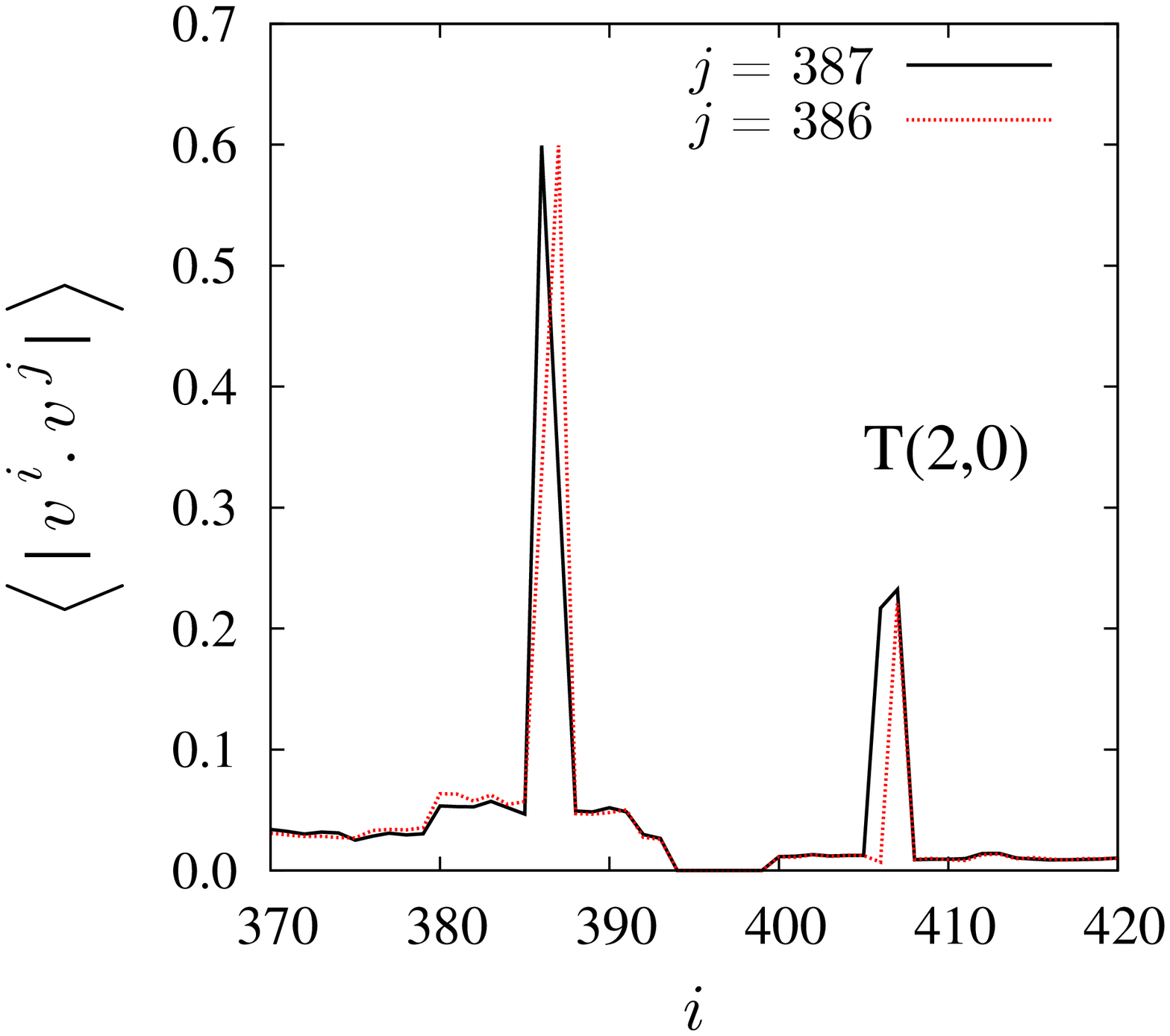}
\end{minipage}\hfill
\begin{minipage}[c]{.45\linewidth}
\includegraphics[angle=-90,width=1\textwidth]{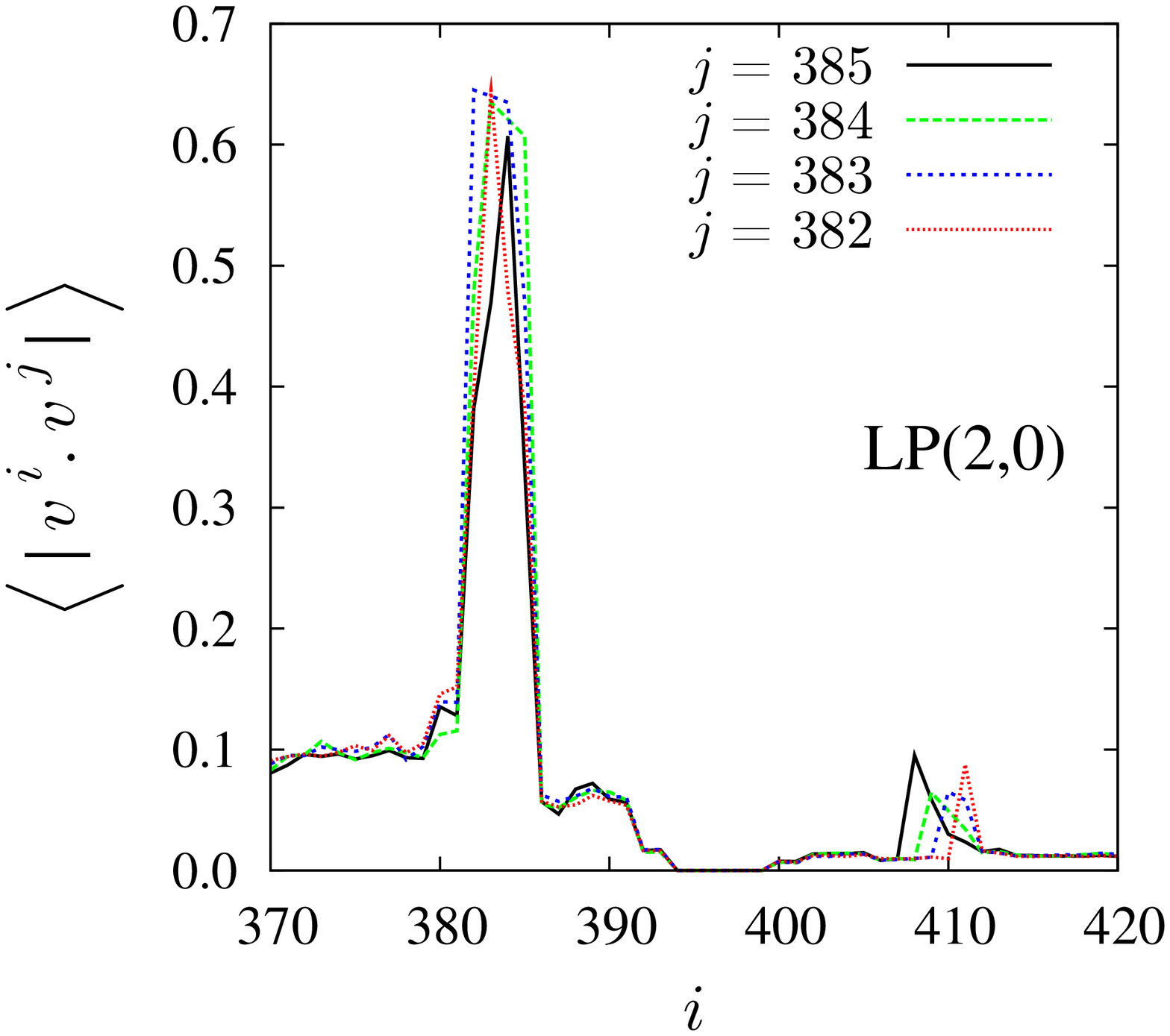}
\end{minipage}\\
\begin{minipage}[c]{.45\linewidth}
\includegraphics[angle=-90,width=1\textwidth]{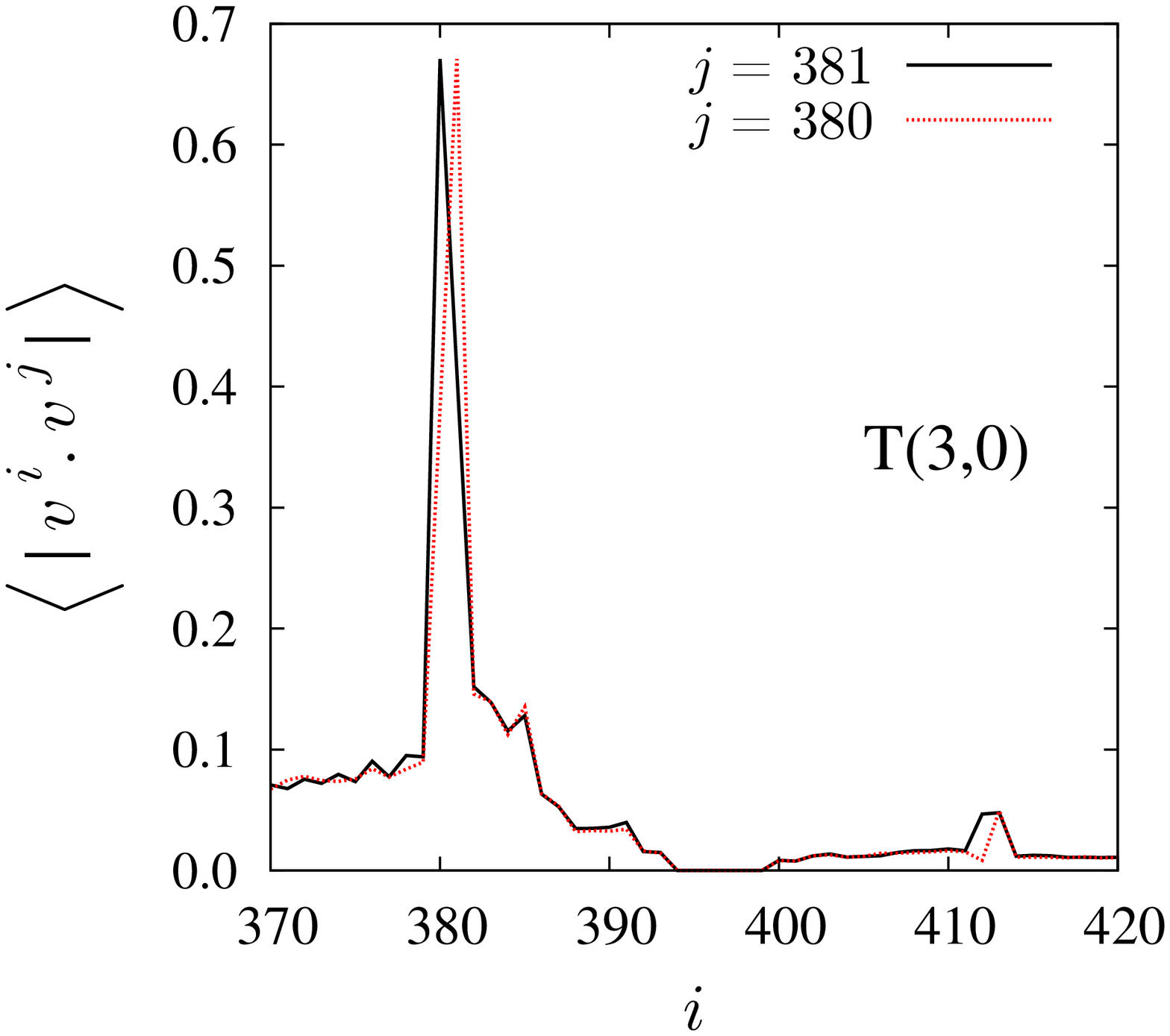}
\end{minipage}\hfill
\begin{minipage}[c]{.45\linewidth}
\includegraphics[angle=-90,width=1\textwidth]{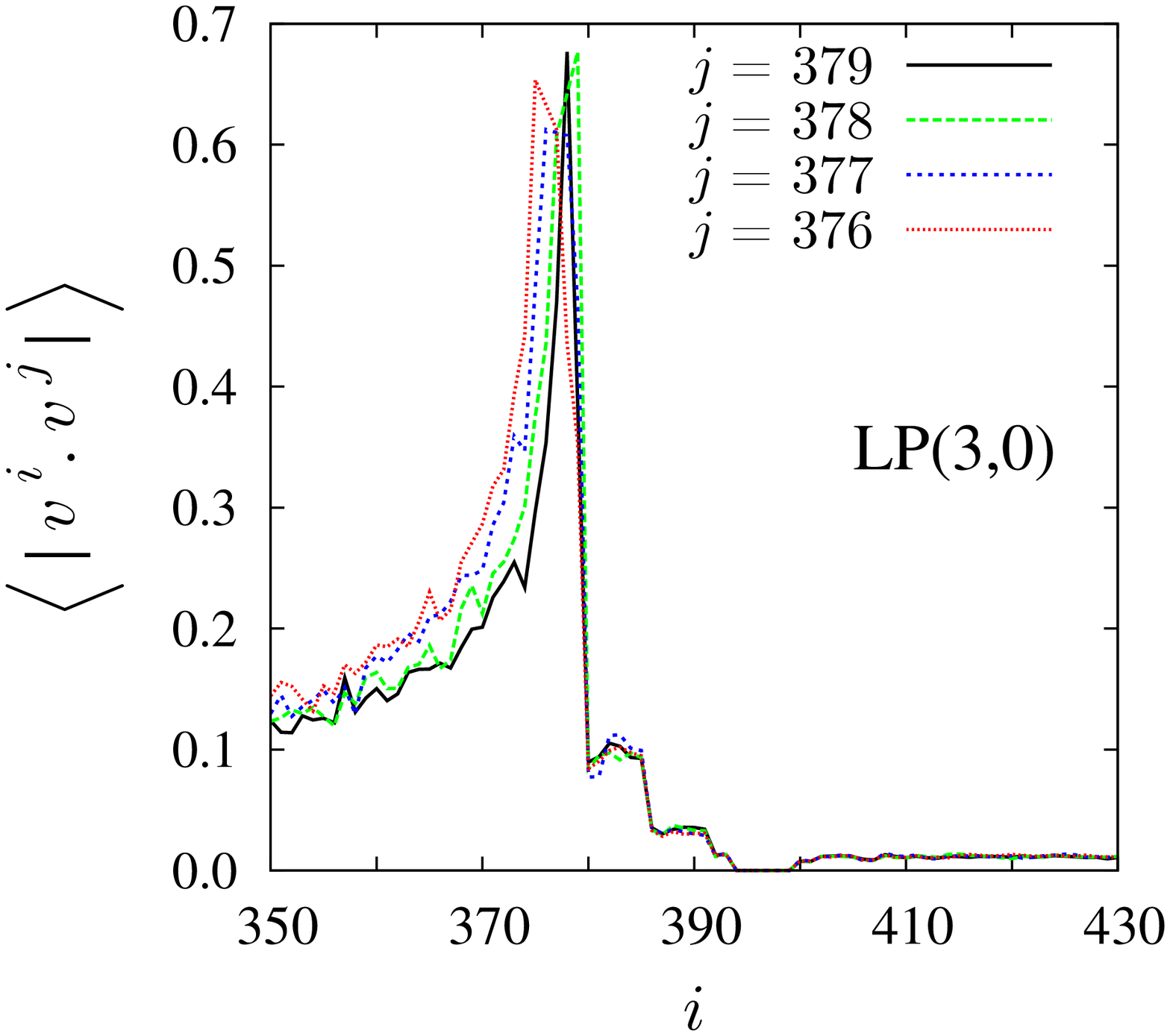}
\end{minipage}
\caption{The lines are time averaged ($100$ events separated by 150 time units) 
absolute values for the scalar product of covariant vectors ${\vec v}^j$
(as specified by the label $j$)  
with all the remaining covariant vectors ${\vec v}_{i\ne j}$ as a function of $i$.
The abscissa is restricted to the mode regime.
$j$ is for modes from the unstable subspace with positive exponents only.
For the respective conjugate modes from the stable subspace, the 
curves in all panels are just the mirror images around the center as in Fig. 
\ref{Figure_13}.}
\label{Figure_14}
\end{figure}
where, as before, time-averaged scalar product norms $\langle | {\vec v}^j\cdot {\vec v}^i | \rangle $   
for $i \ne j$  are plotted as  a function of $ i$. The standard deviation is too small to be included
in the plots. The panels on the left-hand side are for $j$ belonging
to  unstable transversal modes, the panels on the right-hand side for $j$ belonging
to unstable LP pairs. The curves for the conjugate stable modes just look like the mirror images 
around the central index. Each vector representing a T or LP-mode has significant contributions 
to the scalar product only for covariant vectors belonging to the same degenerate exponent and -- to a lesser extent -- the corresponding conjugate (negative) exponents (where the latter is not 
true anymore for the
LP(3,0) modes in the bottom-right panel of Fig.  \ref{Figure_14}, where no peak
around $i=414$ is discernible).   

\begin{figure}[htbp]
\centering
\begin{minipage}[c]{0.8\linewidth}
\includegraphics[angle=-90,width=0.9\textwidth]{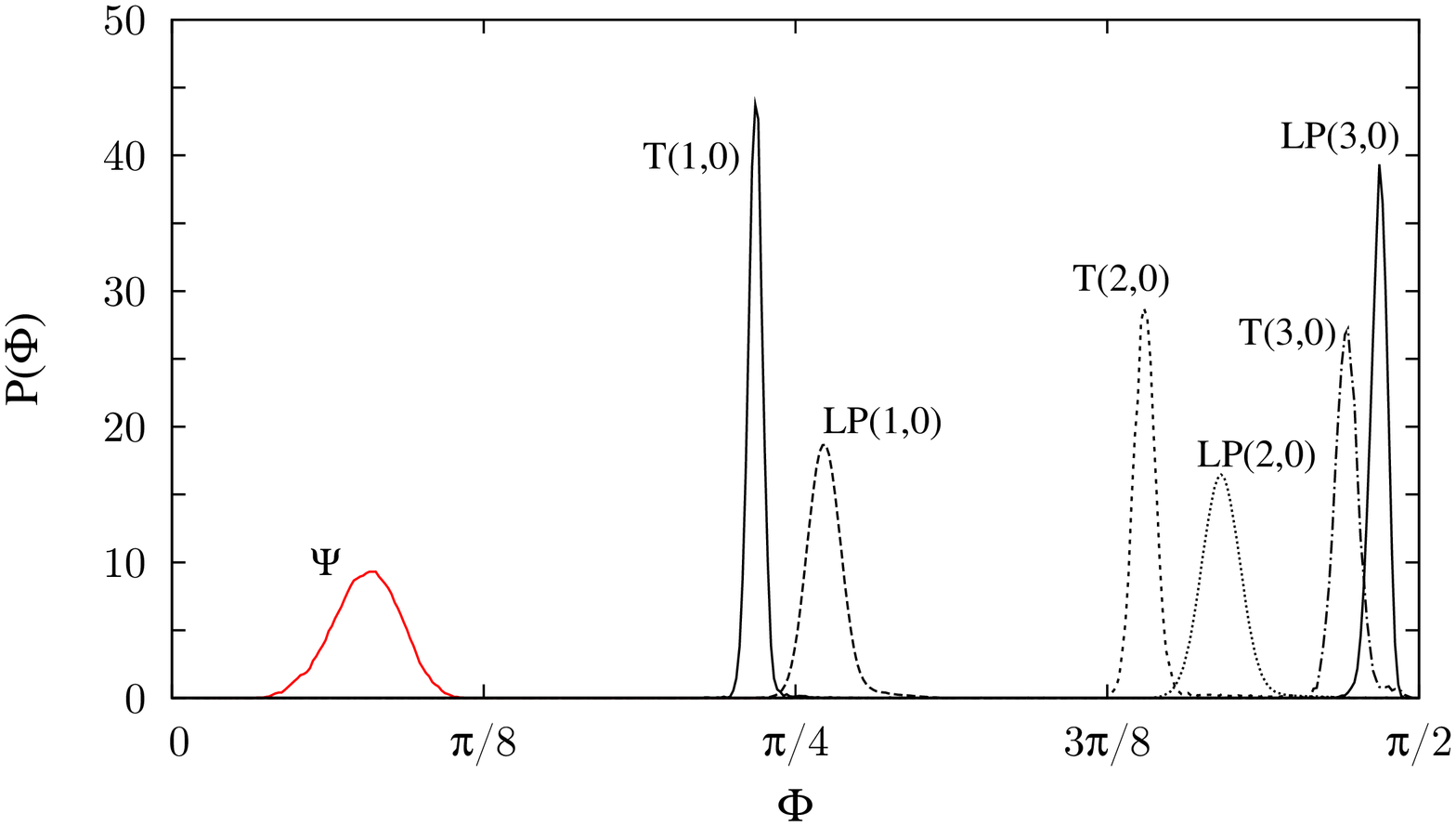}
\end{minipage}
\caption{Probability distributions for the minimum angle between the Oseledec subspace
${\vec E}^{(X)}$ and its conjugate subspace ${\vec E}^{(X)*}$.  Here,
X $\in \{ \mbox{T}(n_x,0), \mbox{LP}(n_x,0) \}$ with $n_x = 1,2,3$ specifies the modes
as indicated by the labels. The probability distribution $\Psi$ for the minimum angle between
the stable and unstable manifolds ${\vec E}^s$ and ${\vec E}^u$ is shown by the red line.}
\label{Figure_15}
\end{figure}
The covariant  vectors belonging to transverse   (or to LP) modes span covariant
Oseledec subspaces ${\vec E}^{(i)}$ with a dimension $m^{(i)}$ equal to 2 (respective 4).
To ease the notation, we refer to them as ${\vec E}^{(X)}$ in the following, where $X$ is 
either  T$(n_x ,0)$ or  LP$(n_x, 0)$ with $n_x \in \{1,2,3\}$. 
The conjugate Oseledec subspaces, ${\vec E}^{(X)*}$,  have the same dimension 
and are spanned by the respective conjugate covariant vectors.
Fig. \ref{Figure_14} shows that the covariant vectors spanning any of the subspaces 
${\vec E}^{(X)}$ or ${\vec E}^{(X)*}$ have a rather small but finite angular distance and, thus,
are transversal. The Oseledec subspaces representing modes are themselves  transversal 
to all other subspaces of the Oseledec splitting, but to a varying degree. 
The angular distances in tangent space are
generally large except between conjugate subspaces ${\vec E}^{(X)}$ and ${\vec E}^{(X)*}$, for which
the scalar products of their spanning vectors may become surprisingly large. 

To check more carefully for transversality even in this case, we show in  Fig. \ref{Figure_15}
 the probability distribution for the minimum angle between the conjugate subspaces 
 ${\vec E}^{(X)} $ and  ${\vec E}^{(X)*} $, where $X$ stands for the T and LP modes  as
indicated by the labels.   This angle $\Phi$ is computed from the smallest principal angle
between the two subspaces \cite{Kuptsov,Bjoerck}. If the covariant vectors belonging to 
 ${\vec E}^{(X)}$  and   ${\vec E}^{(X)*} $ are arranged as the column vectors of matrices 
 ${\vec V}$ and ${\vec V}^*$, respectively, the QR decompositions ${\vec V} = {\bf Q} {\bf R} $ 
 and  ${\vec V}^* = {\bf Q}^* {\bf R}^* $ of the latter provide matrices ${\bf Q}$ and ${\bf Q^*}$, with which 
 the matrix ${\bf M} = {\bf Q}^T {\bf Q}^*$ is constructed. The singular values of ${\bf M}$ are
 equal to the cosines of the principal angles, of which $\Phi$ is the minimum angle. Since
 $\Phi$ is never very small, this method works well and does not
 need more complicated refinements \cite{Kuptsov,Bjoerck,Knyazev}.    
 It is seen that  all distributions are well bounded
away from zero indicating transversality for the respective subspaces. 

      Finally,  we concentrate on the minimum angle between the full unstable subspace
${\vec E}^{u} = {\vec v}^1 \oplus \cdots \oplus {\vec v}^{393}$
and its conjugate  stable counterpart      ${\vec E}^{s} = {\vec v}^{400} \oplus \cdots \oplus {\vec v}^{792}$,
using the same method as before.  These subspaces include the mode-carrying vectors
studied before. The probability distribution for the minimum angle  is denoted by
$\Psi$ and is also shown  in Fig. \ref{Figure_15} (red line).  
Also this distribution is well bounded away from zero  and indicates transversality between ${\vec E}^{s}$ and 
${\vec E}^u$. We conclude that for finite $N$ the hard-disk systems are 
(partially) hyperbolic in phase space. 

\begin{figure}[ht]
\centering
\includegraphics[angle=-90,width=0.5\textwidth]{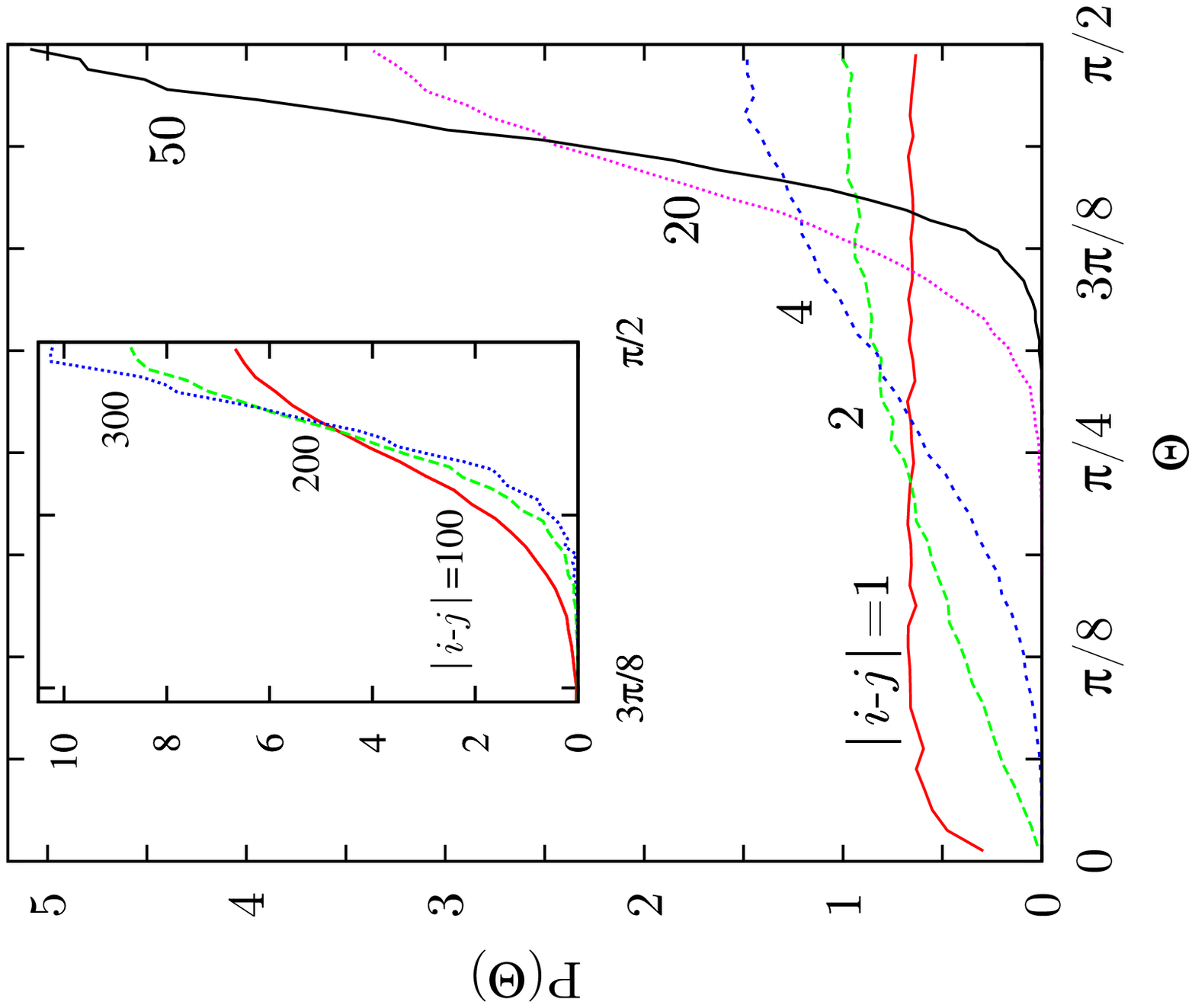}
\caption{Probability distributions for the angles $\Theta$ between all covariant vectors
${\vec v}^i$ and ${\vec v}^j$ from ${\vec F}^{(370)} = {\vec v}^1 \oplus \cdots \oplus {\vec v}^{370}$ with prescribed separation $i-j$ of their indices as indicated by the labels.}
\label{probability}
\end{figure}
In Fig. \ref{Figure_13} it was observed that the scalar products between 
covariant vectors with adjacent indices are rather large and possibly may allow tangencies.
To study this point more carefully, we follow a suggestion of G. Morriss and
consider the angle $\Theta = \cos^{-1} | {\vec v}^i \cdot {\vec v}^j |$ between the vectors ${\vec v}^i \in {\vec F}^{(J)}$ and   ${\vec v}^j \in {\vec F}^{(J)}$, for which $i-j$ is a specified positive integer. 
The probability distributions for angles with $i-j = 1, 2, 4, 20, 50$ are shown in Fig. \ref{probability},
and for $i-j = 100, 200, 300 $ in the inset of the same figure. Whereas the probabilities
for $i-j > 1$ are bounded away from zero, the distribution for $i-j = 1$ seems to converge to zero  for
$\Theta \to 0$.

\begin{figure}[ht]
\centering
\includegraphics[angle=-90,width=0.6\textwidth]{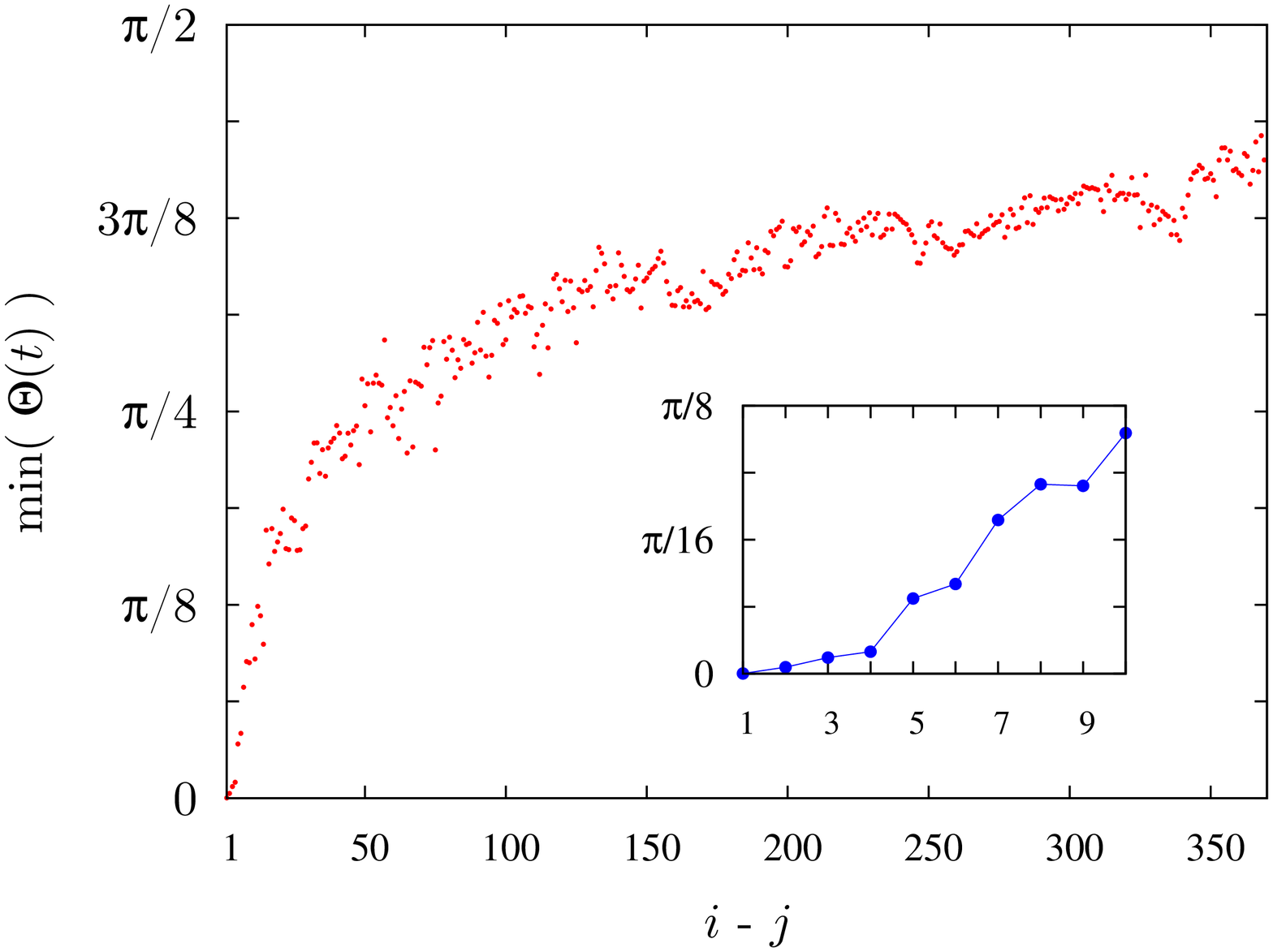}
\caption{Plot of the minimum angle between different covariant vectors  ${\vec v}^i$  and ${\vec v}^j$ from the unstable subspace without the mode-affected directions, 
${\vec F}^{(370)} = {\vec v}^1 \oplus \cdots \oplus {\vec v}^{370}$,  as a function of their
index difference $i-j$.}

\label{Figure_17}
\end{figure}
    An even more demanding test is given in Fig \ref{Figure_17},
where the minimum of $\Theta$ for given $i-j > 0$ is plotted as a function of $i-j$.  The inset
provides a magnification of the most  interesting region. One observes that the minimum of the angle
$\Theta$ between covariant vectors specifying Oseledec subspaces with $i-j = 1$ may indeed become very small, but this happens with extremely small probability. Our 
numerical evidence is consistent with the assumption that the angle becomes zero with vanishing probability.

\section{Conclusion}
\label{summary}

    A comparison of the covariant vectors with corresponding orthonormal Gram-Schmidt vectors reveal
similarities, but also significant differences. The vectors associated with the maximum Lyapunov 
exponent are identical, ${\vec v}^1 = {\vec g}^{1}$, and also the leading vectors in the 
central manifold agree, ${\vec v}^{2N-2} = {\vec g}^{2N-2}$. All the other corresponding vectors
generally point into
different tangent space directions. Whereas the GS vectors are pairwise orthogonal by construction,
the covariant vectors are not. Most notably, the perturbation contributions from the
particles' positions and momenta are significantly different and even 
exhibit a different symmetry between vectors from the stable and unstable manifold
as in Fig. \ref{Figure_6}. For the covariant vectors these contributions
agree in accordance with the time-reversal symmetry required for them, whereas for the
Gram-Schmidt vectors these contributions are interchanged. Another significant
difference is the degree of localization in physical space for the non-degenerate
perturbations. As Fig. \ref{Figure_5} shows, the covariant vectors are 
much more localized than the GS vectors in accordance with the fact that they are
not dynamically constrained by re-orthogonalization.

     From a theoretical point of view, an interesting result is that no
tangencies occur between the respective unstable and stable manifolds ${\vec E}^u$ and 
${\vec E}^s$. In Fig. \ref{Figure_15} the probability distribution $\Psi$ of the  minimum
angle between stable and unstable subspaces (including the Lyapunov modes)
is well bounded away from zero, and even more so for the vectors belonging to unstable
respective stable modes. Thus, a hard disk system with $N = 198$ particles  as
in our case is (partial) hyperbolic  for all points in phase space. 
  We even find
that  all Oseledec subspaces are pairwise transversal with non-vanishing angles between them. 

   We speculate  that for $N \to \infty $ the distribution for the minimum angle between 
${\vec E}^u$ and ${\vec E}^s$ may possibly reach the origin in Fig. \ref{Figure_15}. To clarify this
point  further studies are required \cite{Hadrien}. 
 
     The concept of hyperbolicity is closely linked with  the notion of  dominated Oseledec splitting
for all phase space points  \cite{Bochi:2002}.  We may rewrite Eq. (\ref{ccclya}) for the
Lyapunov exponents, expressed in terms of the covariant vectors, according to
\begin{equation}
 \lambda_{\ell} = \lim_{N \rightarrow  \infty} \frac{1}{N} \sum_{n=0}^{N-1}\frac{1}{\tau}
     \ln \big\|  D \phi^{\tau} \vert_{{\bf \Gamma}(t_n)} \; {\vec v}^{\ell}({\bf \Gamma}(t_n) )\big\|,
     \label{intcov}
\end{equation}     
where $t_n \equiv n \tau$, and $\tau$ is the short time interval between consecutive 
re-normalizations of the covariant vectors. Here, $\lambda_{\ell}$  is expressed as a time average
of a quantity
\begin{equation}
\Lambda_{\ell}^{\mbox{cov}}({\vec \Gamma}(t_n)) =   \frac{1}{\tau}
     \ln \big\|  D \phi^{\tau} \vert_{{\bf \Gamma}(t_{n-1})} \; {\vec v}^{\ell}({\bf \Gamma}(t_{n-1}) )\big\|,
\end{equation}
which is referred to as local or (time-dependent) Lyapunov exponent, and is a function of the
instantaneous phase point ${\vec \Gamma(t)}$. The Oseledec splitting is said to be dominated,
if the local Lyapunov exponents, when averaged over a 
finite time $\Delta$, do not change their order in the spectrum for any $\Delta$ larger 
than some {\em finite} $\Delta_0 > 0$ \cite{Hadrien}. This is a very strong condition on the
fluctuations of the local exponents \cite{YR2008,BPDH2010}. For symplectic systems it is known that
the domination of the splitting implies that the system is (partially) hyperbolic \cite{Bochi:2002}. But 
it is not clear whether the converse is true in our case. The discussion of this point is deferred to a forthcoming 
publication \cite{Hadrien}.

   The number and the dimension of the Oseledec  subspaces are 
constant in phase space. There is no entanglement of subspaces, which has been
identified as one of the main reasons for the occurrence of  well established Lyapunov modes
\cite{YR2008}.  We refer to Ref. \cite{BPDH2010} for a discussion of a simple 
but physically-relevant model, for which the dimensions of the stable and unstable
manifolds frequently change along the trajectory.   
   
      An interesting extension of this work is  the study of rough hard particles allowing for energy exchange
 between translational and rotational degrees of freedom \cite{Hadrien}. Arguably, this is
 the simplest model of a molecular fluid. No 
 Lyapunov modes are found in this case \cite{vMP2009}. An analysis in terms of
 covariant vectors is presently under way and will be published separately.

\section{Acknowledgements}  We dedicate this work to Peter H\"anggi on the occasion of
his 60th birthday. His insight and enthusiasm for science is a continuous source of inspiration. 
We also gratefully acknowledge stimulating discussions with 
Francesco Ginelli, Gary Morriss, Antonio Politi, G\"unter Radons, and Hong-liu Yang.
Our work was supported by the Austrian Wissenschaftsfonds (FWF),
grant P 18798-N20.


\begin{thebibliography}{10}

\bibitem{Ginelli} F. Ginelli, P. Poggi, A. Turchi, H. Chat\'e, R. Livi, and A. Politi,  
                        Phys, Rev. Lett. {\bf 99}, 130601 (2007). 
\bibitem{HHP:1990} W. Hoover, C. Hoover, and H.A. Posch, Phys. Rev. A {\bf 41}, 2999 (1990).          
\bibitem{Szasz} D. Sz\'asz, editor, {\em Hard Ball Systems and the Lorentz Gas},
              Encyclopedia of Mathematical Sciences  {\bf 101},  Springer, Berlin,  2000.          
\bibitem{Barker} J.A. Barker and  D. Henderson, J. Chem. Phys/ {\bf 47}, 4714 (1967).
\bibitem{Hansen} J.-P. Hansen and I. R. McDonald, {\em Theory of simple 
                 liquids}, (Academic Press, London, 1991).
\bibitem{Dellago}  Ch. Dellago, {\em    Using Lyapunov weighted path sampling to identify 
                   rare chaotic and regular
                   trajectories in dynamical systems}, preprint (2010); this volume.
\bibitem{Eckmann:2005} J.-P. Eckmann, Ch. Forster, H.A. Posch, and E. Zabey,
            J. Stat. Phys. {\bf 118}, 813-847 (2005).             
\bibitem{Henk} A. de Wijn and H. van Beijeren, Phys. Rev. E {\bf 70}, 016207 (2004).  
\bibitem{Oseledec:1968} V.I. Oseledec, Trudy Moskov. Mat. Obsc. {\bf 19}, 179, (1968). 
                         English transl. Trans. Moscow Math. Soc. {\bf 19}, 197 (1968).
\bibitem{Ruelle:1979} D. Ruelle, Ergodic theory of differentiable dynamical systems,
             Publications Math\'ematiques de l'IH\'ES {\bf 50}, 27-58 (1979).
\bibitem{Eckmann:1985} J.-P. Eckmann and D. Ruelle, Rev. Mod. Phys. {\bf 57}, 617 (1985).
\bibitem{Benettin} G. Benettin, L. Galgani, A. Giorgilli, and J.-M. Strelcyn,
              Meccanica {\bf 15}, 21 (1980).
\bibitem{Shimada} I. Shimada and T. Nagashima, A numerical approach to ergodic
             problem of dissipative dynamical system, Prog. Theor. Phys. {\bf 61}, 1605 (1979).
\bibitem{recipes} W.H. Press, S.A. Teukolsky, T. Vetterling, and B.P.Flannery,
              {\em Numerical Recipes in Fortran 77: The Art of Scientific Computing}, 2nd Edition,
                 Cambridge University Press, Cambridge, 1999. 
\bibitem{Ershov} S.V. Ershov and A.B. Potapov, Physica D {\bf 118}, 167 (1998).
\bibitem{Legras} B. Legras and R. Vautard, Proceedings of the Seminar on
               Predictability, Vol. 1, ECWF Seminar, edited by T. Palmer, p. 1 (EC MWF Reading, UK, 1996).
\bibitem{Henon}  M. H\'enon, Comm. Mathem. Phys. {\bf 50}, 69 (1976).
\bibitem{BP2010} H. Bosetti and H.A. Posch, in preparation.
\bibitem{DPH1996} Ch. Dellago, H.A. Posch, and W.G.Hoover,  Phys. Rev. E {\bf 53}, 1485 (1996).
\bibitem{PH2000} H.A. Posch and R. Hirschl, {\em Hard Ball Systems and the Lorentz Gas
             (Encyclopedia of Mathematical Sciences vol. 101}, edited by D. Szasz
              (Springer Berlin, 2000), p. 279.
\bibitem{FHPH2004} Ch. Forster, R. Hirschl, H. A. Posch, and Wm. G. Hoover,
             Physica D {\bf 187}, 294 (2004).
\bibitem{TM2003a} T. Taniguchi and G.P. Morriss, Phys. Rev E {\bf 68}, 026218 (2003).              
\bibitem{TM2003b} T. Taniguchi and G.P. Morriss, Phys. Rev E {\bf 68}, 046203 (2003).   
\bibitem{DP1997} Ch. Dellago and H.A. Posch,  Physica A, {\bf 240}, 68  (1997).
\bibitem{MP2002}    Lj. Milanovi\'c and H.A. Posch, J. Mol. Liquids {\bf 96} - {\bf 97}, 221 (2002).
\bibitem{PF2002} H.A. Posch and Ch. Forster, {\em Lecture Notes on Computational Science
       -- ICCS 2002}, ed. P.M.A. Sloot, C.J.K. Tan, J.J.Dongarra, and A.G. Hoekstra, p.1170 (Springer
         Verlag, Berlin, 2002).
\bibitem{FP2005} Ch. Forster and H. A. Posch, New Journal of Physics, {\bf 7}, 32 (2005).
\bibitem{Manneville} P. Manneville, {\em Lecture notes in Physics} {\bf 230}, 319 (Springer-Verlag,
             Berlin, 1985).
\bibitem{LR1989} R. Livi and S. Ruffo, {\em Nonlinear Dynamics}, ed. G. Turchetti, p. 220,
        World Scientific, Singapore, 1989.          
\bibitem{FMV1991} M. Falcioni, U.M.B. Marconi, and A. Vulpiani, Phys. Rev. A {\bf 44} 2263 (1991).
\bibitem{Astrid}  R. van Zon and H. van Beijeren, Journal of Statist. Phys. {\bf 109}, 641 (2002).
\bibitem{TMXXX} T. Taniguchi  and G.P.Morriss, Phys. Rev. E {\bf 73}, 036208 (2006).
\bibitem{Pikovsky} A. Pikovsky and A. Politi, Phys. Rev. E {\bf 63}, 036207 (2001).
\bibitem{Hadrien} Hadrien Bosetti, Ph.D.-Thesis, University of Vienna (2010).
\bibitem{Gaspard} P. Gaspard, {\it Chaos, Scattering, and  Statistical
         Mechanics}, Cambridge University Press, Cambridge, 1998. 
\bibitem{HPFDZ2002} Wm.G. Hoover, H.A. Posch, Ch. Forster, Ch. Dellago, and M. Zhou,
               J. Statistical  Physics, {\bf 109}, 765 (2002). 
\bibitem{MPH1998a} Lj. Milanovi\'c, H.A. Posch and Wm. G. Hoover, Mol. Phys. {\bf 95}, 281 (1998).
\bibitem{MPH1998b} Lj. Milanovi\'c, H.A. Posch and Wm. G. Hoover,  Chaos {\bf 8}, 455 (1998)
\bibitem{RY2004} G. Radons and H.-L. Yang, arXiv:nlin.CD/0404028.
\bibitem{YR2004} H.-L. Yang and G. Radons, Phys. Rev. E {\bf 71}, 036211 (2005).
\bibitem{EG2000} J.-P. Eckmann and O. Gat, J. Stat. Phys. {\bf 98}, 775 (2000).
\bibitem{TM2002} T. Taniguchi and G.P. Morriss, Phys. Rev. E {\bf 65}, 056202 (2002).
\bibitem{TDM2002} T. Taniguchi, C.P. Dettmann, and G.P. Morriss, J. Stat. Phys. {\bf 109}, 747 (2002).
\bibitem{NM2001} S. McNamara and M. Mareschal, Phys. Rev E {\bf 64}, 051103 (2001).
\bibitem{MN2004} M. Mareschal and S. McNamara, Physica D {\bf 187}, 311 (2004).
\bibitem{Bochi:2002} J. Bochi and M. Viana, Ann. I. H. Poincar\'e, {\bf 19}, 1 (2002). 
\bibitem{Alessandro:1995} M. D'Alessandro and A. Tenenbaum, Phys. Rev. E {\bf 52}, R2141 (1995).
\bibitem{Alessandro:2000} M. D'Alessandro, A. D'Aquino, and A. Tenenbaum, Phys. Rev. E {\bf 62}, 4809 (2000).
\bibitem{Kuptsov} P.V. Kuptsov and S.P. Kuznetsov, Phys. Rev. E {\bf 80}, 016205 (2009).
\bibitem{Bjoerck} A. Bj\"orck and G.H. Golub, Mathem. of Computation {\bf 27}, 579 (1973).
\bibitem{Knyazev} A.V. Knyazev and E.M. Argentati, SIAM J. Sci. Comput. {\bf 23}, 2008 (2002).
\bibitem{YR2008}  H.-L. Yang and G. Radons,  Phys. Rev. Lett. {\bf 100}, 024101 (2008). 
\bibitem{BPDH2010} H. Bosetti, H.A. Posch, C. Dellago, and Wm.G. Hoover, submitted (2010);
                arXive:1004.4473.
\bibitem{vMP2009} J. van Meel and H.A. Posch, Phys. Rev. E {\bf 80}, 016206 (2009).


\end{thebibliography}
\end{document}